# Application of Information Spectrum Method on Small Molecules and Target Recognition

Milan Senćanski*, Neven Šumonja, Vladimir Perović, Sanja Glišić, Nevena Veljković and Veljko Veljković

*Laboratory for Bioinformatics and Computational Chemistry, Institute of Nuclear Sciences Vinca, University of Belgrade, Mihajla Petrovica 12-14, 11001 Belgrade, Serbia*

*Corresponding author

email: sencanski@vin.bg.ac.rs

**Abstract**

Current methods for investigation of receptor-ligand interactions in drug discovery are based on three-dimensional complementarity of receptor and ligand surfaces, and they include pharmacophore modelling, QSAR, molecular docking etc. Those methods only consider short-range molecular interactions (distances <5Å), and not include long-range interactions (distances >5Å) which are essential for kinetic of biochemical reactions because they influence the number of productive collisions between interacting molecules. Previously was shown that the electron-ion interaction potential (EIIP) represents the physical property which determines the long-range properties of biological molecules. This molecular descriptor served as a base for the development of the informational spectrum method (ISM), a virtual spectroscopy method for investigation of protein-protein interactions. In this paper, we proposed a new approach to treat small molecules as linear entities, allowing the study of the small molecule - protein interaction by ISM. We analyzed here 21 sets of KEGG drug-protein interactions. We showed that this new approach allows efficient discrimination between biologically active and inactive ligands and consistency with AA regions of their binding site on the target protein.

Keywords: ISM method, CIS spectra, small molecules, smiles notation, target-ligand recognition, protein target regions

## 1. Introduction

Receptor - ligand interactions in biological systems are considered as any non-covalent interactions between protein and small molecule, antibody-antigen, protein-peptide, enzyme-substrate etc. They play a key role all in biological processes in a living cell, for they mediate the whole signal paths. Therefore, the research of those phenomena is highly important as well from biophysical view, to chemical, pharmacological and medicinal. The foundations of the modern understanding of receptor-ligand interactions are based on their short contacts and surface complementarity (short-range molecular interactions), where specific non-covalent interactions form and trigger the important processes in the cell. Modern drug design methods that deal with receptor-ligand interactions are mostly oriented towards three-dimensional (3D) approaches, based on direct interactions of ligand and receptor molecules. They are divided into two groups: structure-based and ligand-based, depending on if the structure of receptor is known or not, respectively. Approaches such as 3D QSAR and pharmacophore modelling belong to ligand-based, while molecular docking and molecular dynamics belong to structure-based drug design methods [Silverman and Holladay al, 2014]. They are all founded on Emil Fisher's [Fischer E., 1894] theory, where interactions between the ligand molecule and the target receptor are based on the complementarity of surfaces between receptor and ligand molecules. They occur at distances lower than 5Å and, as stated, represent the current concept of short-range intermolecular interactions. According to collision theory, thermal motions achieve the first contact between interacting molecules accidentally and such generated interactions are considered as short-range intermolecular contacts. However, there are strong disagreements between theory and experimental measurements. Comparison of theoretically calculated rate constants and experimental data yields prior values lower by several orders of magnitude [Smoluchowski, 1916; Northrup and Erickson, 1992]. To overcome those disagreements, several models were proposed, such as reducing the dimensionality of diffusion [Wiegel and DeLisi, 1982; McCloskey and Poo, 1986; Peters, 2005], protein association based on hydrodynamic steering [Brune et al, 1994], desolvation-mediated protein-protein binding [Camacho et al, 2000], etc. One of the most acceptable considerations for this issue was reported by the physicist Frönlich [Fröhlich, 1968; Fröhlich; 1970, Fröhlich, 1975]. From a very general theoretical consideration, he acclaimed that biomolecules are capable to excite dipole vibrations. Based on that vibrations, in polar medium biomolecules mutually develop attractive long-range forces, characterized by specific frequency. Because of common vibration frequencies, interacting molecules produce a larger number of effective collisions than by random encounter. Long-range intermolecular interactions play a key role in receptor-ligand molecular recognition and determine processes in vivo and overall biological activities. Complementary to the structure-based methods, long-range molecular interactions are important on distances above 1000 Å [Leckband and Israelachvili, 2001]. Based on Fröhlich's theory [Fröhlich, 1968; Fröhlich; 1970, Fröhlich, 1975] and general pseudopotential model [Veljkovic V. and Slavic I., 1972], Electron-Ion Interaction Potential (EIIP) descriptor [Veljkovic V., 1980] was developed for small molecules. The EIIP descriptor considers valence electrons on small molecule, and based on molecular formula, calculates two values, the average quasi valence number and EIIP value. Both values are considered as a two-dimensional molecular descriptor, which can be used to cluster similar molecules through targeting the same receptor. However, the EIIP descriptor is limited to small molecules and development of approaches for larger molecules, mostly proteins were demanded. As an extension of EIIP descriptor methodology, the Informational spectrum method (ISM) [Veljkovic et al, 1985], a virtual spectroscopy method was developed. In ISM method, each amino acid (AA) reside in the protein sequence is coded with corresponding EIIP value, and corresponding data graph EIIP value vs. ordinal number of AA is constructed. Applying the Discrete Fourier Transformation on this data one obtains corresponding amplitudes and frequencies as output data, which is used to construct ISM spectrum. According to ISM [Veljkovic et al, 1985], two biological macromolecules (proteins) interact if multiplication of their ISM spectra (consensus informational spectrum, CIS) yield one or more common frequencies, which determines amino acid regions in both proteins, responsible for molecular recognition on long distances. Thus, one can explain whether two proteins interact or not, or it can be used for prediction and peptide design for a specific target. The ISM method has been fully developed for proteins and nucleic acids [Veljkovic et al, 1985]. For small molecules, candidate prediction for a specific target is limited to their selection according to ranges of EIIP descriptor values, determined from molecular formulas of known target ligands. Using EIIP descriptor one can select small molecule candidates from numerous databases under criteria if they fall in the appropriate value range. This approach is important in design and selection of new drug candidates. It can be considered as an initial step in drug design because reducing the number of potential candidates saves computational time and effort in 3D structure-based discovery processes coupled in one common long-range-short-range virtual screening protocol. It is also useful in drug re-purposing discovery because drugs multi-targeting can be predicted [Veljkovic et al, 2015]. However, a question on small molecule raises through the improvement of EIIP method and possible applying of ISM method. Therefore, for the first time, we propose a new approach to treat molecules as quasi-linear entities, analogous to peptides and application of ISM method to predict potential candidates for a specific target. We based our research on GPCR drugs of Golden dataset and used amino acid sequences of corresponding receptors. The ISM treatment of small molecules as pseudo peptides shown consistency with the treatment of peptides, yielding specific binding regions in GPCR. Physically, we showed that the absence of 3D structure information in small molecules is irrelevant for the estimation of their targeting to a specific receptor protein.

## 2. Methods

### 2.1 Linear notation of the small molecule

The idea of linear notation of the small molecule dates from 1988 and was originally reported by Anderson and Weininger [Anderson et al, 1987; Weininger, 1988], through line notation simplified molecular-input line-entry system (SMILES). The idea is based on a fact that each atomic group in linear or chain molecule can be written as one letter, denoting its type. Thus, we have alkyl groups, aromatic carbon atoms, halogen, hydroxyl group etc. Based on that idea, analogously to peptide, a small molecule can be written as a sequence of atomic groups, with specific labelling. For instance, the structural formula of simple hydrocarbon pentane is:

CH3CH2CH2CH2CH3 (2.1.1)

The corresponding smiles notation is:

CCCCC (2.1.2)

If we denote CH3 group in (2.1.1) as P (primary carbon atom) and CH2 group as S (for secondary carbon atom), and if we use the sequence of (2.1.2) to apply those notations on pentane molecule, we obtain:

PSSSP (2.1.3)

There, we wrote a pentane molecule in a sequence. Similar can be derived for chain molecules or those including rings. For instance, smile notation of simple chain molecule of 2-methyl butane is

CC(C)CC (2.1.4)

where (C) denotes methyl group substituent on the second carbon atom. If we label C atoms in (2.1.4), we obtain:

$C^1C^2(C)^3C^4C^5$ (2.1.5)

Where atom superscripts denote their ordinal numbers. Thus, we can write a molecule (2.1.5) as:

PTPSS (2.1.6)

Where T signifies tertiary carbon atom. In the case of ring-containing molecules, which is the vast majority of chemical compounds, for instance, the molecule 3-(piperidine-4-yl)phenol, corresponding smiles notation is:

Oc1cc(ccc1)C2CCNCC2 (2.1.7)

Using the smiles sequence for pseudo peptide generation, if we denote NH group as Y, quaternary carbon atom as Q and OH group as A, we obtain:

AQTQTTTTSSYSS (2.1.8)

Thus, we can use smiles to write small molecules in one dimension and, using specific notations for each atomic group, we can write it as a pseudo peptide. The very first question that arises from and notations (2.1.7) and (2.1.8) is the absence of three-dimensional information of the molecule, i.e. parenthesis from formula (2.1.7). However, through this paper, we will show that loss of a molecular 3D structural information does not affect the physical interpretation of a method proposed here.

**2.2 Electron – ion interaction potential and ISM method**

The concept of electron-ion interaction potential (EIIP) is derived from the “general model pseudopotential” . It assumes that the number of valence electrons and their main energy term of the valence determine long-range properties of the molecule. The EIIP descriptors are easily calculated using the following formulas:

$Z^* = \sum_{i=1}^{m} n_i Z_i / N$ (2.2.1)

$EIIP = 0.25 Z^* \sin(1.04 \pi Z^*) / 2\pi$ (2.2.2)

Where:

- i, Type of the chemical element;
- Z, Valence of the i-th chemical element;
- n, Number of the i-th chemical element atoms in the compound;
- m, Number of types of chemical elements in the compound;
- N, the total number of atoms.

The EIIP values calculated according to the Equation (2.2.2) are in Rydbergs (Ry = 13.6 eV). As stated in the Introduction, the EIIP descriptor is fully developed for small organic molecules. Based on EIIP descriptors, the ranges of their values are borders of 2D chemical space where molecules that target specific receptor belong. There is a strong connection between EIIP and Z* molecular descriptors of small molecules and their biological activities (carcinogenicity, antibiotic activity, antiviral activity, toxicity, etc.) [Veljkovic et al, 2016]. However, EIIP descriptor holds only for small organic molecules, and for larger (peptides for instance) the different method is demanded. As its extension, Informational Spectroscopy Method (ISM) was developed. The ISM is a virtual spectroscopy method for calculation of the long-range properties of biological macromolecules. It is based on a model that assigns to each amino acid value of electron-ion interaction potential (EIIP) descriptor. [Veljkovic et al, 1985]

ISM method consists of three necessary steps:

1. Transformation of protein primary sequence into an array of numbers representing EIIP of each AA residue.

2. Conversion of the numerical array by fast Fourier Transformation into information spectrum, which yields dominant frequency peaks of the molecule

3. Consensus Information Spectrum (CIS) analysis between ISM spectra of two interacting molecules, which generates functional locus of the interaction of two molecules.

Frequencies of peaks in CIS are common for interacting biomolecules. The measure of their similarity to individual IS of interacting molecules is the signal-to-noise ratio (S/N), the ratio between the signal intensity in IS and the CIS spectra. The method has been successful in the identification of functional protein domains representing candidate therapeutic targets for anti-HIV drugs [Veljkovic et al, 2007], anthrax [Doliana et al, 2008], and human influenza viruses [Veljkovic et al, 2009a; Veljkovic et al, 2009b; Perovic et al, 2013]. and is used used in more than 100 research centres worldwide [Veljkovic et al, 2011]. Analogously to EIIP values of AA for ISM method, we isolated from smile notation atomic groups and calculated corresponding EIIP values, which would be used in the calculation of small molecule ISM spectra, and they are presented in Table 1.

Table 1. Atomic groups with corresponding pseudo amino acid labels and EIIP values. Note that duplicate values are due to different element or atomic group representation in smiles.

| Atomic group in smiles | Label for ISM | EIIP value |
|---|---|---|
| 2H | H | 0.004987 |
| 3H | H | 0.004987 |
| As | As | 0.56762 |
| C | C | 0.076673 |
| CH | CH | 0.094603 |
| CH2 | CH2 | 0.01979 |
| CH3 | CH3 | 0.03731 |
| 11CH3 | 11CH3 | 0.03731 |
| N | N | 0.116936 |
| NH | NH | 0.043942 |
| NH2 | NH2 | 0.090387 |
| NH3 | NH3 | 0.01979 |
| O | O | 0.163424 |
| OH | OH | 0.126007 |
| OH2 | OH2 | 0.06933 |
| S | S | 0.163424 |
| SH | SH | 0.126007 |
| cH | CH | 0.094603 |
| n | N | 0.116936 |
| nH | NH | 0.043942 |
| B | B | 0.043942 |
| BH | BH | 0.01979 |
| BH2 | BH2 | 0.049281 |
| Au | Au | 0.429924 |
| Hg | Hg | 0.476523 |
| Sn | Sn | 0.547176 |
| Se | Se | 0.576031 |
| se | Se | 0.576031 |
| P | P | 0.116936 |
| BH3 | BH3 | 0.058626 |
| Si | Si | 0.076673 |
| PH | PH | 0.043941 |
| I | I | 0.004986 |
| 125I | 125I | 0.004986 |
| C=O=O | COO | 0.20993 |
| N=C=S | NCS | 0.116936 |
| S=C=N | SCN | 0.116936 |
| S=O=O | SOO | 0.163424 |
| O=N=O | ONO | 0.074149 |
| N=O=O | NOO | 0.074149 |
| N#N=N | NNN | 0.116936 |
| C=N | CN | 0.151176 |
| N=C | NC | 0.151176 |
| C=O | CO | 0.116936 |
| O=C | OC | 0.116936 |
| c=O | CO | 0.116936 |
| O=c | OC | 0.116936 |
| C=S | CS | 0.116936 |
| c=S | CS | 0.116936 |
| F=O | FO | 0.126007 |
| I=O | IO | 0.126007 |
| N=O | NO | 0.168618 |
| P=O | PO | 0.168618 |
| P=S | PS | 0.168618 |
| C#N | CN | 0.151176 |
| N#C | NC | 0.151176 |
| Br | Br | 0.004987 |
| Cl | Cl | 0.004987 |
| Sc | Sc | 0.043941 |
| Sn | Sn | 0.547176 |
| Se | Se | 0.576031 |
| se | Se | 0.576031 |
| C | C | 0.076673 |
| F | F | 0.004987 |
| I | I | 0.004987 |

| | | | | | | |
|---|---|---|---|---|---|---|
| N | N | 0.116936 | | n | N | 0.116936 |
| O | O | 0.163424 | | o | O | 0.163424 |
| P | P | 0.116936 | | s | S | 0.163424 |
| S | S | 0.163424 | | B | B | 0.043942 |
| c | C | 0.076673 | | PH | PH | 0.043941 |

**2.3 Small molecule databases and protein sequences**

For our research, we used Golden dataset receptor-drug matrix, the standard in current bioinformatics [Yamanishi et al, 2008, Wolf & Grünewald 2015, Vagmita & Li 2012, Rosenbaum et al 2014, Carrieri et al 2001, Scheer et al 1996, Evers & Klabunde 2005, Fraser et al 1989]. However, we limited only to GPCR receptor-drug data, consisting of 222 KEGG drugs and 95 receptors. The number of drugs showing activity per receptor in matrix varies from 1 to 35, and we had to limit the number of receptors to an arbitrary value. We selected only the ones with 10 and more drugs per receptor, yielding 21 receptors. From this matrix, we constructed 21 positive and 21 negative training sets, selecting 10 random drugs per protein of each class. ISM spectra of selected proteins and drugs were calculated using internal laboratory software. Drug spectra of each set were multiplied with spectra of receptor to obtain characteristic frequencies. Protein AA regions were calculated on a basis of characteristic frequencies using laboratory software. Data on binding site AAs for each protein were taken from point mutation experiments literature [Shi L & Javitch 2002]. The amino acid sequences of proteins, as well as drug structures, were downloaded from KEGG database [Kanehisa et al 2017, Kanehisa et al  2016, Kanehisa & Goto 2000]. The drug structural mol files were converted to smiles notation. Explicit hydrogen atoms were added to smiles using Open Babel program [O'Boyle et al, 2011], followed by conversion into canonical form. AQVN and EIIP descriptors values of drugs were also calculated. All data can be found in Supplementary material.

## 3. Results and discussion

The results of ISM spectra multiplications and determination of corresponding AA regions in receptors are presented in Table 2.

Table 2. Dominant ISM frequencies of drug-receptor CIS spectra of 21 GPCR, with relevant protein domains – positive sets

| No | Receptor KEGG code | Receptor name | Major ISM F /approximative F within the error range | Major ISM frequency amplitude | Binding site AAs from crystal structures or point mutations data | Corresponding AA region | Window size |
|---|---|---|---|---|---|---|---|
| 1 | hsa:146 | Alpha-1D adrenergic receptor | 0.062 | 5.56E-13 | 156 | 277 - 292 | 16 |
| 2 | hsa:147 | Alpha-1B adrenergic receptor | 0.055/0.062 | 1.23E-12 | 125 130 | 210 - 225 | 16 |
| 3 | hsa:148 | Alpha-1A adrenergic receptor | 0.062 | 1.22E-13 | 86 106 167 185 | 192- 207 | 16 |
| 4 | hsa:150 | Alpha-2A adrenergic receptor | 0.055/0.062 | 1.07E-13 | 79 117 130 200 201 204 373 | 62 - 87 | 16 |
| 5 | hsa:151 | Alpha-2B adrenergic receptor | 0.048/0.047 | 1.21E-14 | 109 (unchanged effect) | 109-141 | 33 |
| 6 | hsa:152 | Alpha-2C | 0.4375 | 3.73E-15 | 189 381 | 194-209 | 16 |
| | | adrenergic receptor | | | (unchanged effect) | | |
| 7 | hsa:153 | Beta-1 adrenergic receptor | 0.039/0.031 | 2.26E-14 | 104 110 322 | 209-225 | 17 |
| 8 | hsa:154 | Beta-2 adrenergic receptor | 0.043/0.047 | 2.11E-14 | 63 64 69 72 93 109 113 114 117 118 191 192 193 199 200 203 207 271 274 275 289 290 293 312 316 326 331 329 332 | 176-208 | 33 |
| 9 | hsa:155 | Beta-3 adrenergic receptor | 0.039/0.031 | 3.51E-14 | 117 | 188-204 | 17 |
| 10 | hsa:1128 | Muscarinic acetylcholine receptor M1 | 0.40625 | 1.03E-14 | 106 109 110 157 192 196 197 378 381 382 404 407 | 206-222 | 17 |
| 11 | hsa:1129 | Muscarinic acetylcholine receptor M2 | 0.060/0.062 | 1.09E-14 | 80 83 103 104 107 155 177 191 193 400 403 404 410 419 422 426 429 430 | 113-128 | 16 |
| 12 | hsa:1131 | Muscarinic acetylcholine receptor M3 | 0.055/0.062 | 1.87E-13 | 114 142 148 149 156 200 240 504 507 508 523 530 534 | 154 - 169 | 16 |
| 13 | hsa:1812 | D(1A) dopamine receptor | 0.060/0.062 | 4.55E-14 | 70 198 199 202 205 286 | 105-120 | 16 |
| 14 | hsa:1813 | D(2) dopamine receptor | 0.058/0.062 | 5.93E-14 | 100 114 118 386 389 390 408 416 | 117-132 | 16 |
| 15 | hsa:1814 | D(3) dopamine receptor | 0.0625 | 8.07E-13 | 110 111 183 345 346 349 | 113-128 | 16 |
| 16 | hsa:3269 | Histamine H1 receptor | 0.334/0.328 | 1.37E-13 | 107 108 112 158 428 431 432 435 | 188-220 | 33 |
| 17 | hsa:3351 | 5-hydroxytryptamine-1B receptor | 0.293/0.297 | 1.15E-13 | 129 130 133 134 137 200 201 216 327 330 331 337 351 359 | 60-92 | 33 |
| 18 | hsa:3352 | 5-hydroxytryptamine-1D receptor | 0.287/0.281 | 1.65E-13 | 189 196 | 312-344 | 33 |
| 19 | hsa:3356 | 5-hydroxytryptamine-2A receptor | 0.342/0.344 | 8.68E-15 | 155 156 228 229 336 339 340 366 370 | 123-139 | 17 |
| 20 | hsa:3358 | 5-hydroxytryptamine-2C receptor | 0.285/0.282 | 9.70E-14 | 134 135 139 208 209 215 327 328 335 350 | 97-129 | 33 |
| 21 | hsa:3577 | High affinity interleukin-8 receptor A/CXCR1 | 0.082/0.078 | 6.93E-17 | 128 178 188 191 192 200 247 251 268 269 270 273 | 218-250 | 33 |

During the interpretation of the results, we had to apply a few conditions. Regarding frequency values, they had to be assigned to the closest value within error range with minimum possible window size, to get the most specific domain in the receptor, responsible for the ligand-target recognition. Also, it is not meaningful to interpret domains that do not fall into the AA range that corresponds to the membrane region. Therefore, some domains are recognized as “main” although they do not have the highest amplitude values

Our results show good agreement with experimentally determined binding site AAs, obtained from GPCR database [Pándy-Szekeres et al 2017], regarding positive sets of drugs. The frequency values are such they can be approximated to the values corresponding to narrow, hence specific regions of the receptors (window sizes are from 16 to 33 AA residues). They occupy either precise binding site AA residues or at least are structurally nearby. For instance, in case of the beta-2 adrenergic receptor, the main domain perfectly matches the binding site. While, for instance in the case of 3356, the domain spatially suits AA residues from nearby TMs. There are also noticeable patterns of target recognition domain. Thus, the pattern of occupying the ECL2 and TM5 is noticeable in cases of hsa:152, hsa:153, hsa:154 and hsa:155. Other domain patterns are expressed in cases of covering solely TM5 or TM3, as in cases of hsa:146, hsa:147, hsa:148 and hsa:1129, hsa:1131, hsa:1812, hsa:1813 and hsa:1814, respectively. In some cases, binding domains covers ICL2 (hsa:151 and hsa:1129) and other ICL or ECLs. What also drives attention is that receptors from the same class have a similar pattern of target recognition domain and similar main F values range (Table 1). Therefore, from the biophysical point, this method of determination of the receptor binding site, taking as input information only receptor AA sequence and smiles notation of small molecule is fully justified.

Regarding negative sets (Table 4), it is noticeable that F values are different and amplitude values are generally from several times to several orders of magnitude lower than values of the positive set. However, there are exceptions, such as hsa:151, hsa:152, hsa:155, hsa:1128, hsa:3356 and hsa:3577. However, we determined binding site domains for these cases. They differ from positive sets by two traits: window sizes are wider hence less specific (for instance window size 65), and corresponding domains are further from the experimentally determined important AA residues.

The reason for a false positive can be explained on the example of hsa:151. The negative set consists of D00760, D00765, D00769, D00780, D00845, D00954, D00965, D00987, D01071 and D01103. A few of these drugs, for instance, target, among other receptors, GPCR hsa:1135, hsa:1136 hsa:1137 and hsa:1138 (cholinergic receptors). It is known that GPCR drugs are “dirty” i.e. due to high homology between different classes, it is usual that one drug targets several GPCRs. As we stated, two conditions for in vivo activity must be met: long-range interaction recognition and short-range compatibility between the structures of binding site and ligand. While in multitargeting one drug meets both criteria, in false-positive cases it meets only the first one. Therefore, the selection of positive and negative sets for this method should be carefully carried, with a possible selection of ligands for negative sets that do not target the same family of receptors.

Table 2. Schematic presentation of ISM determined binding site domains pattern.

| TM | 1 | 2 | 3 | 4 | 5 | 6 | 7 | ECL1 | ECL2 | ECL3 | ICL1 | ICL2 | ICL3 |
|---|---|---|---|---|---|---|---|---|---|---|---|---|---|
| hsa:146 | | | | | X | | | | | | | | |
| hsa:147 | | | | | X | | | | | | | | |
| hsa:148 | | | | | X | | | | | | | | |
| hsa:150 | | X | | | | | | | | | X | | |
| hsa:151 | | | X | X | | | | | | | | X | |
| hsa:152 | | | | | X | | | | X | | | | |
| hsa:153 | | | | | X | | | | X | | | | |
| hsa:154 | | | | | X | | | | X | | | | |
| hsa:155 | | | | | X | | | | X | | | | |
| hsa:1128 | | | | | X | | | | | | | | |
| hsa:1129 | | | X | | | | | | | | | X | |
| hsa:1131 | | | X | | | | | | | | | | |
| hsa:1812 | | | X | | | | | | | | | | |
| hsa:1813 | | | X | | | | | | | | | | |
| hsa:1814 | | | X | | | | | | | | | | |
| hsa:3269 | | | | | X | | | | | | | | |
| hsa:3351 | X | X | | | | | | | | | X | | |
| hsa:3352 | | | | | | X | X | | | X | | | |
| hsa:3356 | | X | | | | | | X | | | | | |
| hsa:3358 | | X | X | | | | | X | | | | | |
| hsa:3577 | | | | | | X | X | | | | | | X |

Table 4. Dominant ISM frequencies of drug-receptor CIS spectra of 21 GPCR, with corresponding protein domains – negative sets

| No | Receptor KEGG code | Receptor name | Major ISM frequency | Major ISM frequency amplitude value | Ratio ISM F negative/ positive set | ISM frequency corresponding AA region | Window size |
|---|---|---|---|---|---|---|---|
| 1 | hsa:146 | Alpha-1D adrenergic receptor | 0.121 | 1.80E-16 | 0.0003 | - | - |
| 2 | hsa:147 | Alpha-1B adrenergic receptor | 0.282 | 1.58E-15 | 0.0013 | - | - |
| 3 | hsa:148 | Alpha-1A adrenergic receptor | 0.273 | 1.24E-15 | 0.0102 | - | - |
| 4 | hsa:150 | Alpha-2A adrenergic | 0.096 | 1.06E-15 | 0.0099 | - | - |
| | | receptor | | | | | |
| 5 | hsa:151 | Alpha-2B adrenergic receptor | 0.0625 | 7.24E-11 | 5983.4711 | 68 - 83 | 16 |
| 6 | hsa:152 | Alpha-2C adrenergic receptor | 0.045 | 2.29E-14 | 6.1394 | 350-382 | 33 |
| 7 | hsa:153 | Beta-1 adrenergic receptor | 0.143 | 1.91E-16 | 0.0085 | - | - |
| 8 | hsa:154 | Beta-2 adrenergic receptor | 0.111 | 2.37E-14 | 1.1232 | - | - |
| 9 | hsa:155 | Beta-3 adrenergic receptor | 0.051 | 1.27E-13 | 3.6182 | 180-244 | 65 |
| 10 | hsa:1128 | Muscarinic acetylcholine receptor M1 | 0.059 | 3.40E-12 | 330.0971 | 205-269 | 65 |
| 11 | hsa:1129 | Muscarinic acetylcholine receptor M2 | 0.221 | 3.03E-16 | 0.0278 | - | - |
| 12 | hsa:1131 | Muscarinic acetylcholine receptor M3 | 0.139 | 6.90E-16 | 0.0037 | - | - |
| 13 | hsa:1812 | D(1A) dopamine receptor | 0.118 | 1.86E-14 | 0.4088 | - | - |
| 14 | hsa:1813 | D(2) dopamine receptor | 0.121 | 2.41E-15 | 0.0406 | - | - |
| 15 | hsa:1814 | D(3) dopamine | 0.228 | 1.92E-16 | 0.0002 | - | - |
| | | receptor | | | | | |
| 16 | hsa:3269 | Histamine H1 receptor | 0.078 | 5.54E-16 | 0.0040 | - | - |
| 17 | hsa:3351 | 5-hydroxytryptamine-1B receptor | 0.045 | 7.99E-15 | 0.0695 | - | - |
| 18 | hsa:3352 | 5-hydroxytryptamine-1D receptor | 0.111 | 1.87E-15 | 0.0113 | - | - |
| 19 | hsa:3356 | 5-hydroxytryptamine-2A receptor | 0.111 | 6.29E-14 | 7.2465 | 407-471 | 65 |
| 20 | hsa:3358 | 5-hydroxytryptamine-2C receptor | 0.045 | 4.70E-17 | 0.0005 | - | - |
| 21 | hsa:3577 | High affinity interleukin-8 receptor A/CXCR1 | 0.043 | 2.65E-14 | 382.3954 | 59-91 | 33 |

All spectra of receptors, their positive and negative sets CIS spectra and corresponding AA regions of the receptor are given in Supplementary Material.

## 4. Conclusion

In this paper, we introduced a new approach in bioinformatics treatment of small molecules and their classification on active and inactive compounds towards a specific target. We acquired standard linear notation of small their structure - SMILES and converted atomic groups into pseudo amino acid residues. The small molecule in such notation can be treated as a peptide, and therefore ISM method can be applied. Its spectra can be multiplied with spectra of protein receptor to obtain proper CIS. Individual CIS spectra of sets active and inactive compounds and corresponding receptor differ in frequency and amplitude values. Their interpretation employing the receptor AA sequence yields specific regions, which contain crucial amino acid residues for ligand-target recognition, based on long-range intermolecular interactions. The essential information that can be extracted from ligand CIS spectra is AA region of the binding site in the corresponding receptor. However, there are notable limitations of this method, which are due to drug sets and statistical probabilities regarding the formation of a proper reliable set.

This method can be used for fast screening of large compound libraries through candidate selection on certain frequency value. Of course, it is just the first step towards the finding of a good candidate for the second part, short-range compatibility must be met, which can be carried by either a ligand-based (QSAR) or structure-based screening (molecular docking etc).

With further development and introduction of advanced data analysis, such as machine learning we believe that those deficiencies will be overcome. This new approach brings new possibilities through developing new techniques in ligand classification and selection of new candidates for a specific target.

## 5. Competing interests

The authors declare that they have no competing interests.

## 6. Funding sources

This work was financially supported by the Ministry of Education, Science and Technological Development of the Republic of Serbia, Contract No 173001.

## 7. References

Anderson, E., Veith, G.D, Weininger, D. (1987). SMILES: A line notation and computerized interpreter for chemical structures (Report No. EPA/600/M-87/021 izd.). U.S. EPA, Environmental Research Laboratory-Duluth, Duluth, MN 55804.

Brune D., S. Kim (1994) Hydrodynamic steering effects in protein association. Proc Natl Acad Sci USA 91, 2930-2934.

Camacho C. J., Kimura S. R.,DeLisi C., Vajda S. (2000) Kinetics of desolvation-mediated protein-protein binding. Biophys J 78, 1094-1105.

Carrieri A, Centeno NB, Rodrigo J, Sanz F & Carotti A (2001) Theoretical Evidence of a Salt Bridge Disruption as the Initiating Process for the alpha 1d -Adrenergic Receptor Activation : A Molecular Dynamics and Docking Study. 394, 382–394.

Doliana R., Veljkovic V., Prljic J., Veljkovic N., De Lorenzo E., Mongiat M., Ligresti G., Marastoni S., Colombatti A. (2008) EMILINs interact with anthrax protective antigen and inhibit toxin action in vitro. Matrix Biol 27(2), 96-106.

Evers A & Klabunde T (2005) Structure-based Drug Discovery Using GPCR Homology Modeling: Successful Virtual Screening for Antagonists of the Alpha1A Adrenergic Receptor. , 1088–1097.

Fischer E., (1894) Einfluss der Configuration auf die Wirkung der Enzyme. European Journal of Inorganic Chemistry, 27(3), 2985-2993.

Fraser CM, Wang CD, Robinson DA, Gocayne JD & Venter JC (1989) Site-directed mutagenesis of m1 muscarinic acetylcholine receptors: conserved aspartic acids play important roles in receptor function, Molecular Pharmacology 36 (6) 840-847;

Fröhlich H. (1968) Long-range coherence and energy storage in biological systems. Int. J Quantum Chem 2, 641-649.

Fröhlich H. (1970) Long range coherence and the action of enzymes. Nature 228, 1093

Fröhlich H. (1975) The extraordinary dielectric properties of biological materials and the action of enzymes. Proc Natl Acad Sci USA 72, 4211-4215.

Kanehisa, Furumichi, M., Tanabe, M., Sato, Y., and Morishima, K.; KEGG: new perspectives on genomes, pathways, diseases and drugs. Nucleic Acids Res. 45, D353-D361 (2017).

Kanehisa, M. and Goto, S.; KEGG: Kyoto Encyclopedia of Genes and Genomes. Nucleic Acids Res. 28, 27-30 (2000).

Kanehisa, M., Sato, Y., Kawashima, M., Furumichi, M., and Tanabe, M.; KEGG as a reference resource for gene and protein annotation. Nucleic Acids Res. 44, D457-D462 (2016).

Leckband D., Israelachvili J. (2001) Intermolecular forces in biology. Quarterly Reviews of Biophysics 34(2) 105–267.

McCloskey M. A., Poo M. M. (1986) Rates of membrane-associated reactions: reduction of dimensionality revisited. J Cell Biol 102, 88-96.

Northrup S. H., Erickson H. P. (1992) Kinetics of protein-protein association explained by Brownian dynamics computer simulation. Proc Natl Acad Sci USA 89, 3338-3342.

O'Boyle N. M., et al. (2011) Open Babel: An open chemical toolbox. Journal of cheminformatics, 3 (1), 33. Open Babel, version 2.3.1, http://openbabel.org (accessed Apr 2017)

Pándy-Szekeres G, Munk C, Tsonkov TM, Mordalski S, Harpsøe K, Hauser AS, Bojarski AJ, Gloriam DE. GPCRdb in 2018: adding GPCR structure models and ligands. 2017, Nucleic Acids Res., Nov 16. 10.1093/nar/gkx1109

Perovic V.R., Muller C.P., Niman H.L., Veljkovic N., Dietrich U., Tosic D.D., Glisic S., Veljkovic V. (2013) Novel phylogenetic algorithm to monitor human tropism in Egyptian H5N1-HPAIV reveals evolution toward efficient human-to-human transmission. PLoS One. 8(4):e61572.

Peters R. (2005) Translocation through the nuclear pore complex: selectivity and speed by reduction-of-dimensionality. Traffic, 6, 421-427.

Rosenbaum DM, Rasmussen SGF & Kobilka BK (2014) The structure and function of G-protein-coupled receptors. Nature 459, 356–363.

Scheer A, Fanelli F, Costa T, Benedetti PG De & Cotecchia S (1996) Constitutively active mutants of the a1B-adrenergic receptor : role of highly conserved polar amino acids in receptor activation. 15, 3566–3578.

Silverman R. B., Holladay M. W. (2014) The organic chemistry of drug design and drug action. Academic press

Smoluchowski M.V. (1916) Versuch einer mathematischen Theorie der Koagulationskinetik kolloider Lösungen. Phys Z 17, 129-168

Vagmita Pabuwal & Li Z (2012) Comparison Analysis of Primary Ligand Binding Sites in Seven-Helix Membrane Proteins. Biopolymers 95, 31–38.

Veljkovic N., Glisic S., Perovic V., Veljkovic V. (2011) The role of long-range intermolecular interactions in discovery of new drugs, Expert Opin Drug Discov, 6 (12) 1263-1270.

Veljkovic V (1980) A theoretical approach to preselection of carcinogens and chemical carcinogenesis. New York: Gordon & Breach

Veljkovic V. and Slavic I., (1972) Simple general-model pseudopotential. Phys Rev Lett. 29, 105–7.

Veljkovic V., Loiseau P.M., Figadere B. et al (2015) Virtual screen for repurposing approved and experimental drugs for candidate inhibitors of EBOLA virus infection, F1000Research 4, 34

Veljkovic V., Niman H.L., Glisic S., Veljkovic N., Perovic V., Muller C.P. (2009a) Identification of hemagglutinin structural domain and polymorphisms which may modulate swine H1N1 interactions with human receptor. BMC Struct Biol. 28 (9), 62.

Veljkovic V., Veljkovic N., Esté J.A., Hüther A., Dietrich U. (2007) Application of the EIIP/ISM bioinformatics concept in development of new drugs. Curr Med Chem, 14(4), 441-53.

Veljkovic V., Veljkovic N., Muller C.P., Müller S., Glisic S., Perovic V., Köhler H. (2009b) Characterization of conserved properties of hemagglutinin of H5N1 and human influenza viruses: possible consequences for therapy and infection control. BMC Struct Biol, 9, 21.

Veljkovic, V., Cosic, I., Dimitrijevic, B., and Lalovic, D. (1985). Is it possible to analyze DNA and protein sequence by the method of digital signal processing? IEEE Trans. Biomed. Eng. 32, 337–341.

Weininger, D. (1988) SMILES, a chemical language and information system. 1. Introduction to methodology and encoding rules. J Chem Inf Comput Sci, 28, 31-36.

Wiegel F. W., DeLisi C. (1982) Evaluation of reaction rate enhancement by reduction in dimensionality. Am J Physiol 243, R475-R479.

Wolf S & Grünewald S (2015) Sequence , Structure and Ligand Binding Evolution of Rhodopsin-Like G Protein- Coupled Receptors : A Crystal Structure-Based Phylogenetic Analysis. PLoS One 10, 1–23.

Yamanishi, Y., Araki, M., Gutteridge, A., Honda, W., & Kanehisa, M. (2008). Prediction of drug–target interaction networks from the integration of chemical and genomic spaces. Bioinformatics, 24(13), i232-i240.

**Appendix.** Supplementary Material

Receptor ISM spectra, CIS spectra of positive set, AA  region, 7TM scheme with marked region and CIS spectra of negative set, respectively:

hsa:146 alpha-1d adrenergic receptor

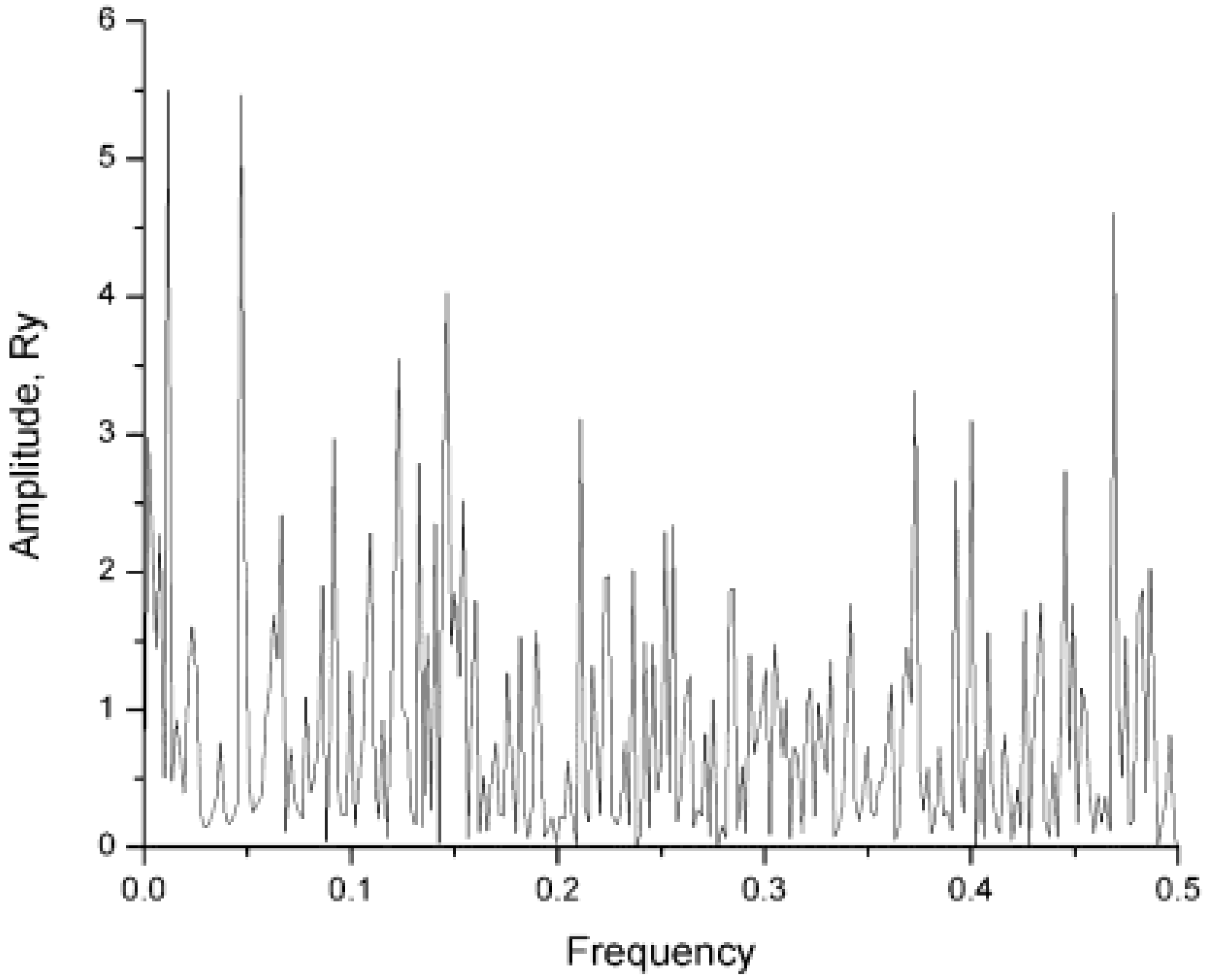

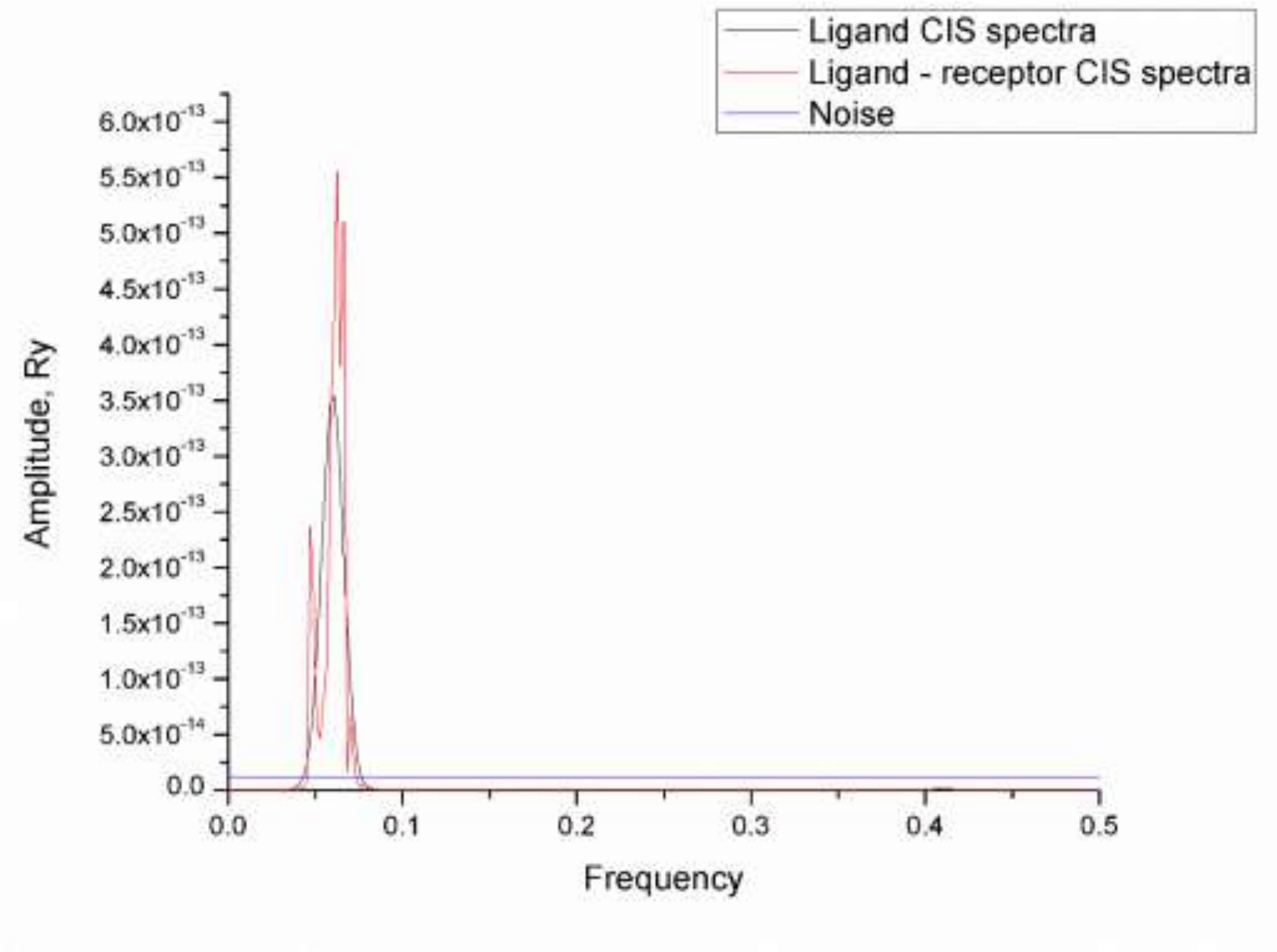

Ligand CIS spectra
Ligand - receptor CIS spectra
Noise
Amplitude, Ry
Frequency

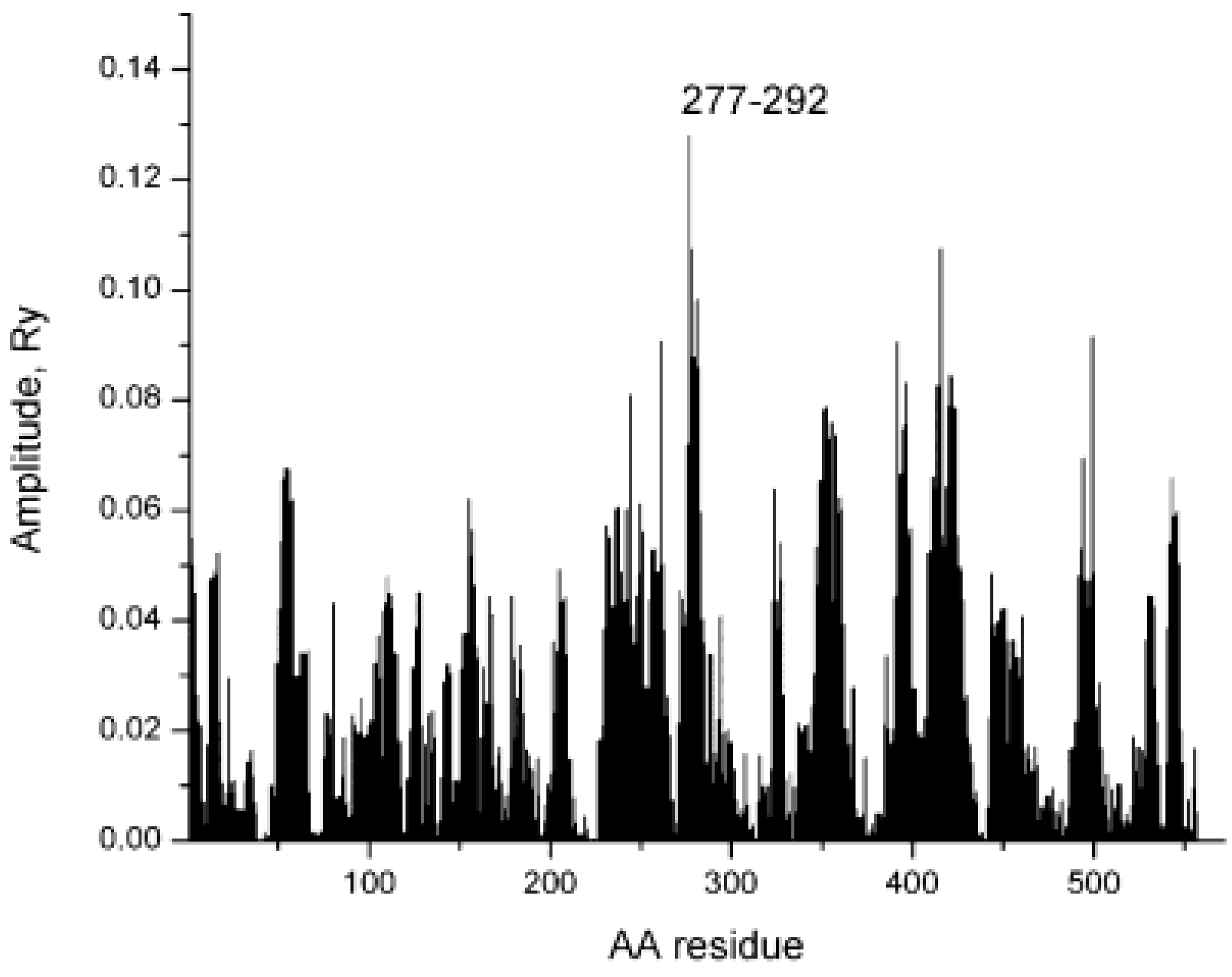

277-292
Amplitude, Ry
AA residue

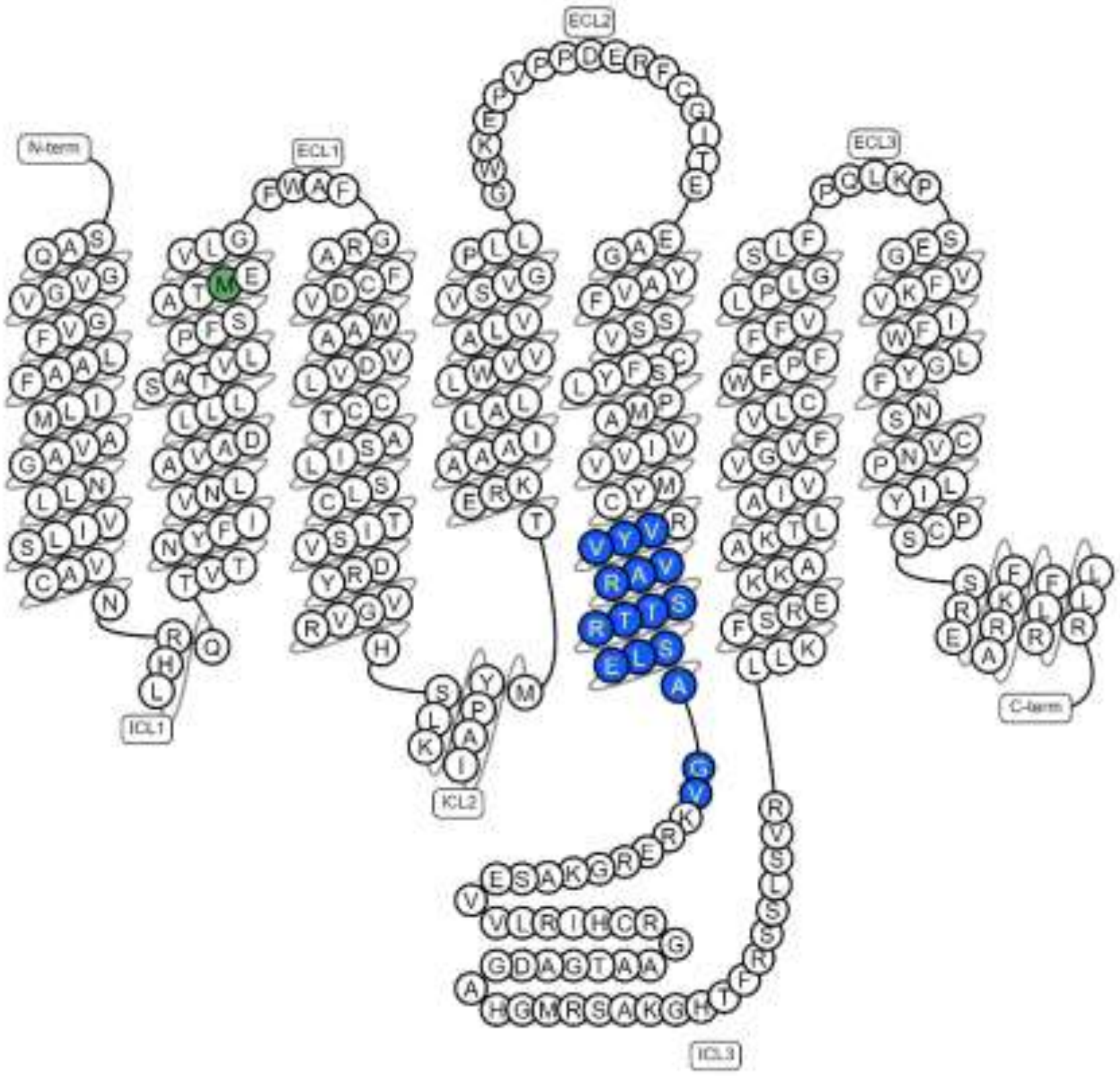

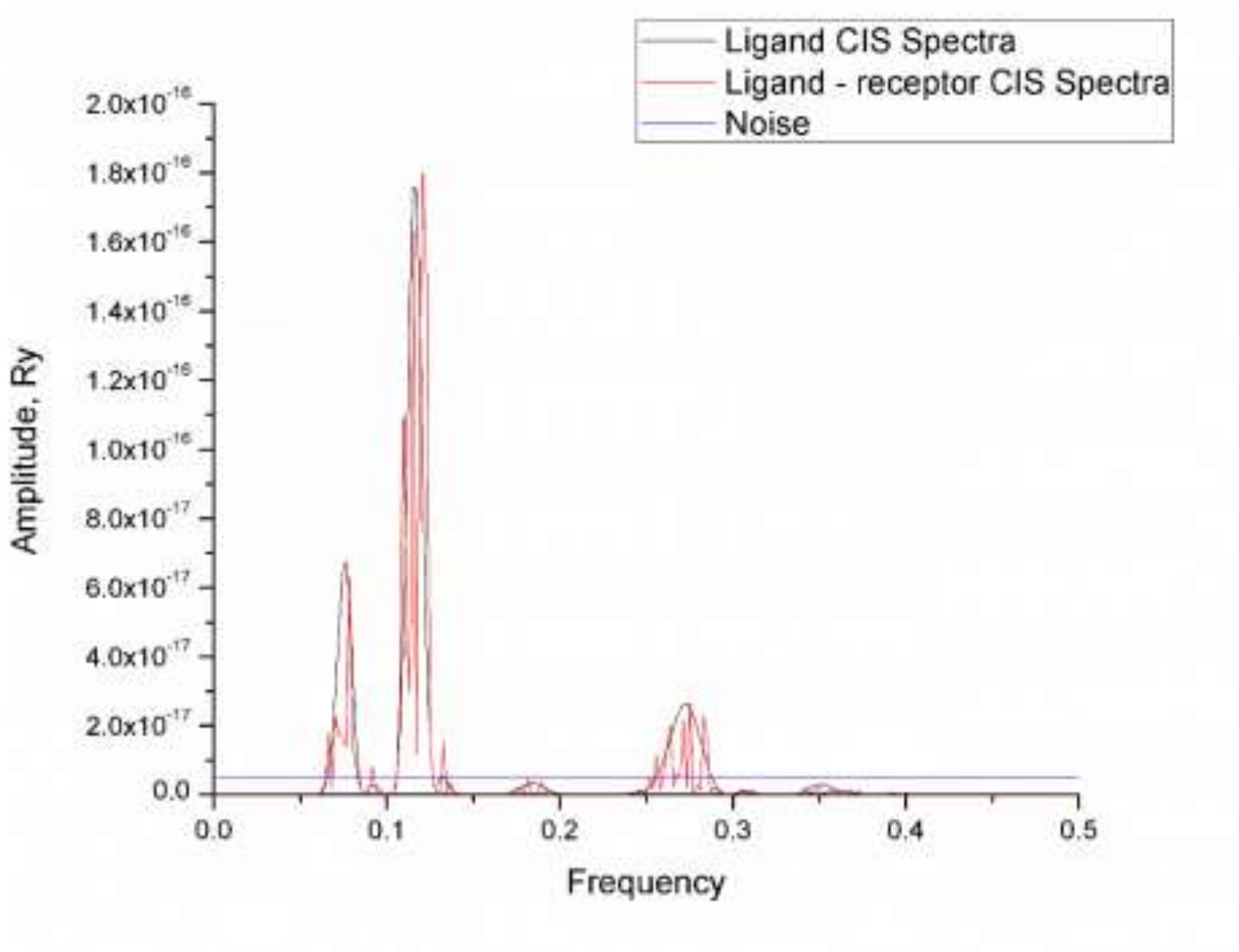

hsa:147 alpha-1b adrenergic receptor

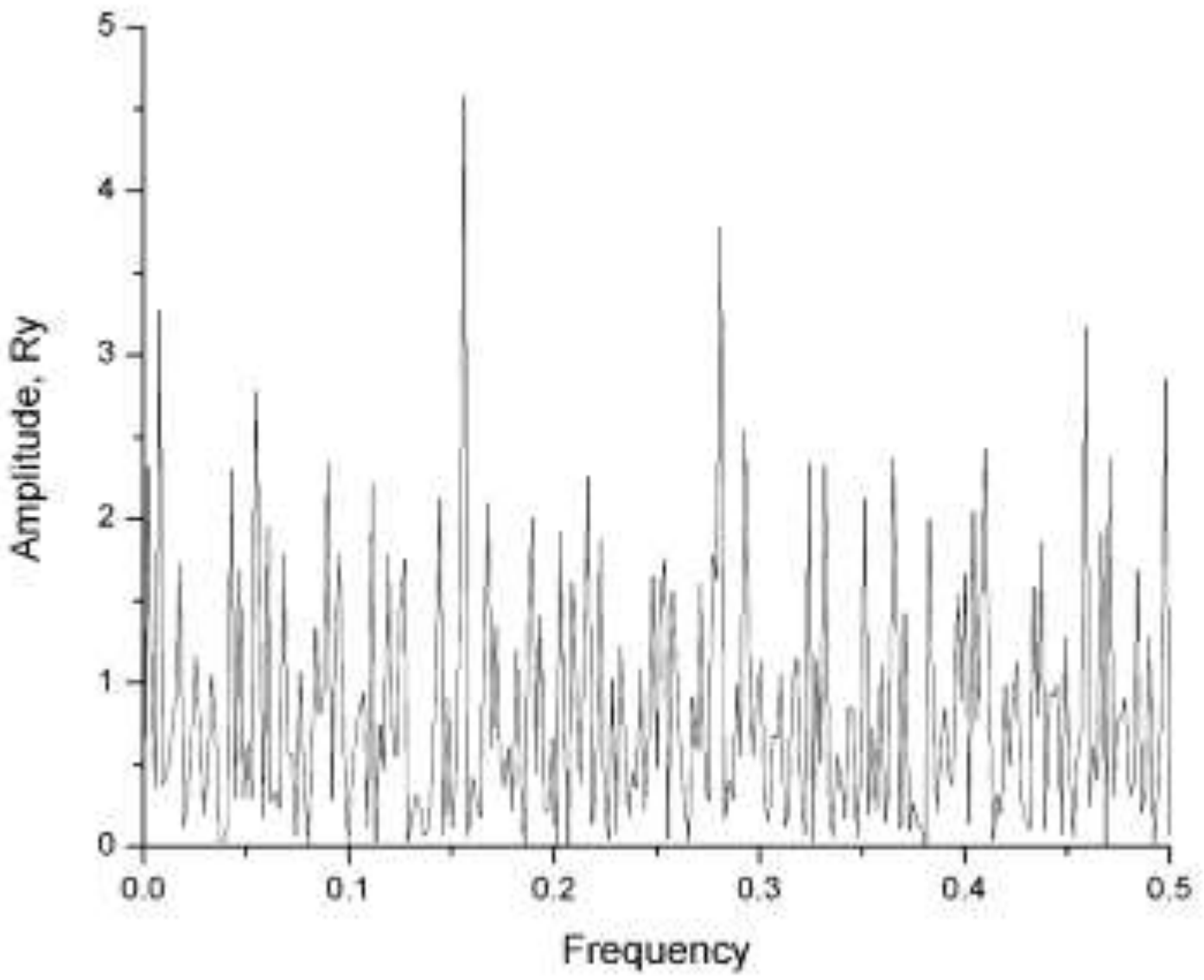

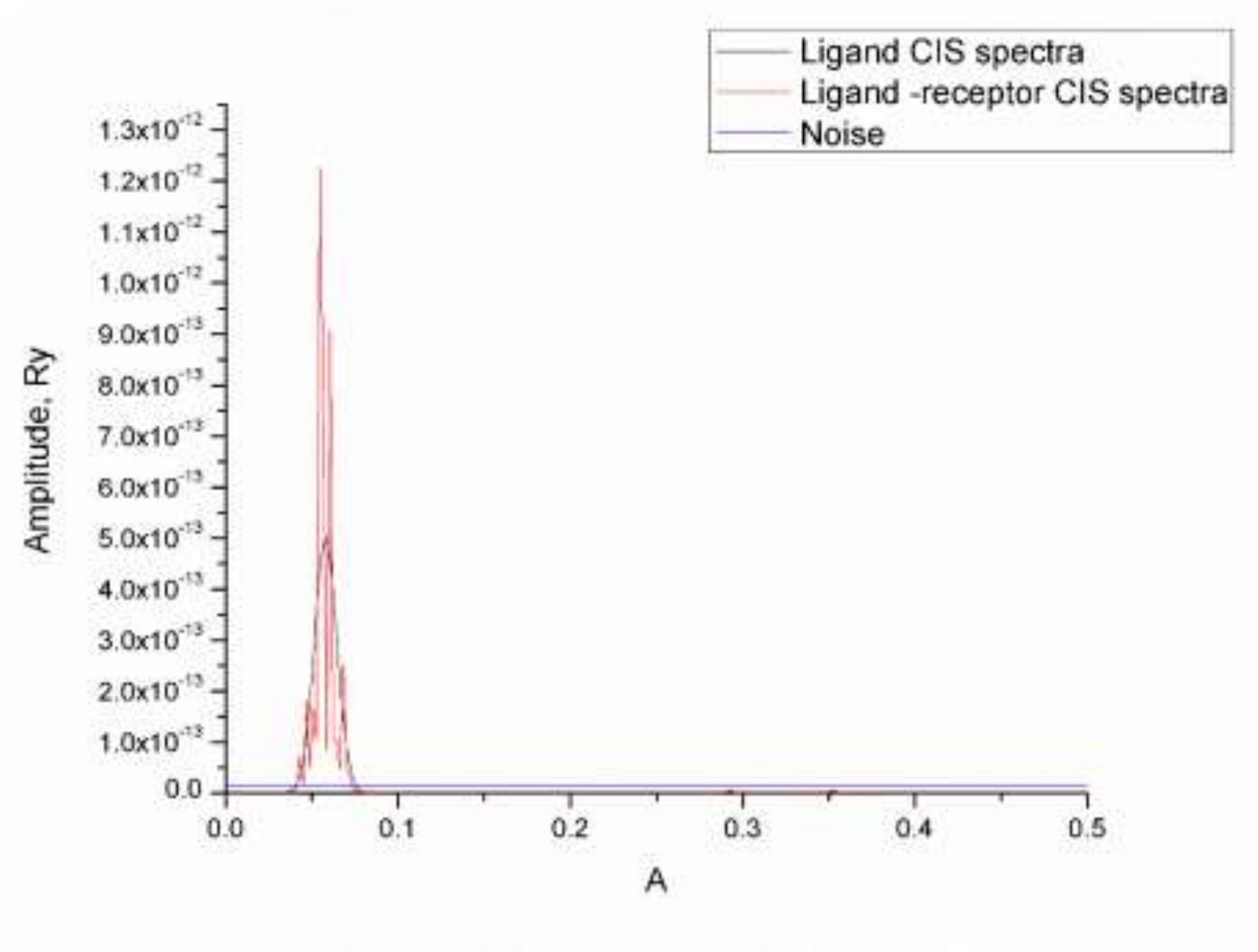

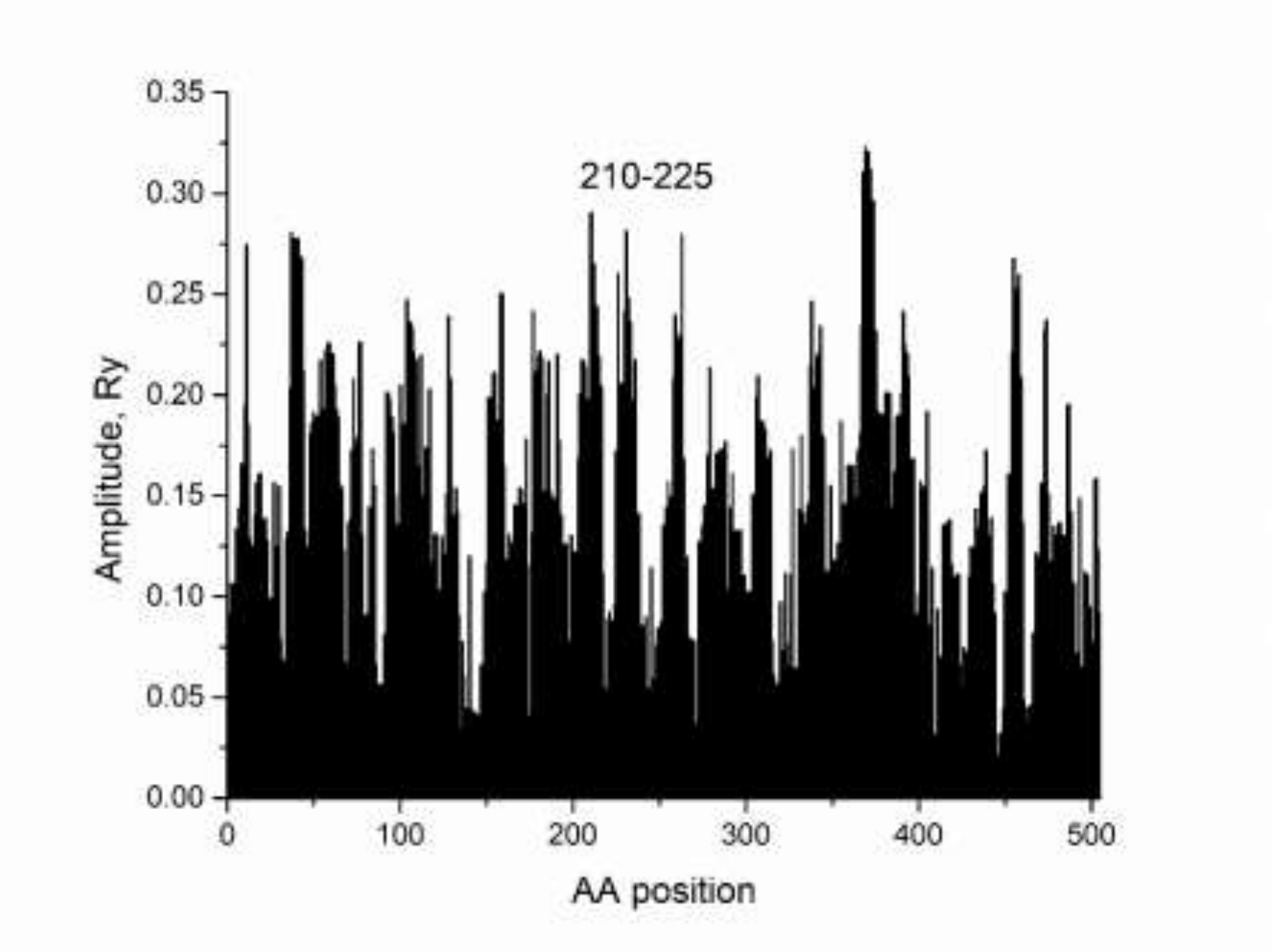

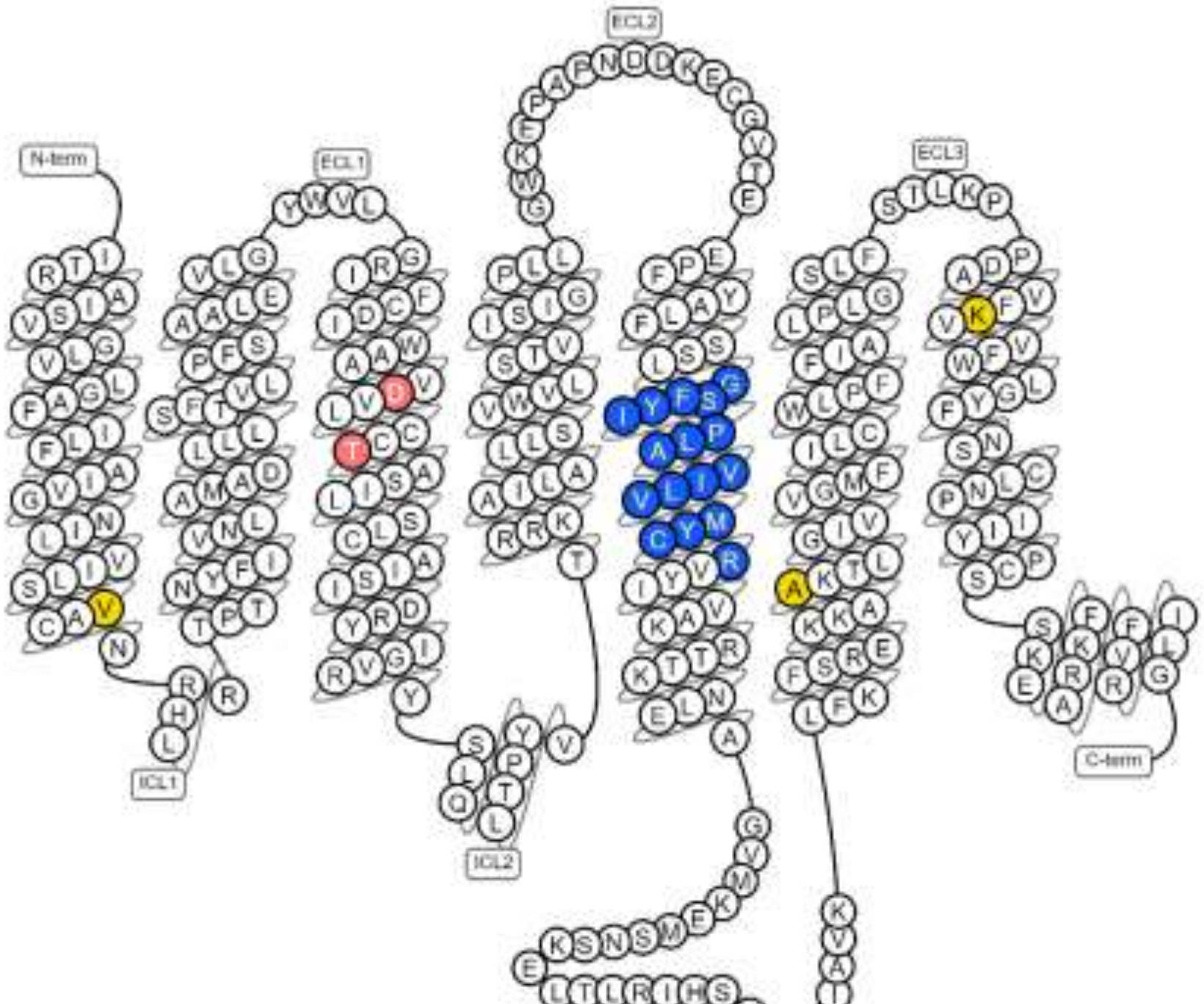

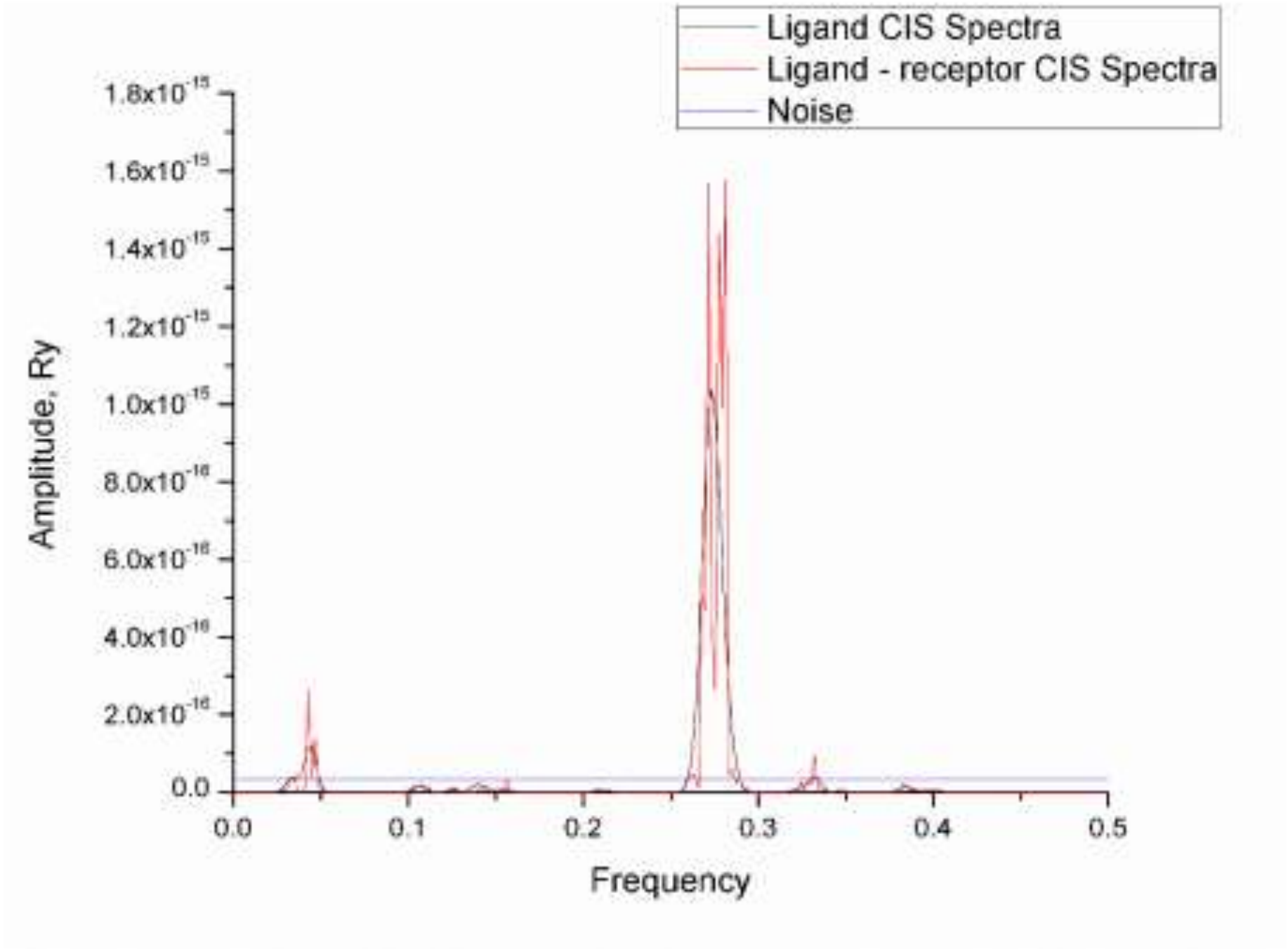

hsa:148 alpha-1a adrenergic receptor

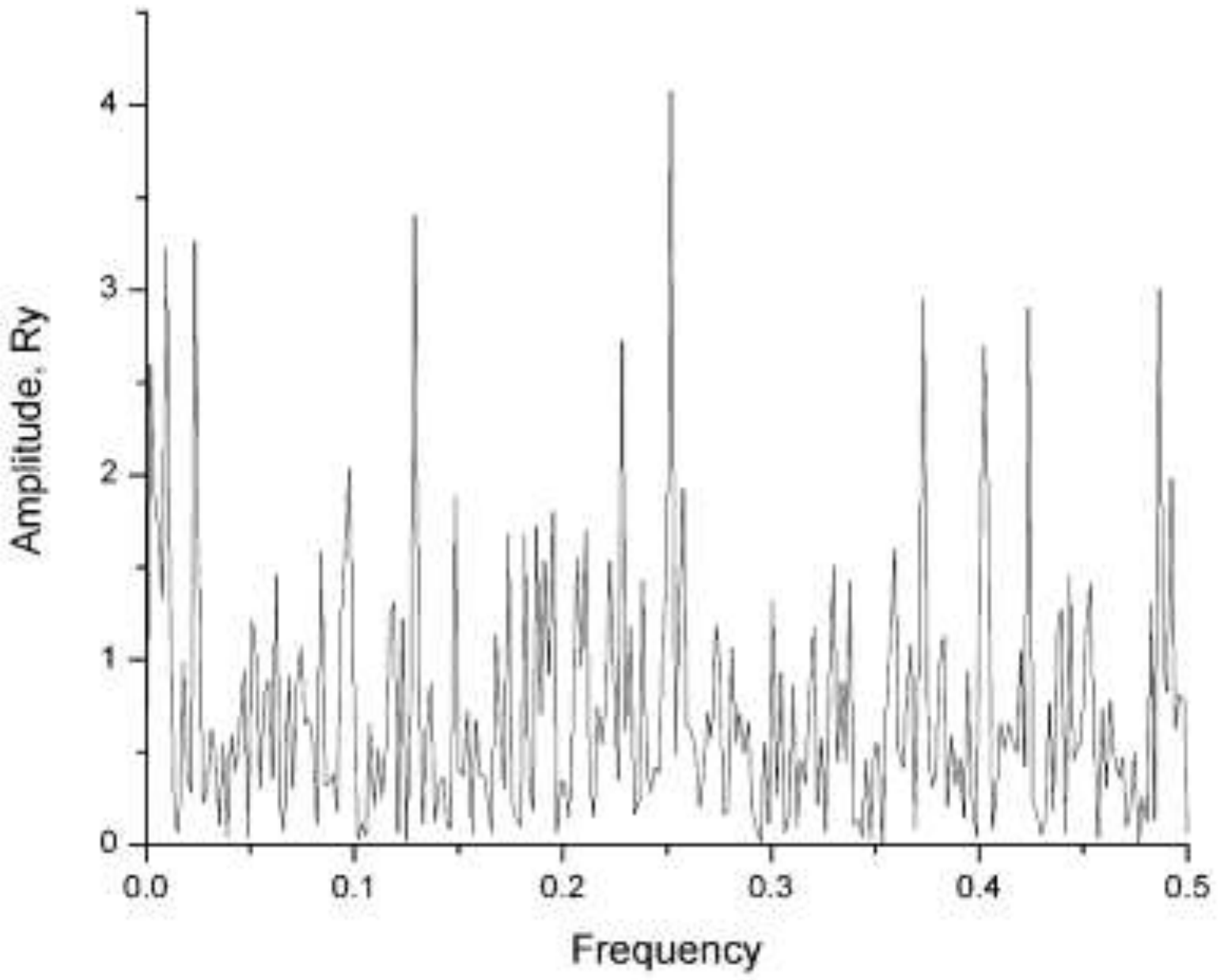

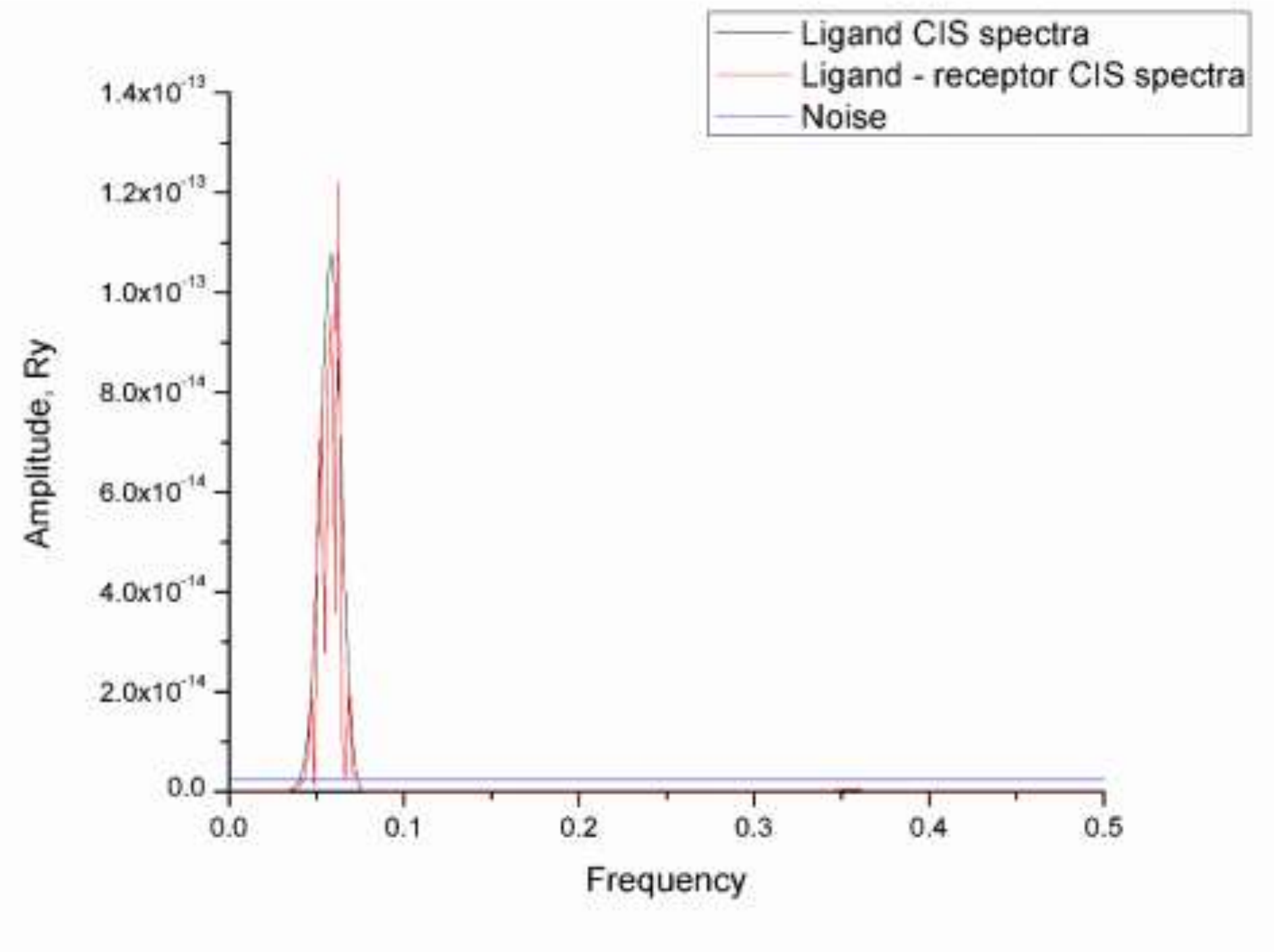

Ligand CIS spectra
Ligand - receptor CIS spectra
Noise
Amplitude, Ry
Frequency

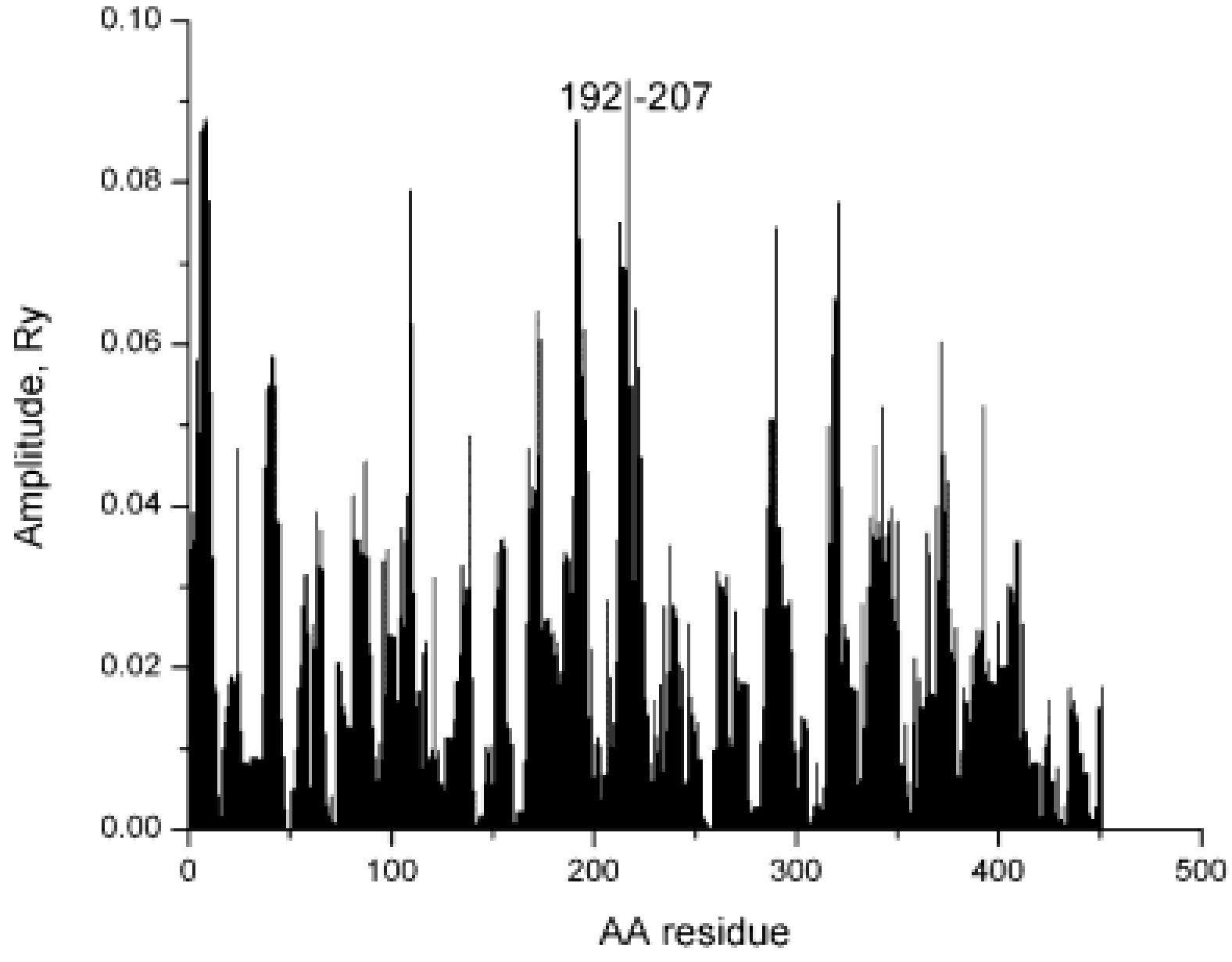

192
207
Amplitude, Ry
AA residue

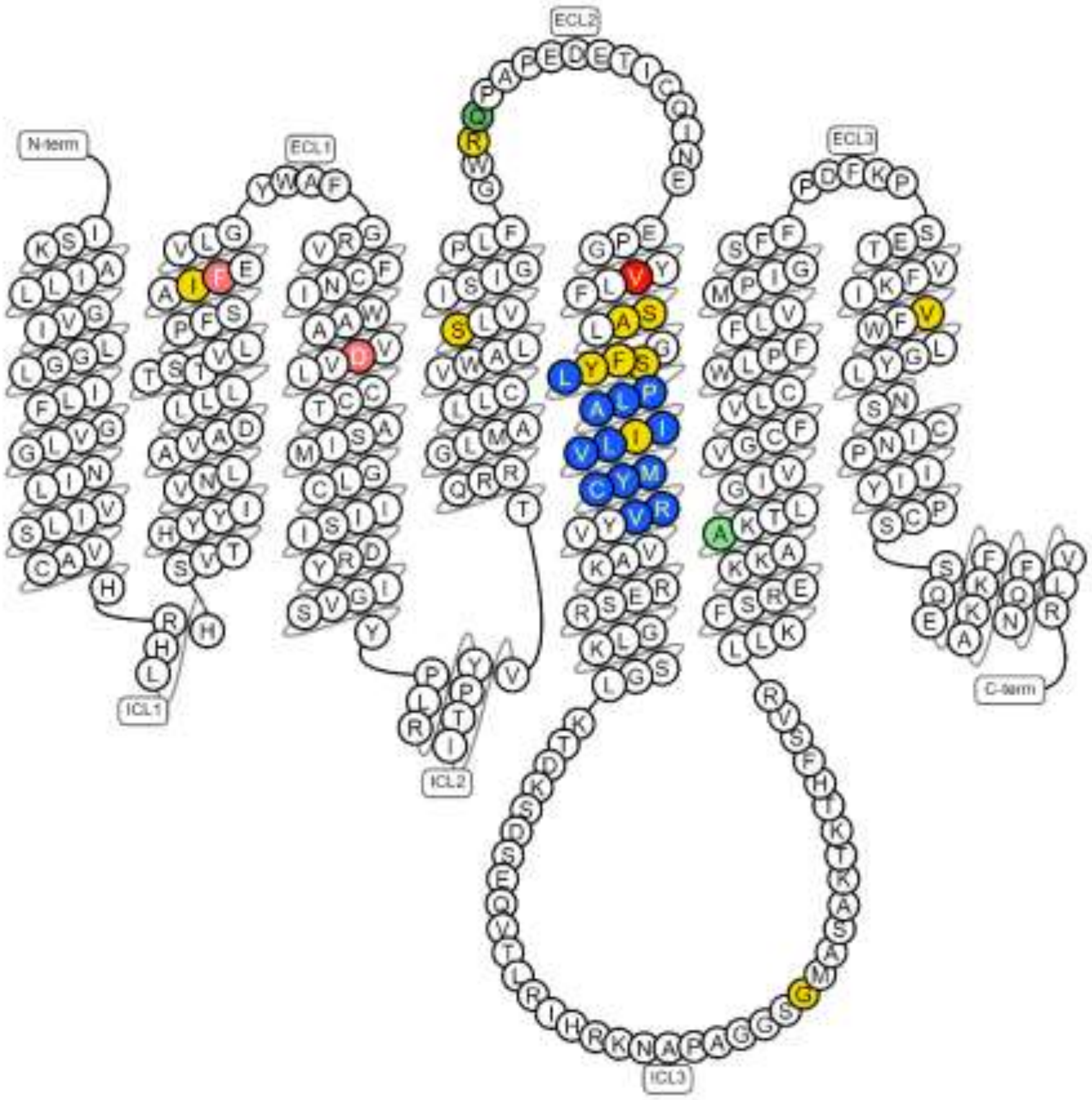

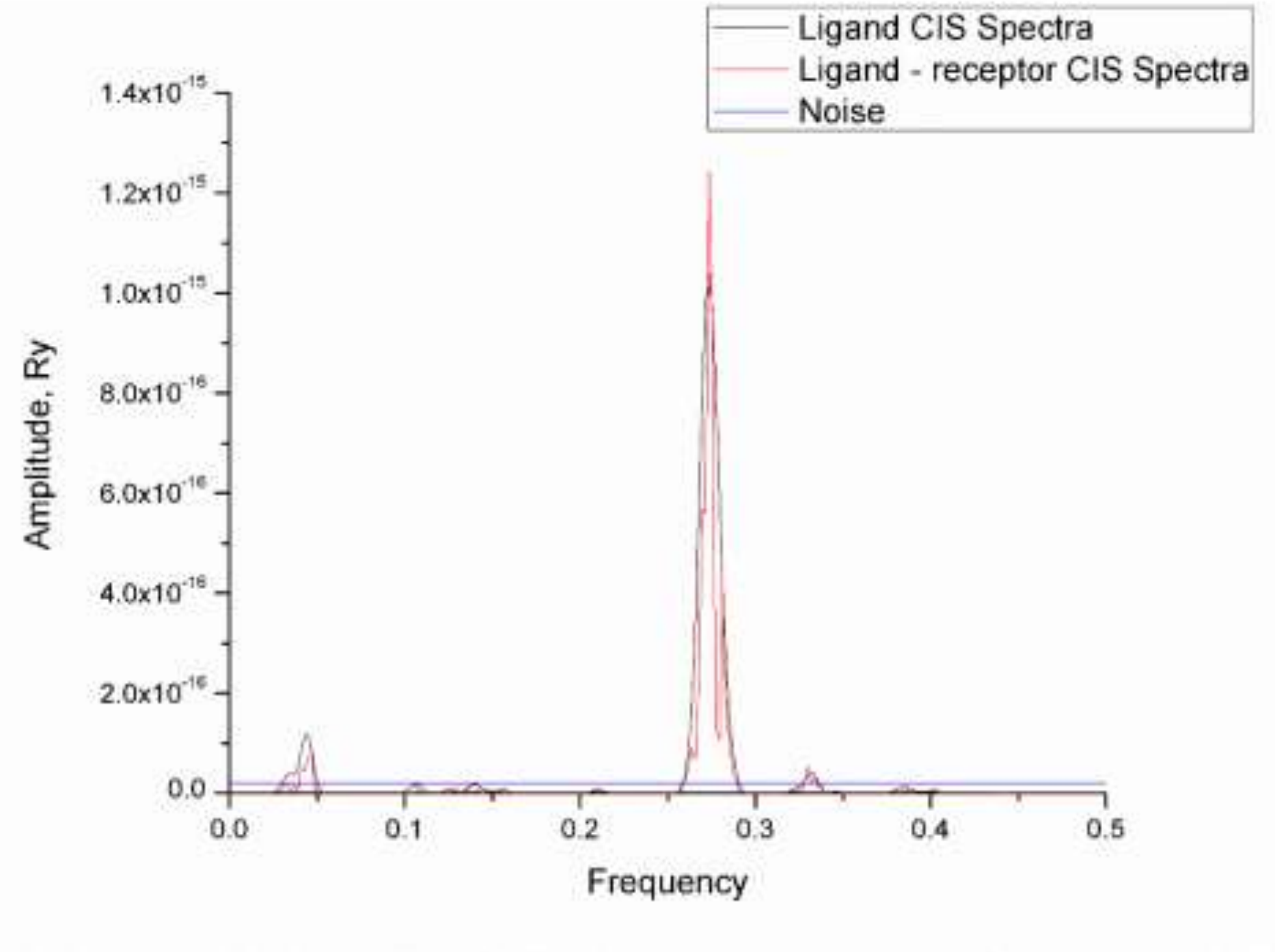

hsa:150 alpha-2a adrenergic receptor

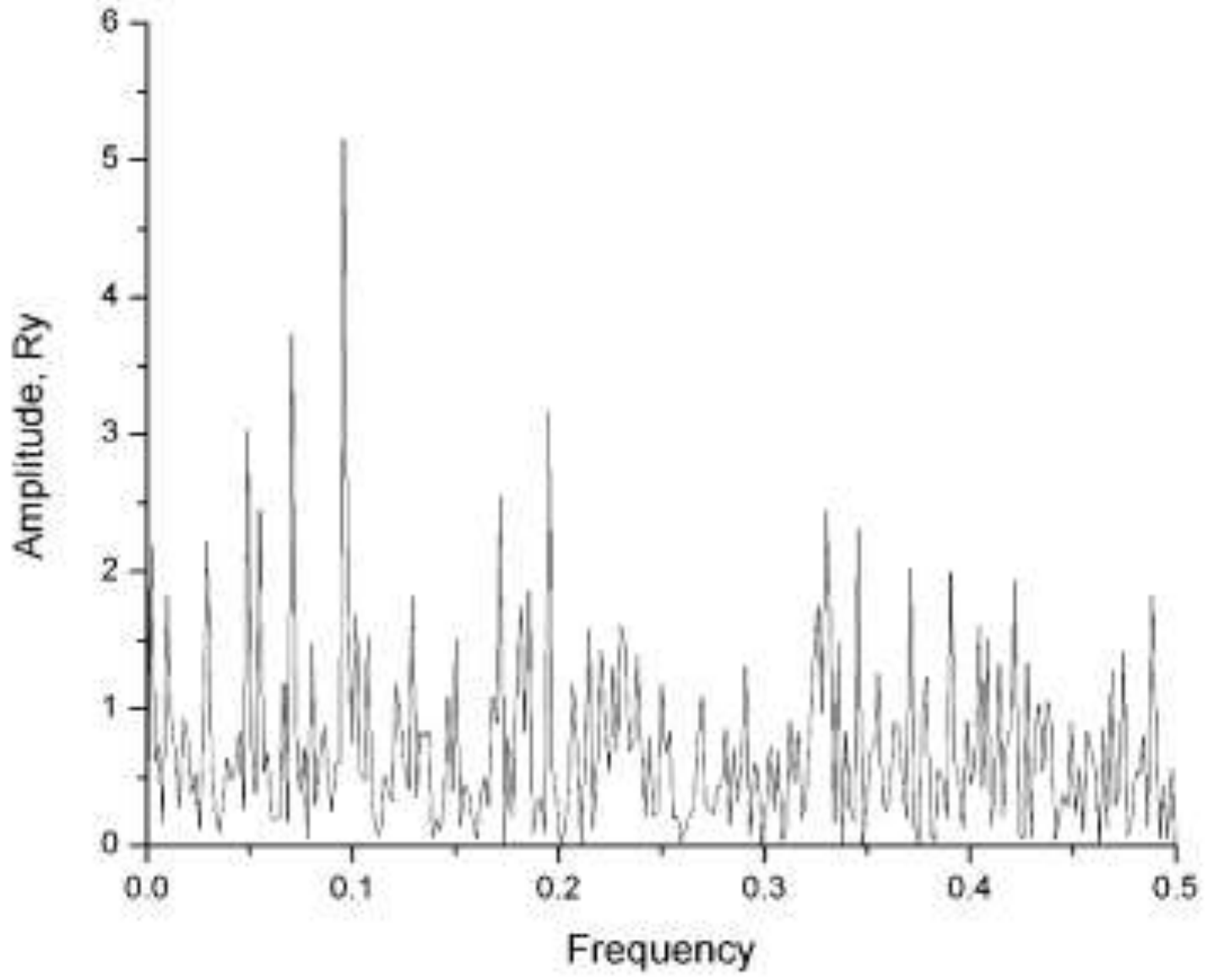

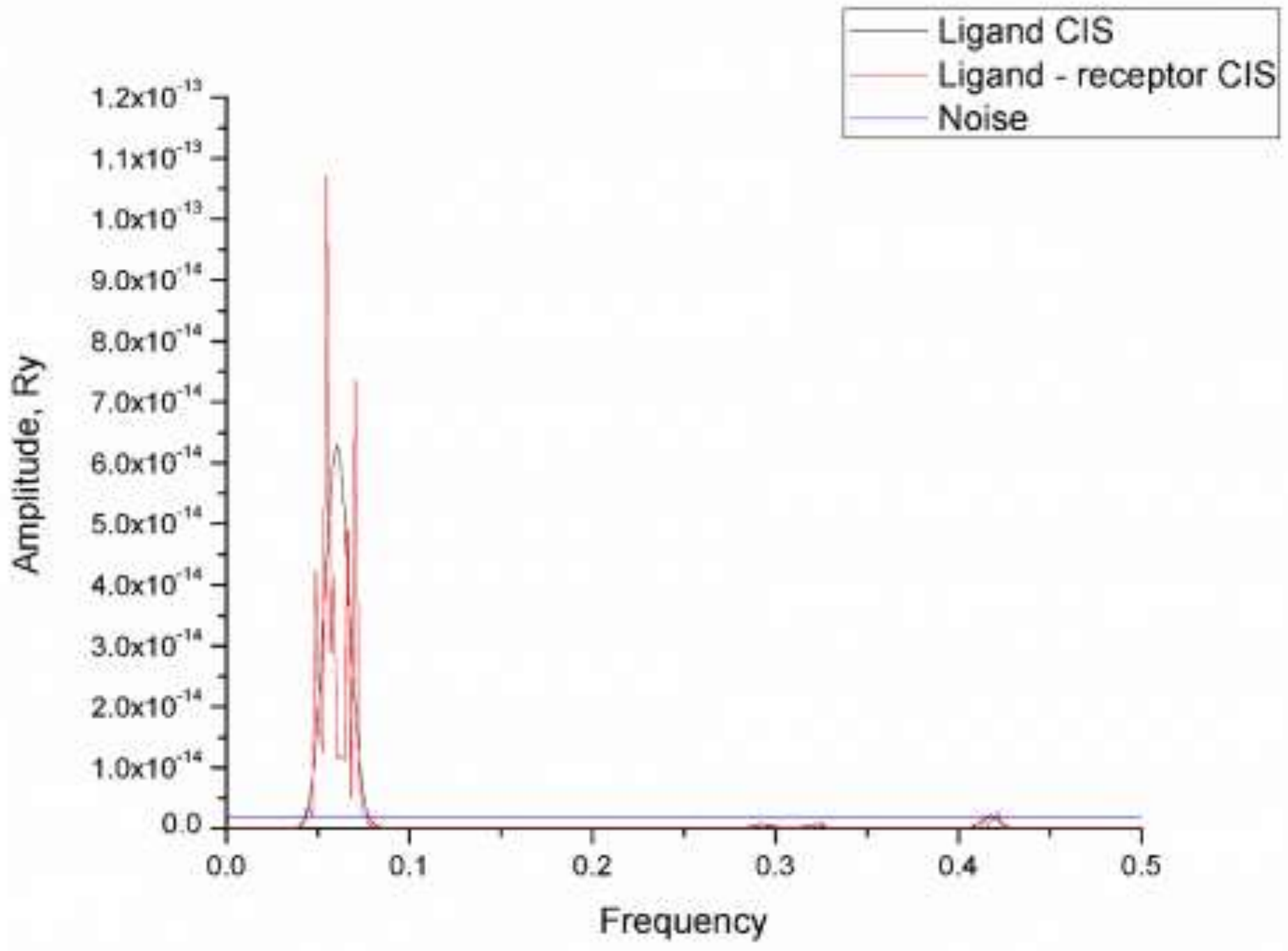

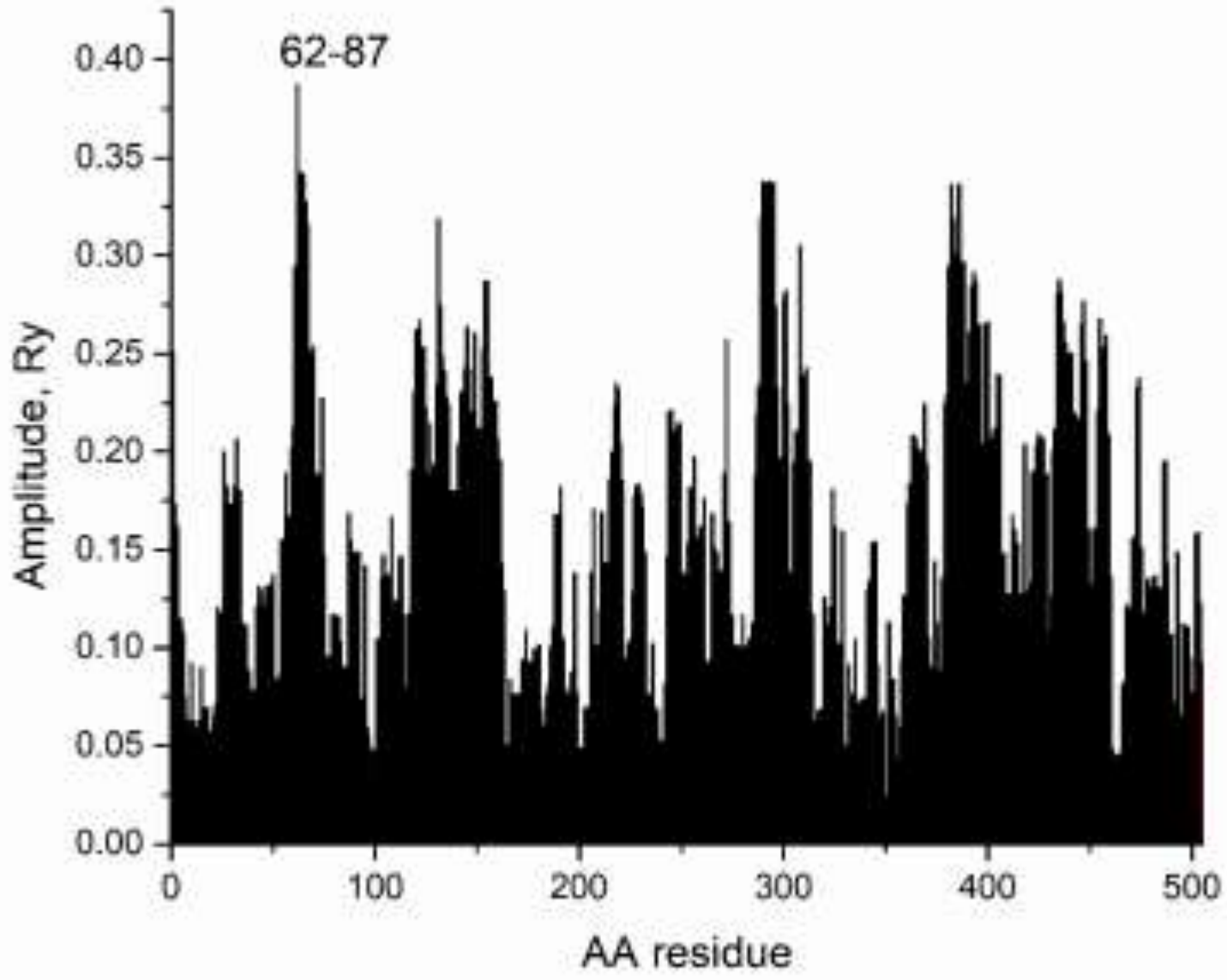

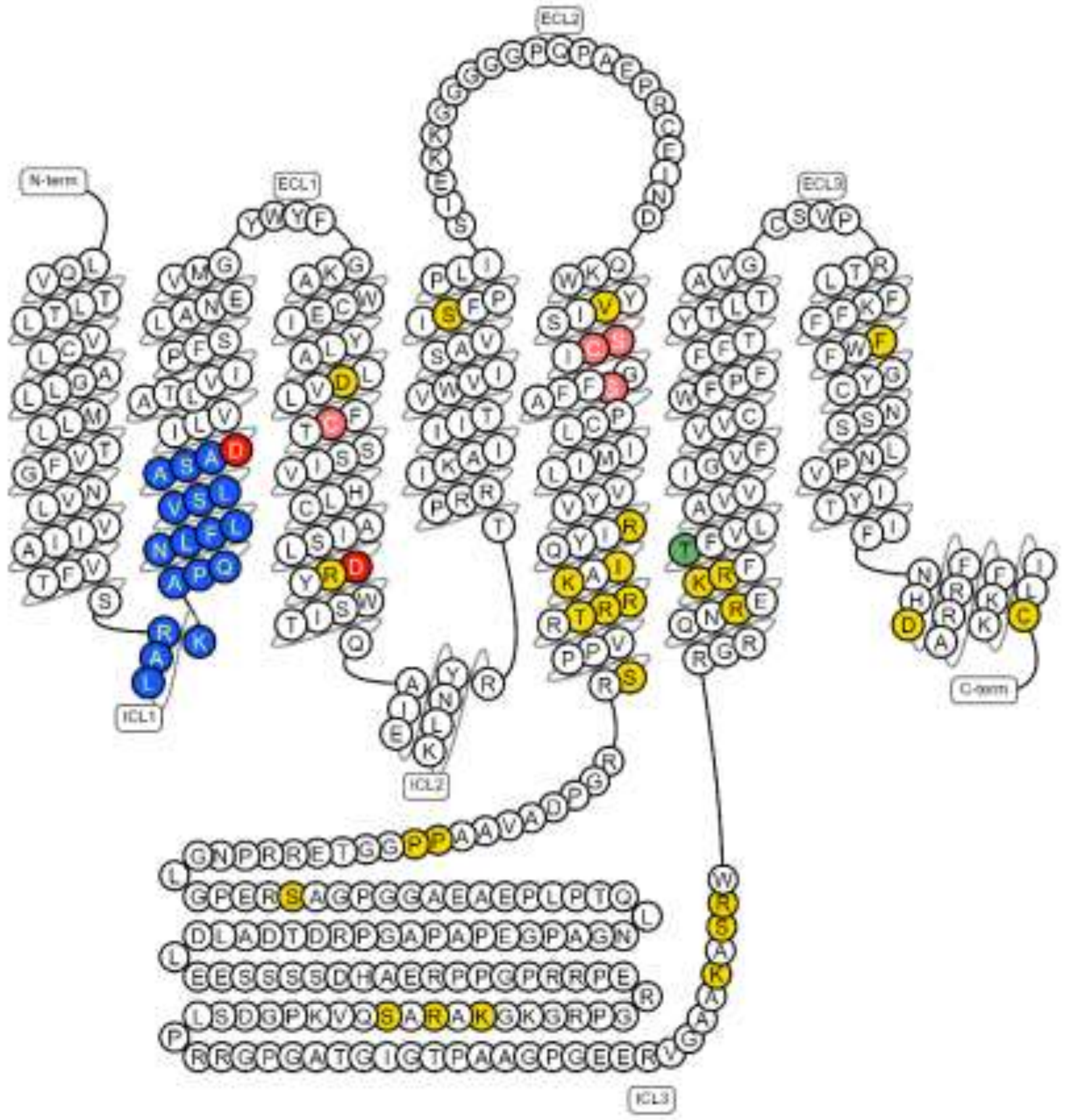

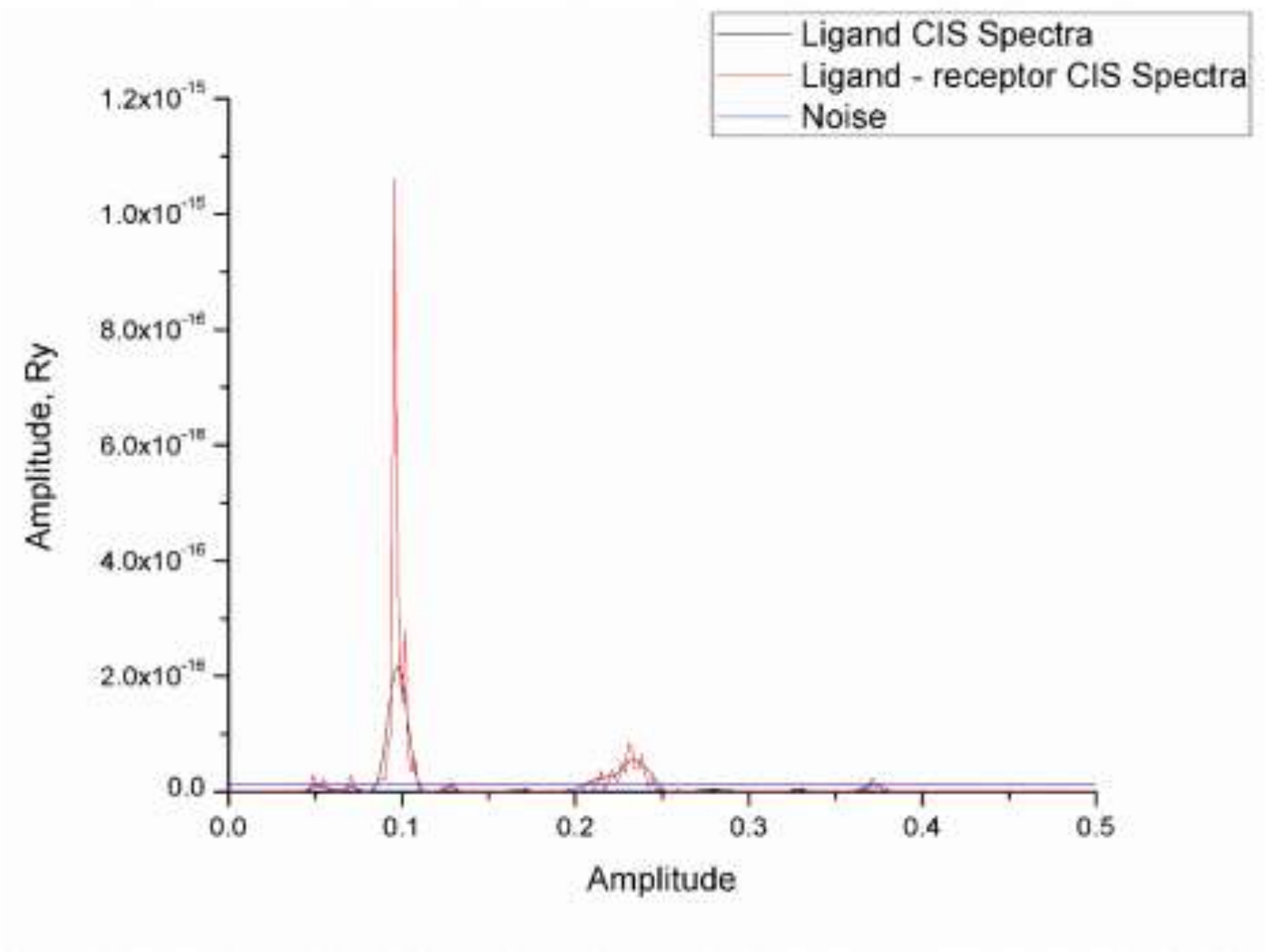

hsa:151 alpha-2b adrenergic receptor

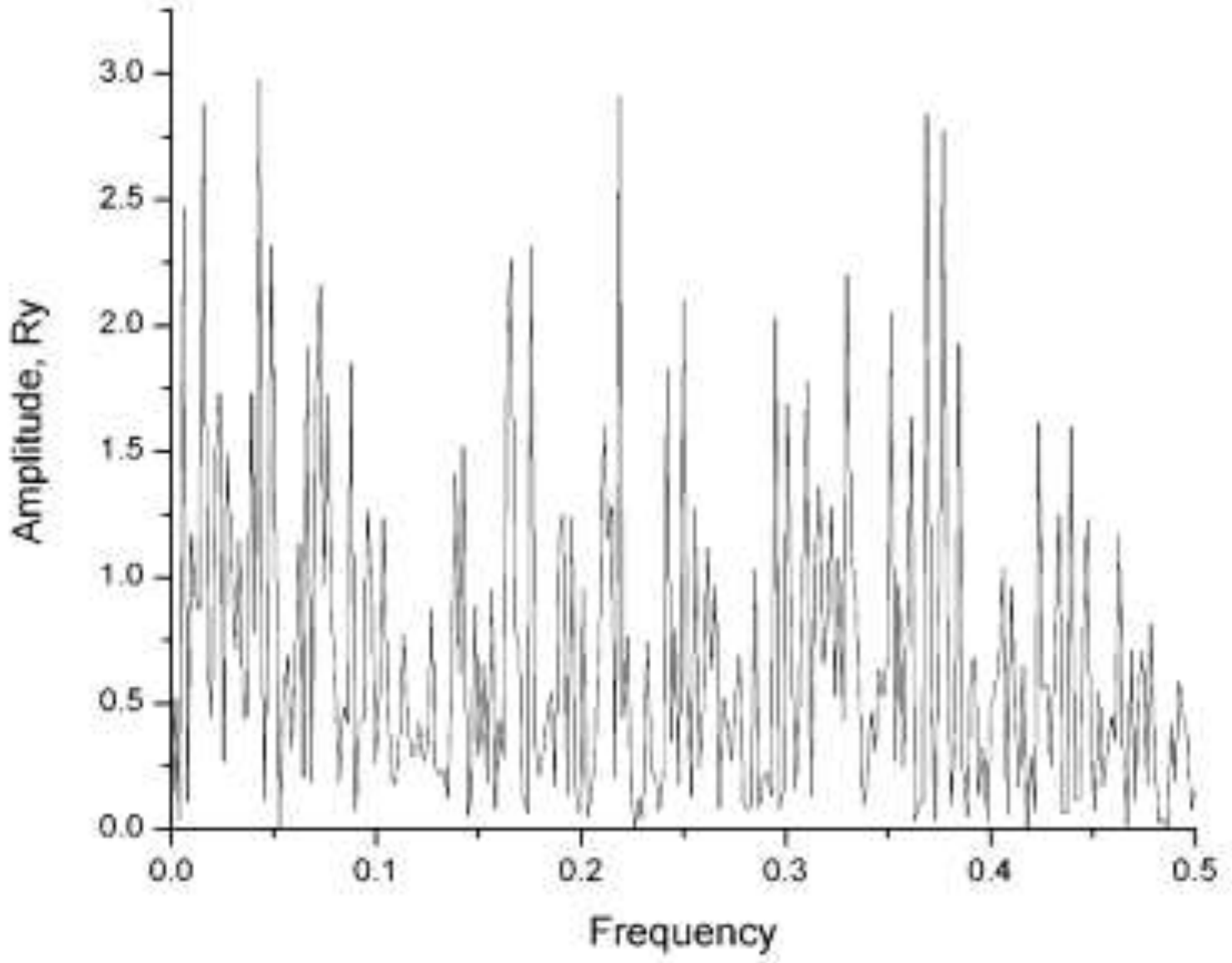

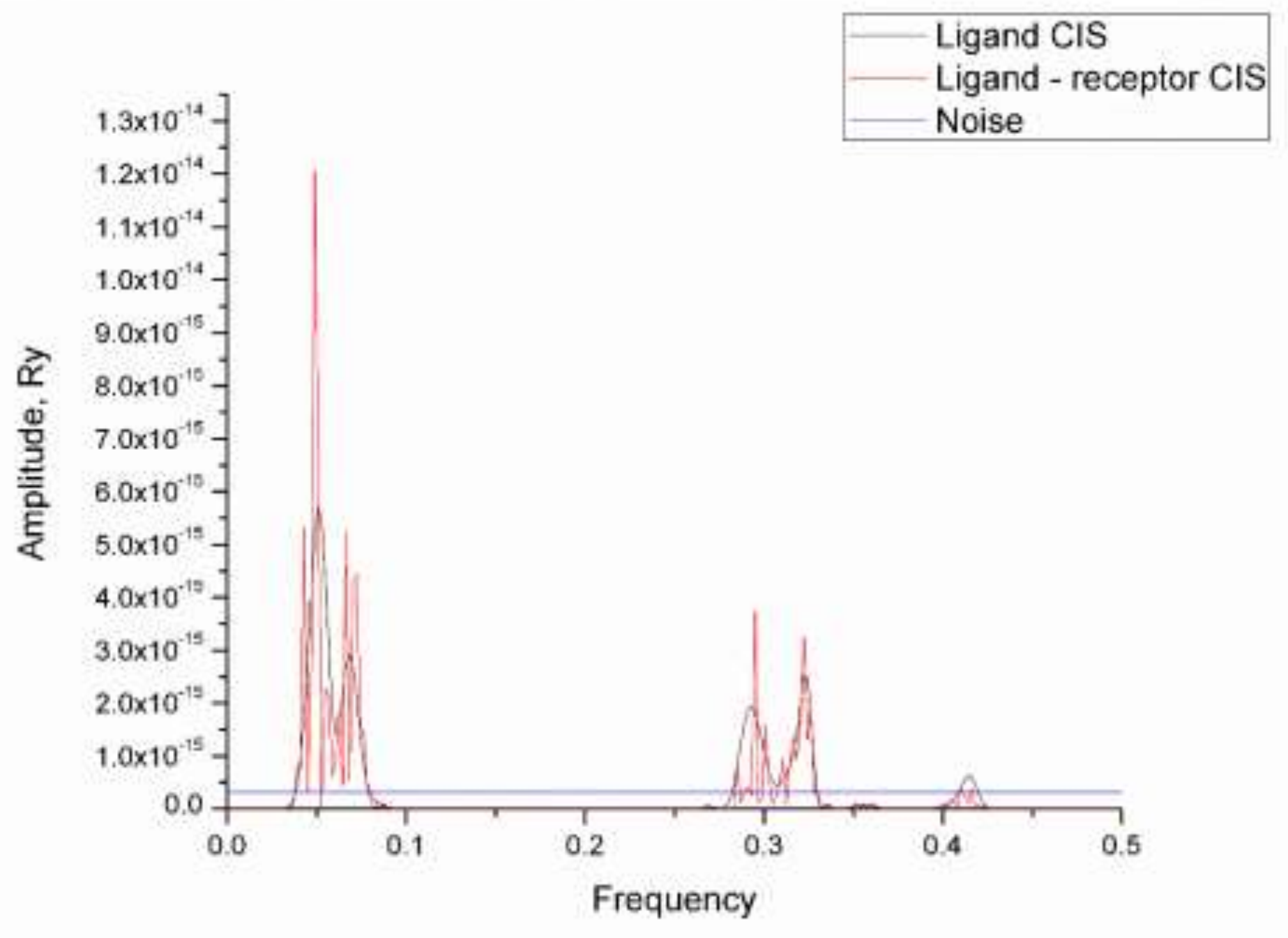

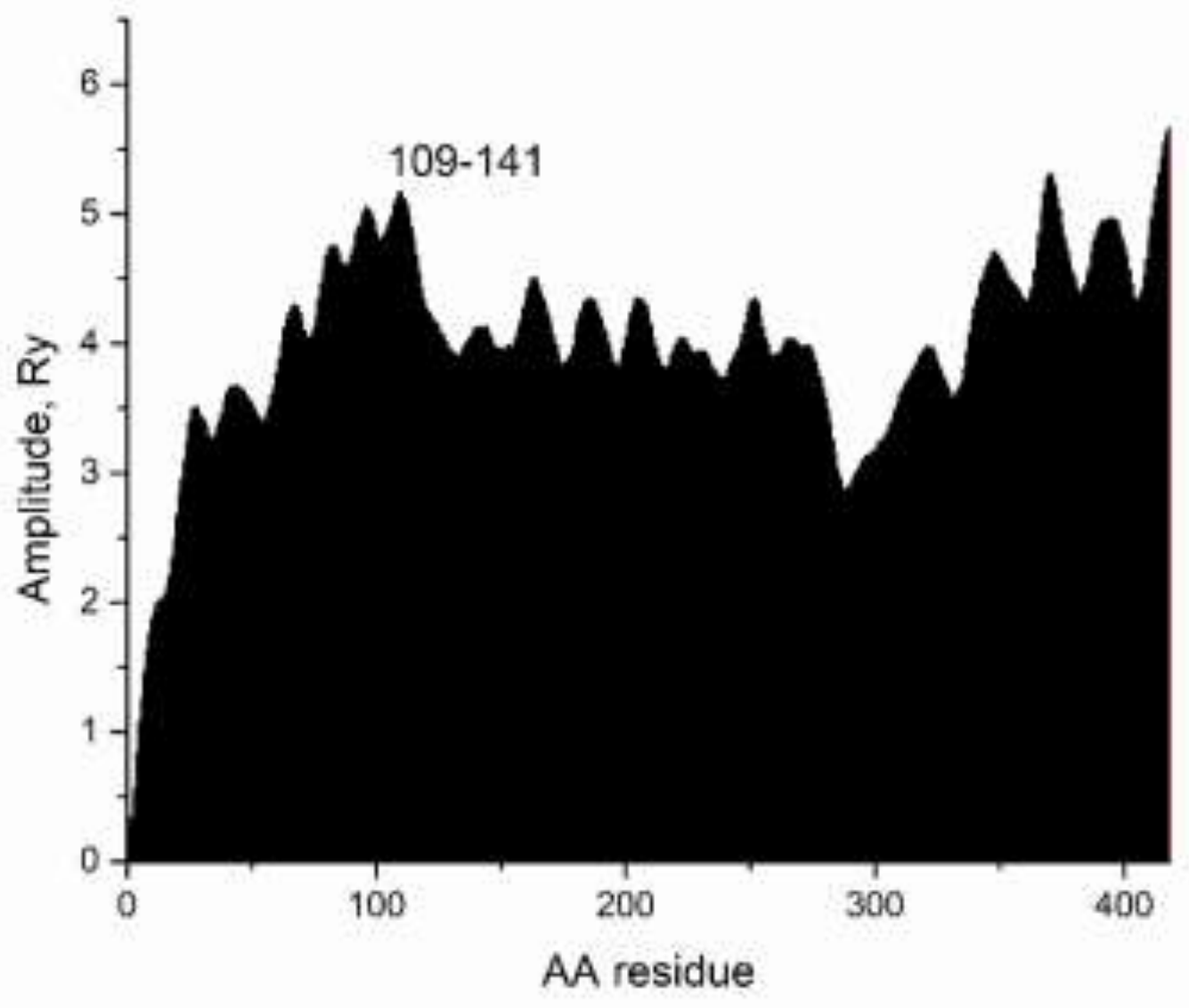

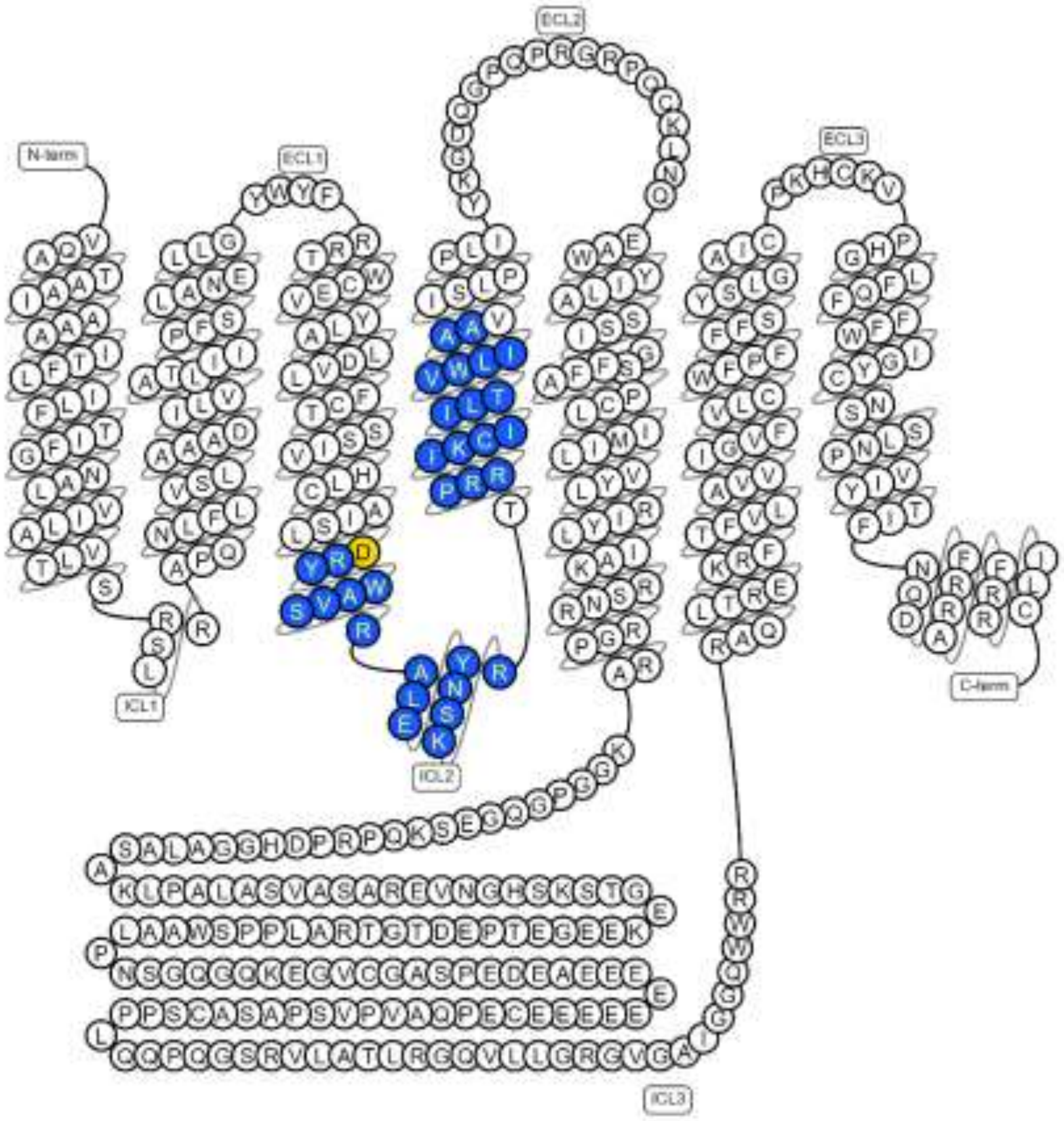

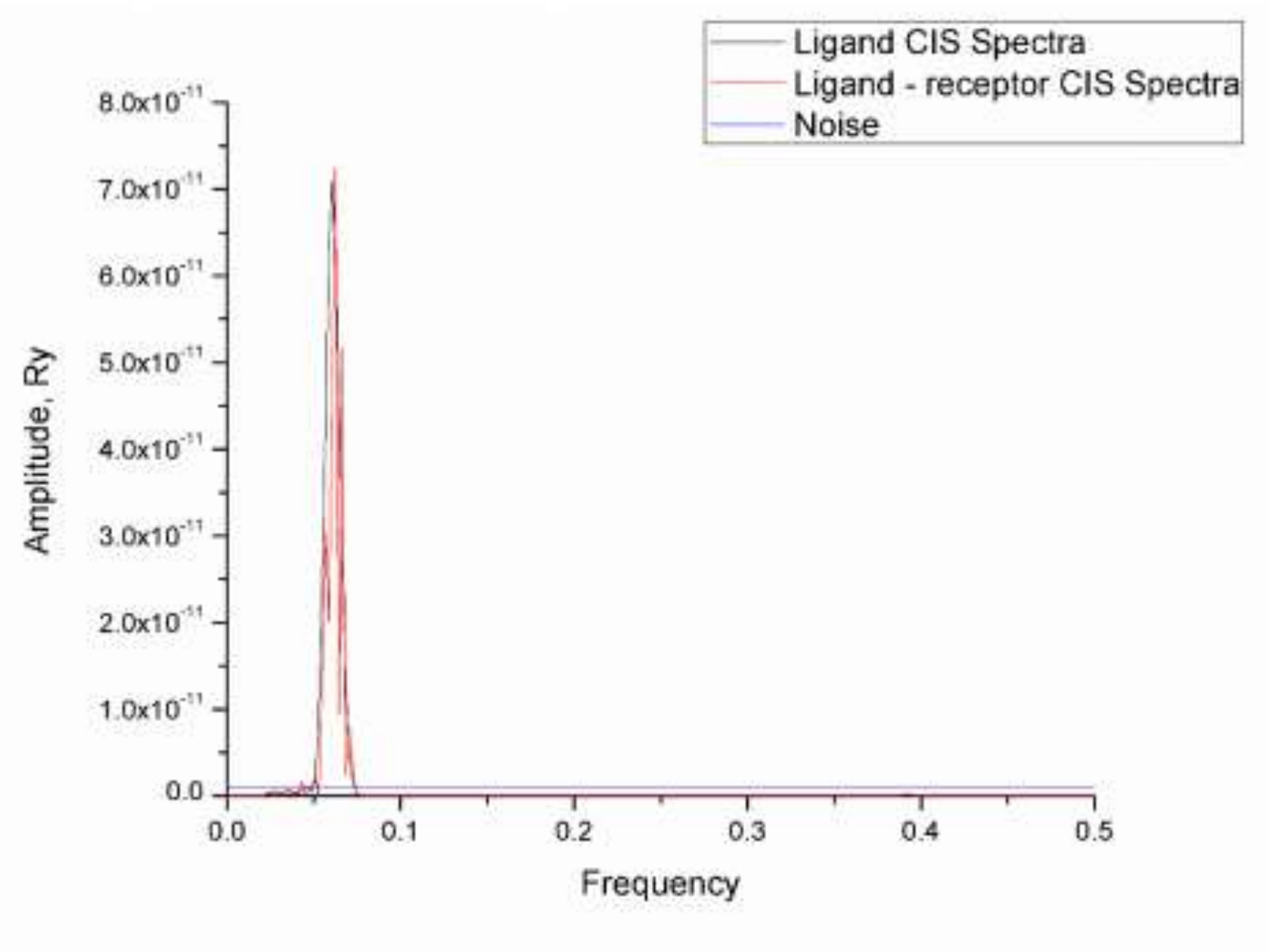

hsa:152 alpha-2c adrenergic receptor

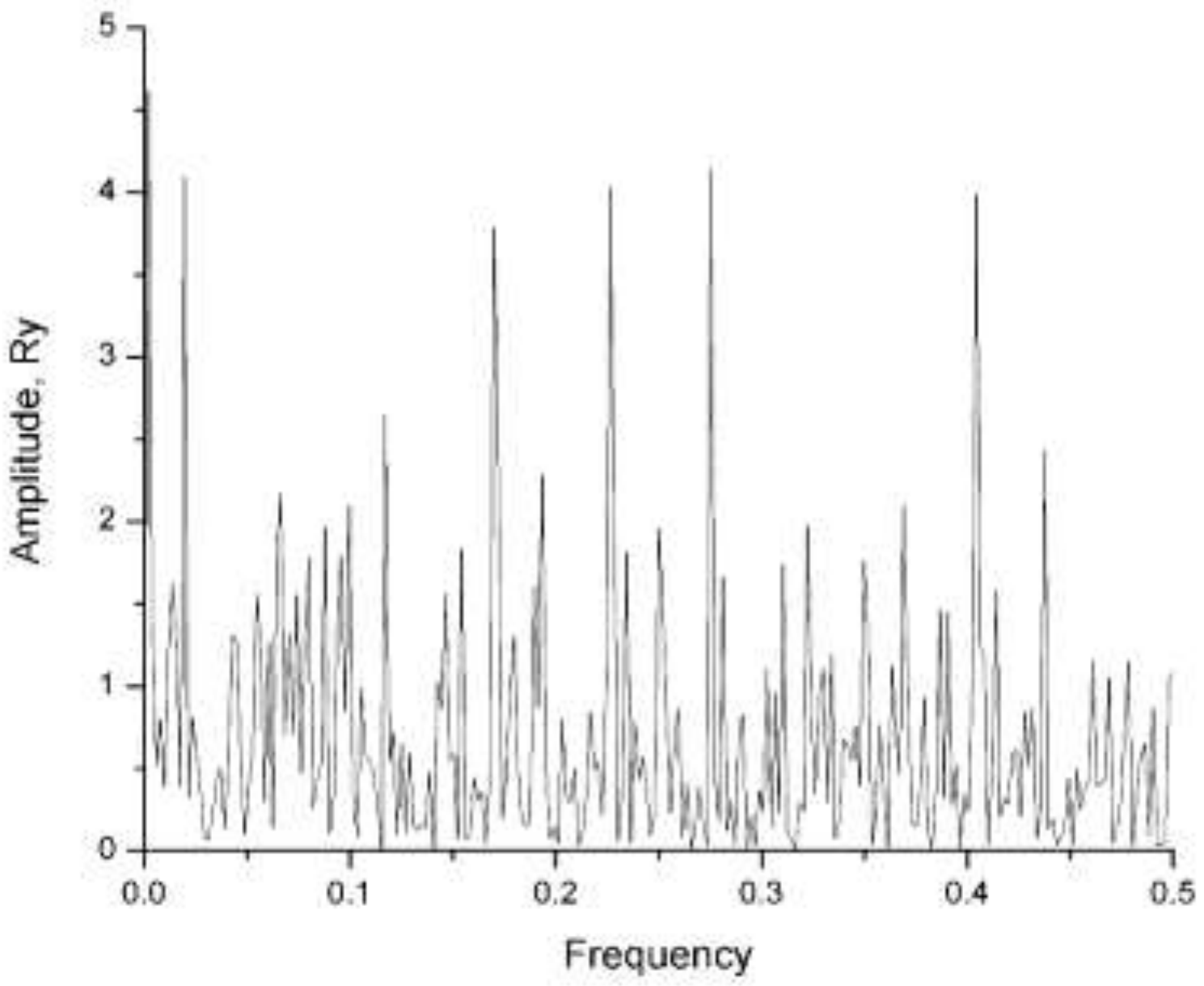

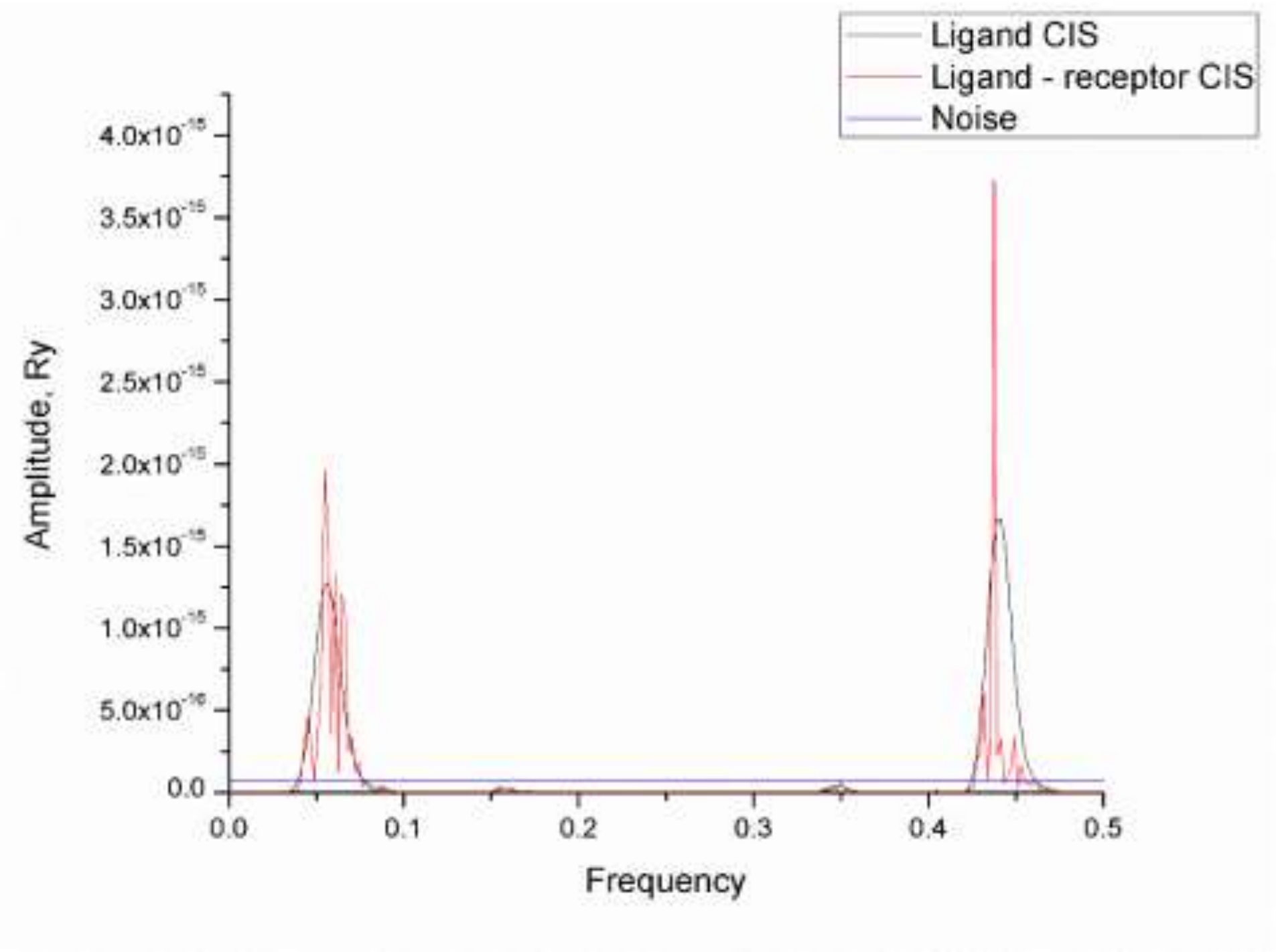

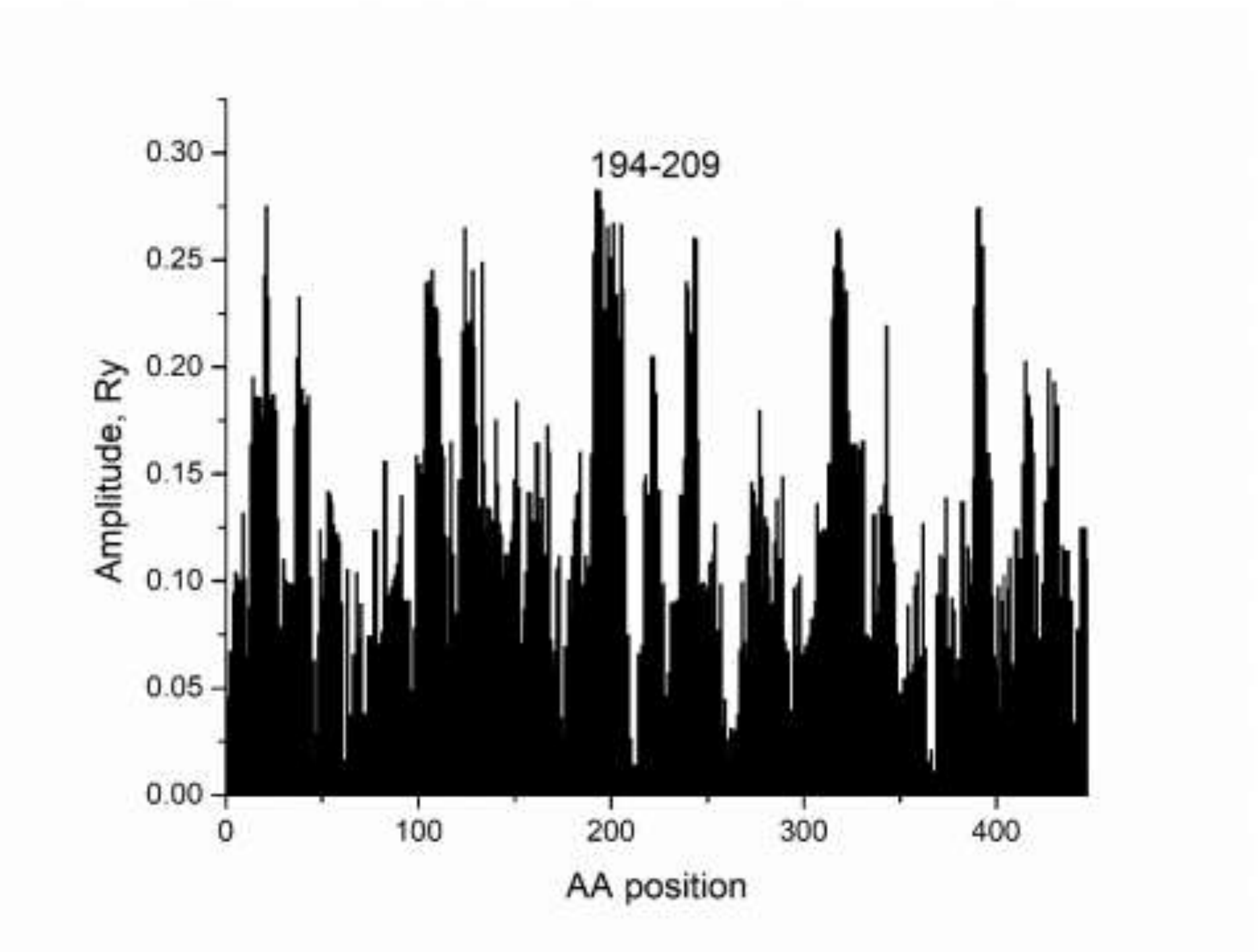

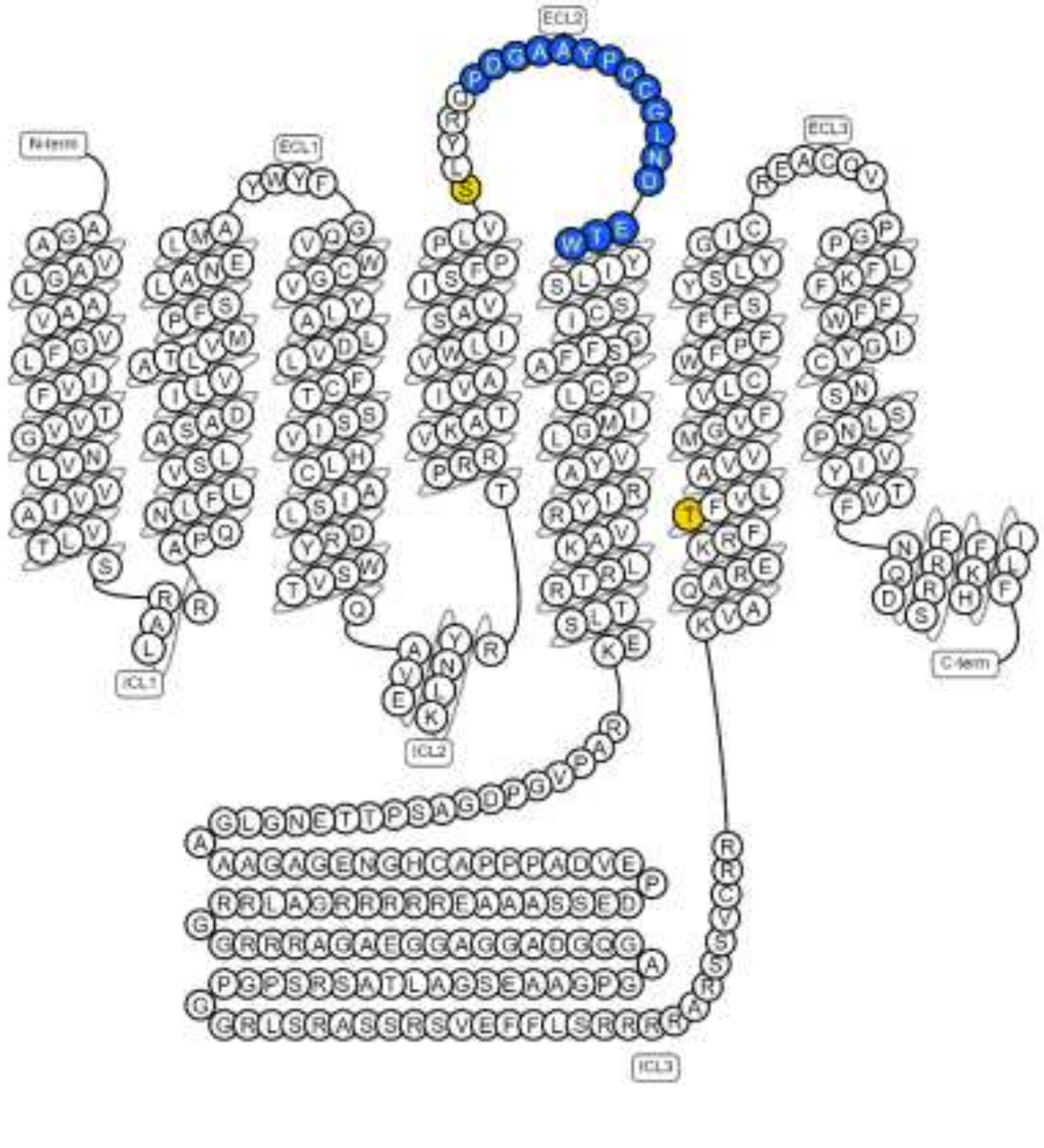

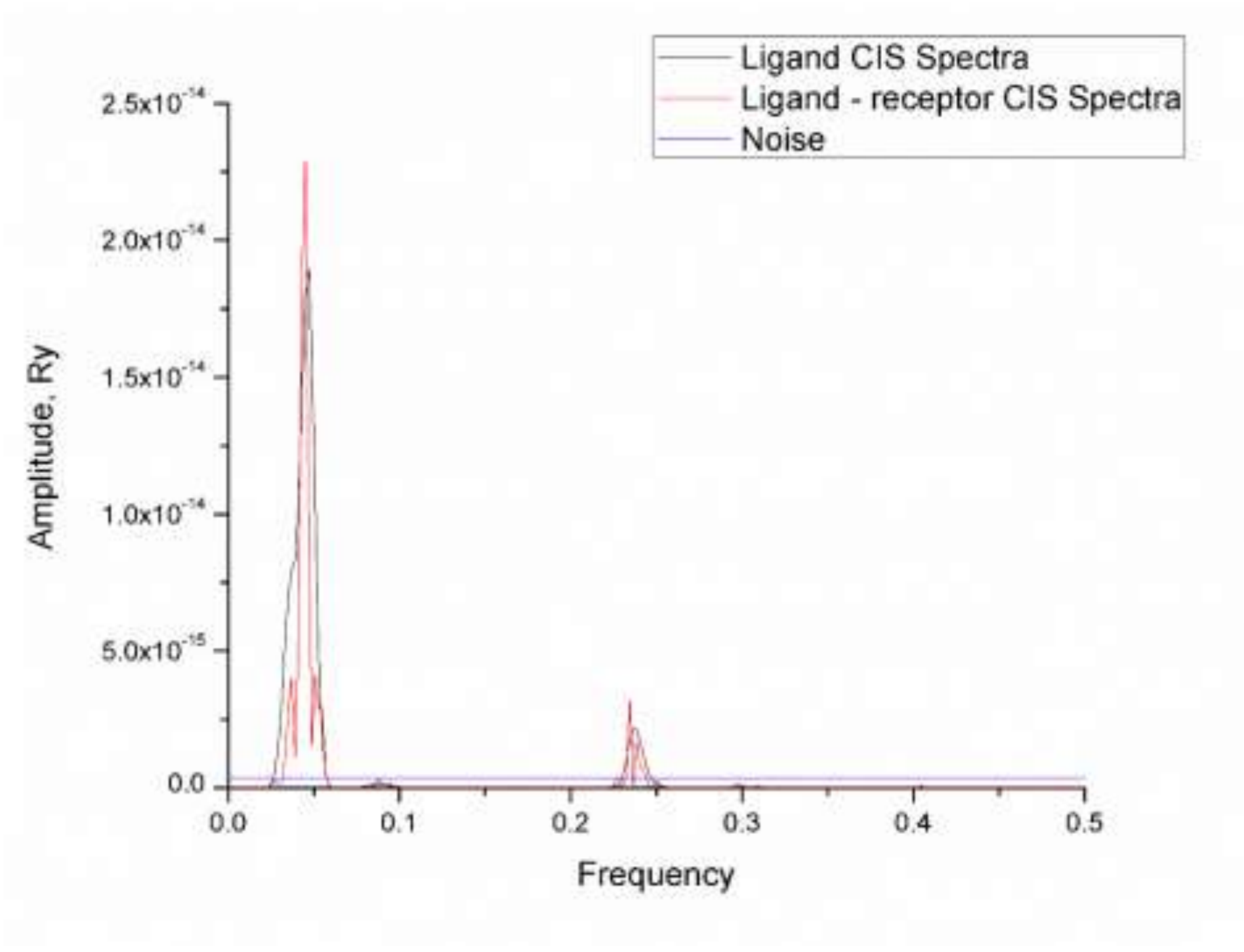

hsa:153 beta-1 adrenergic receptor

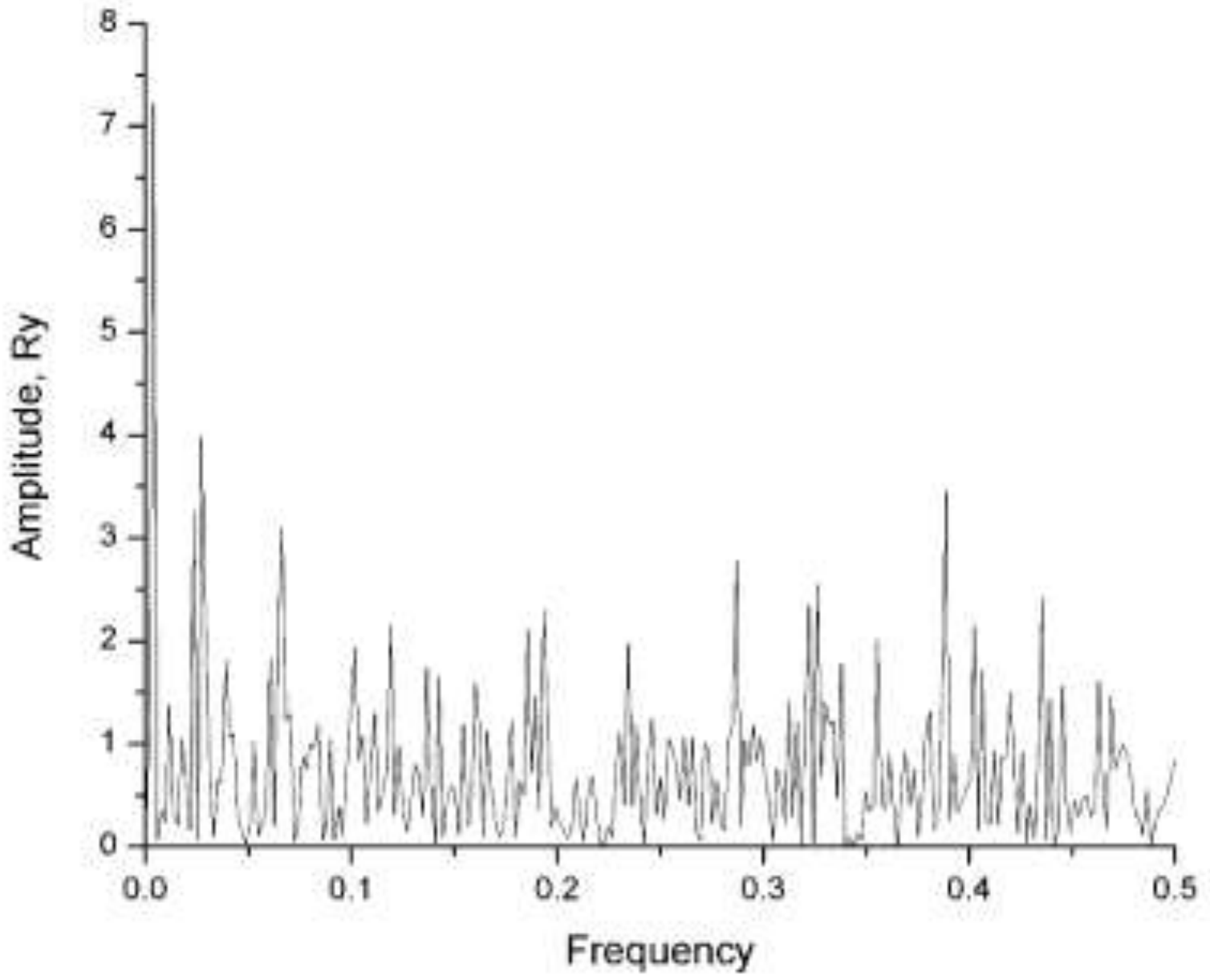

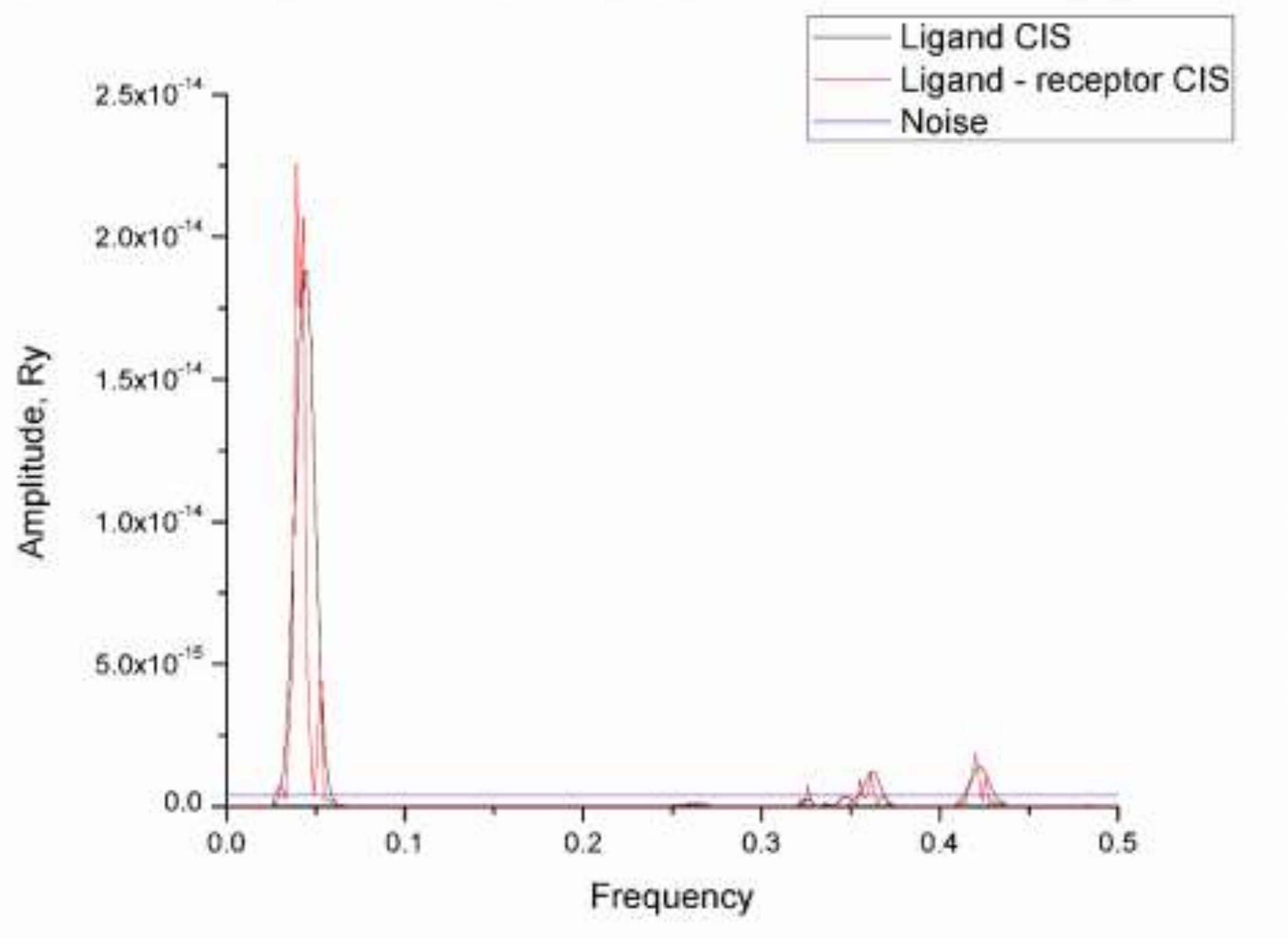

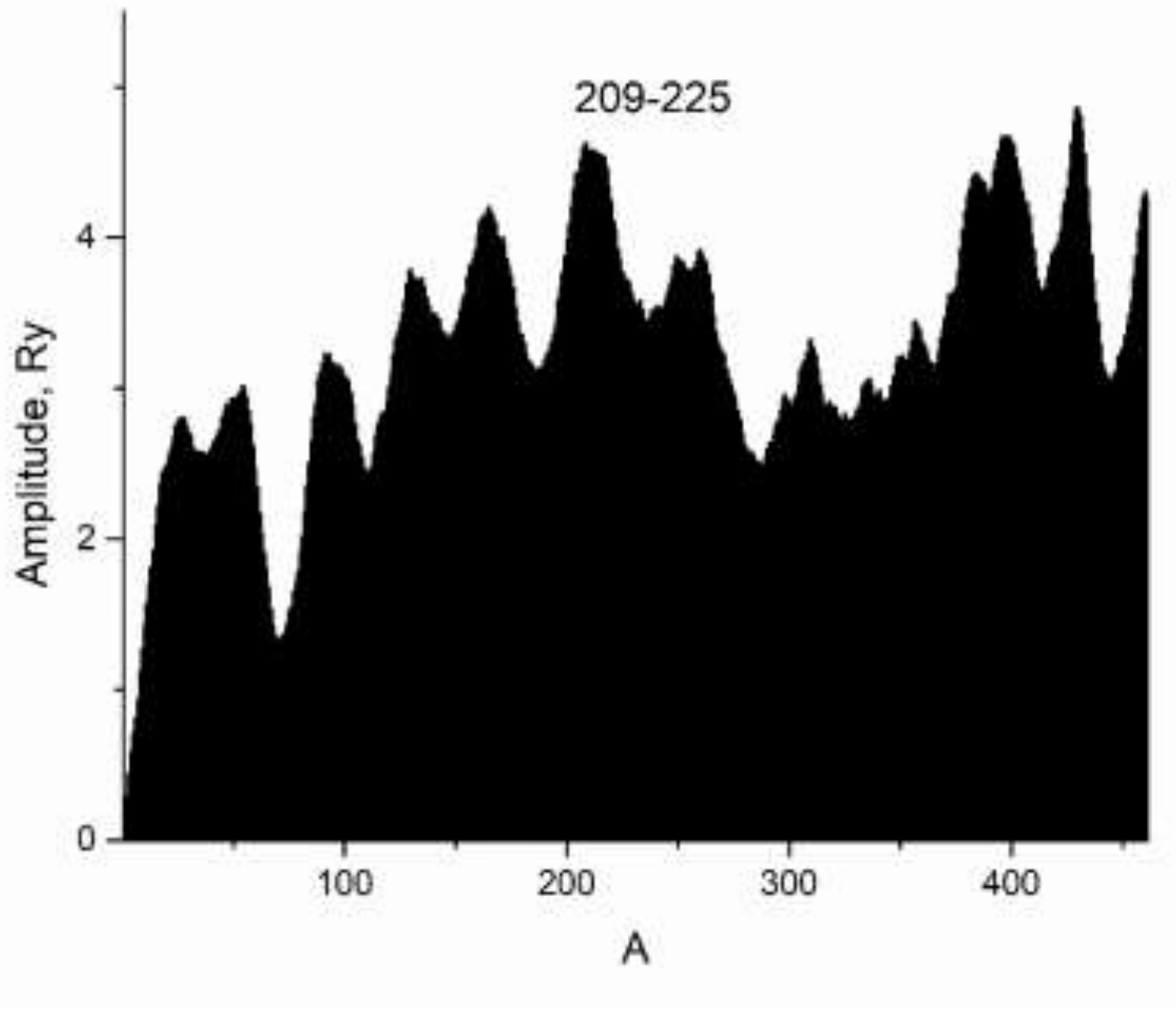

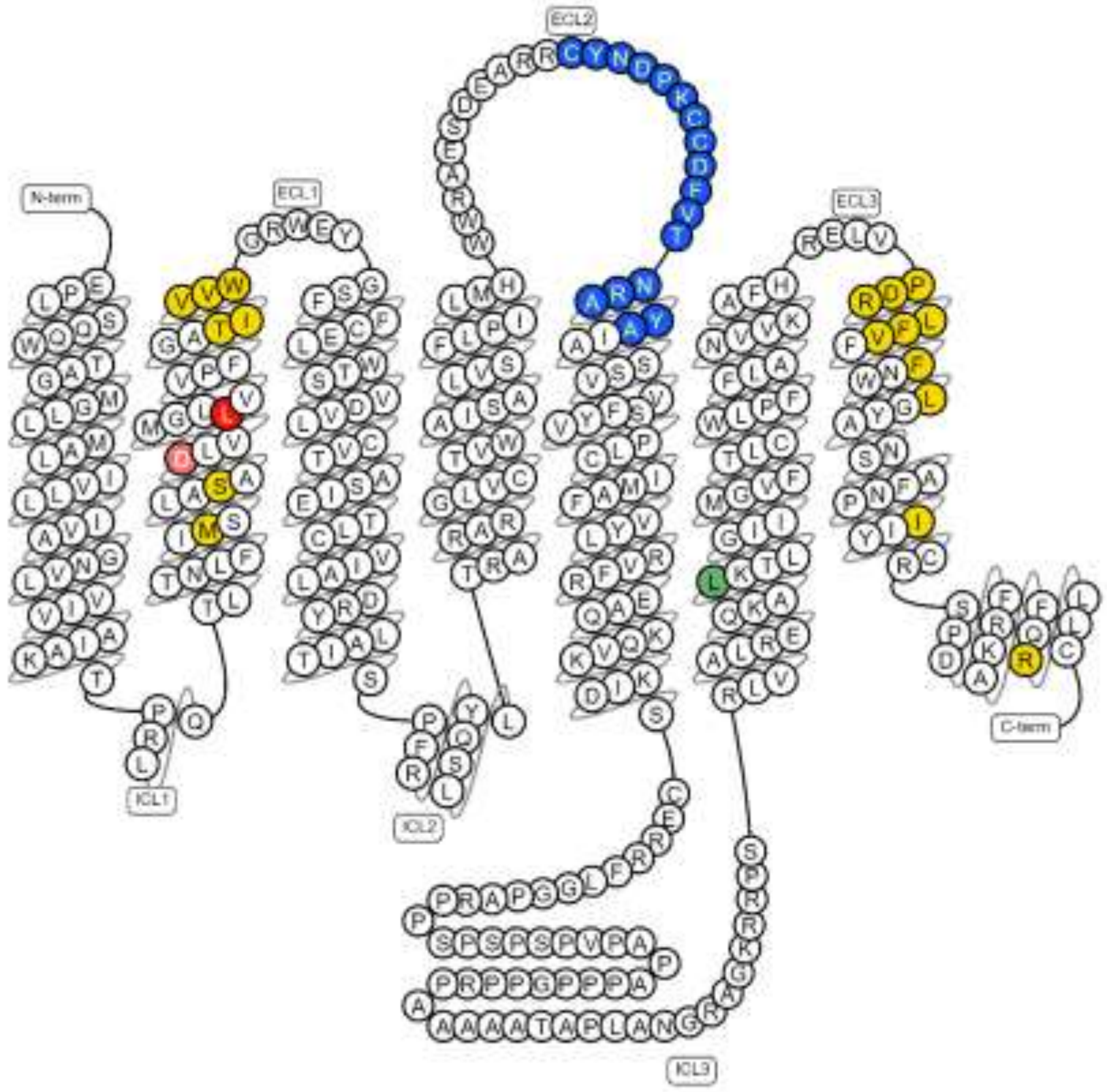

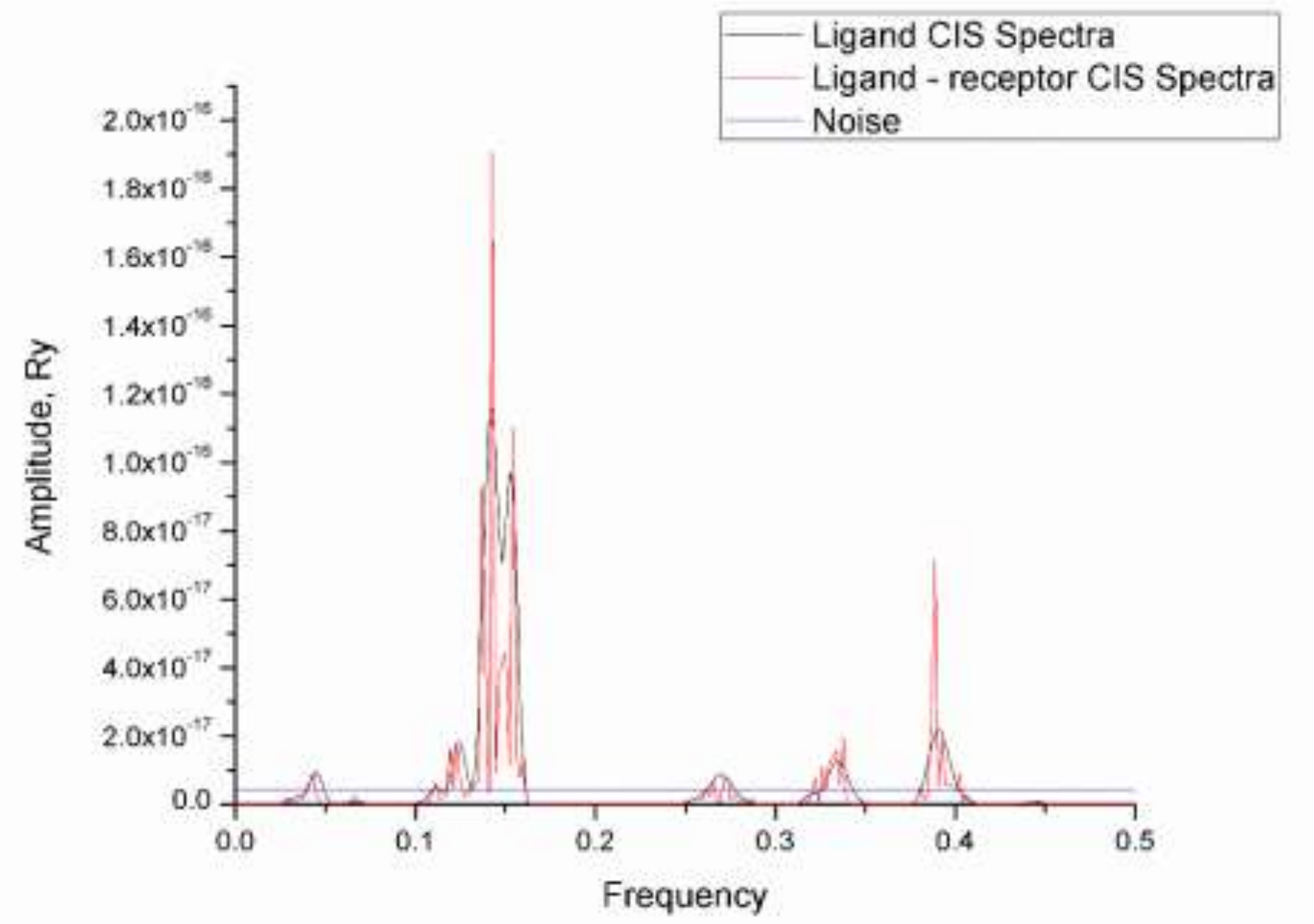

hsa:154 beta-2 adrenergic receptor

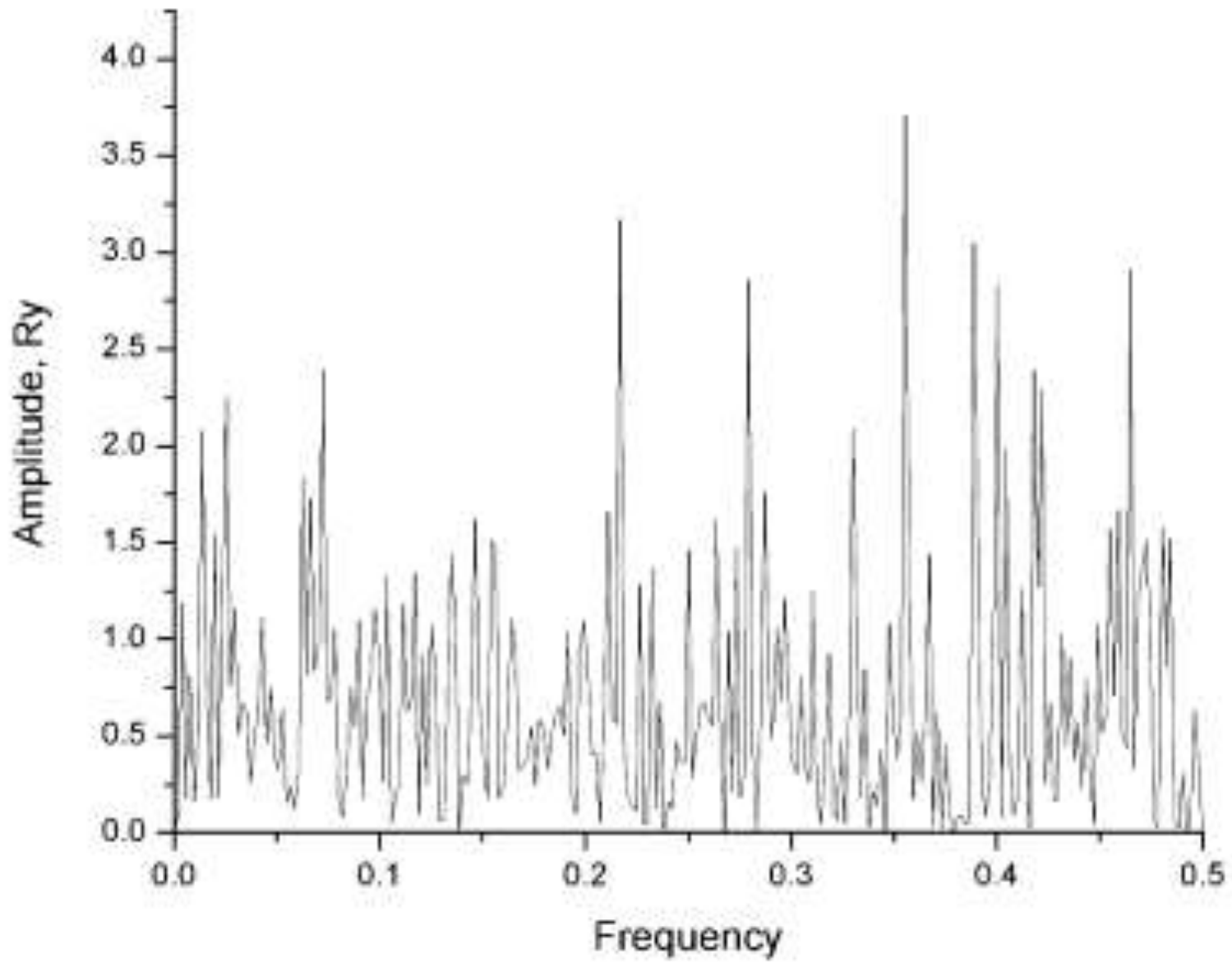

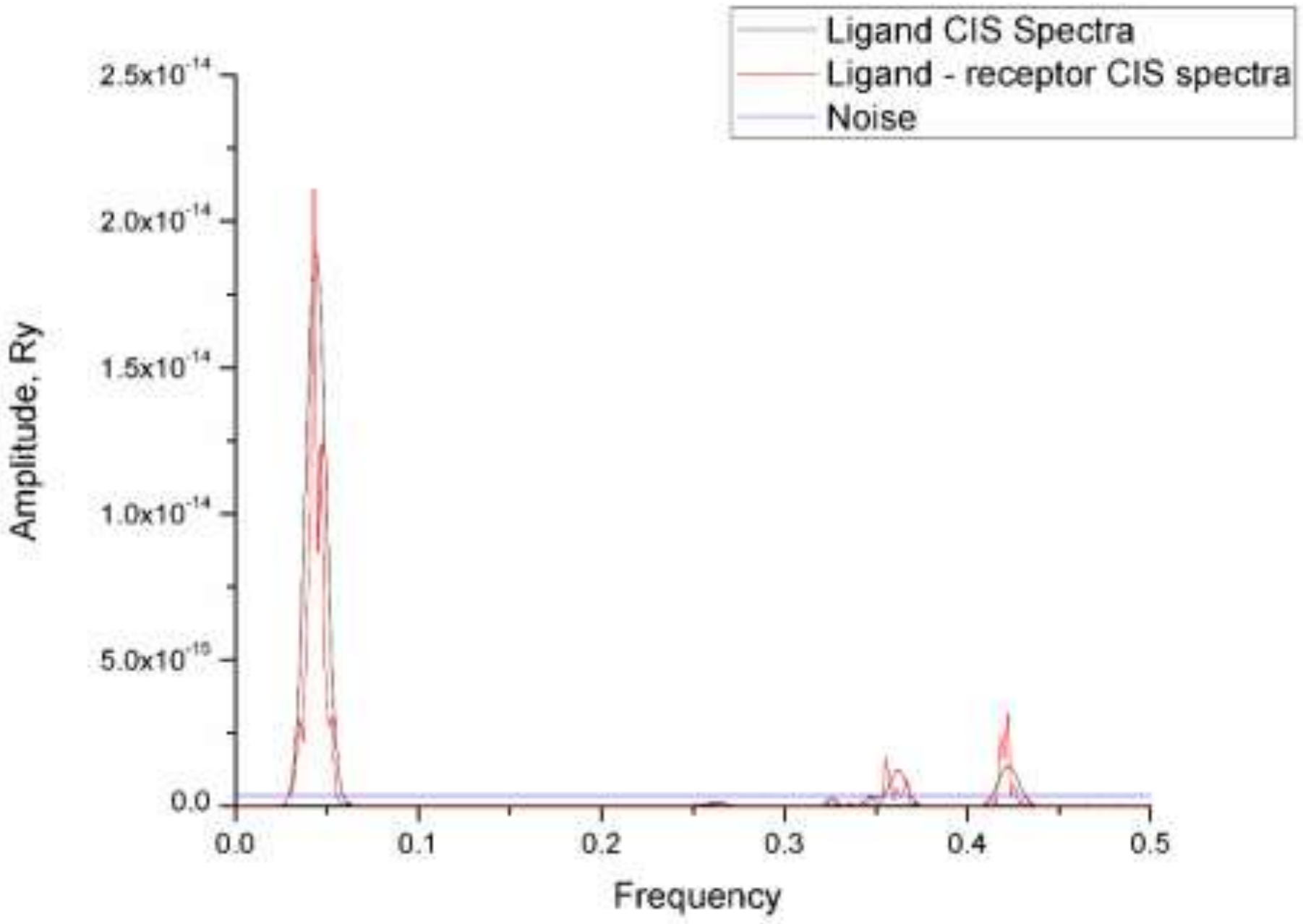

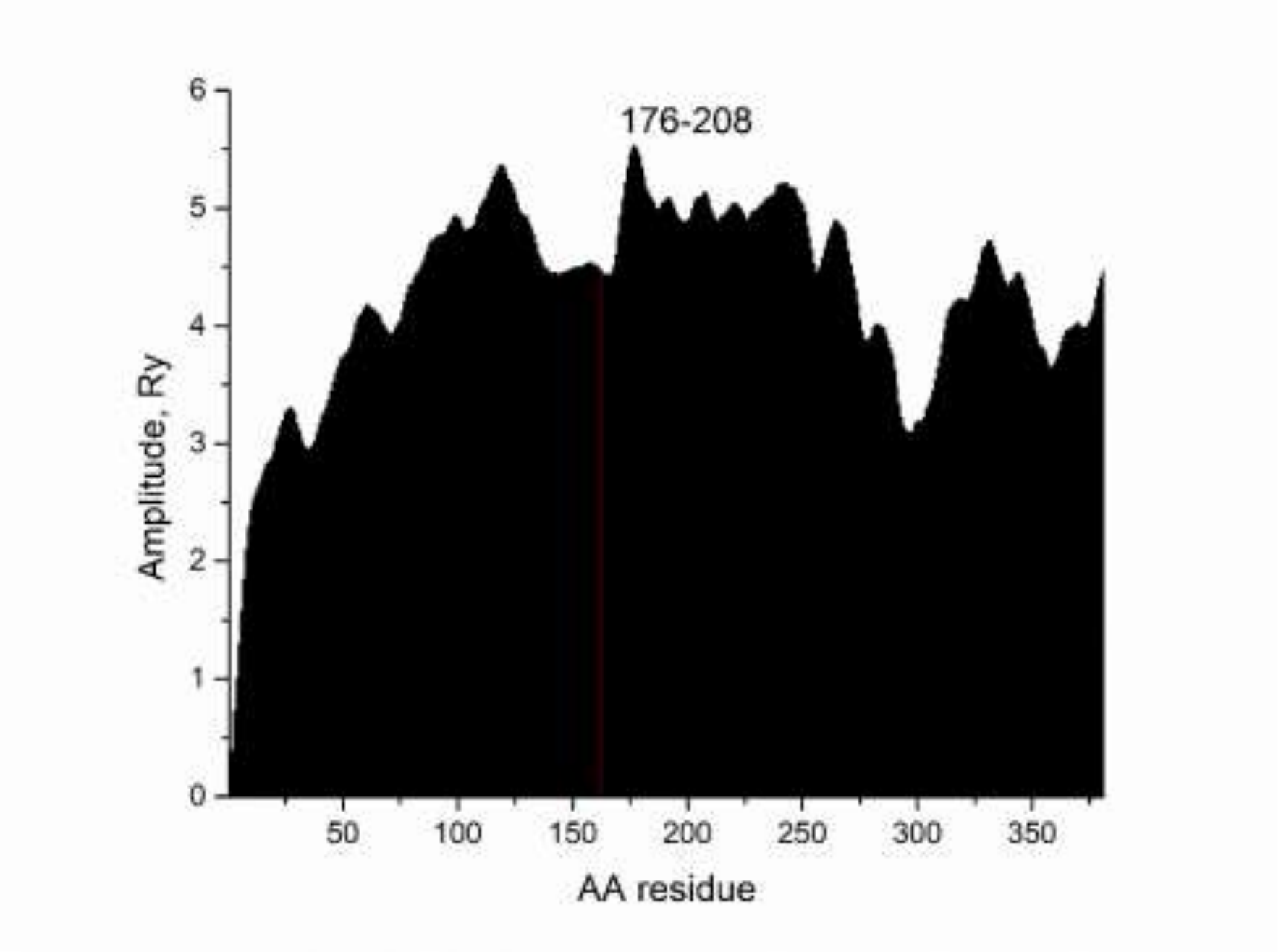

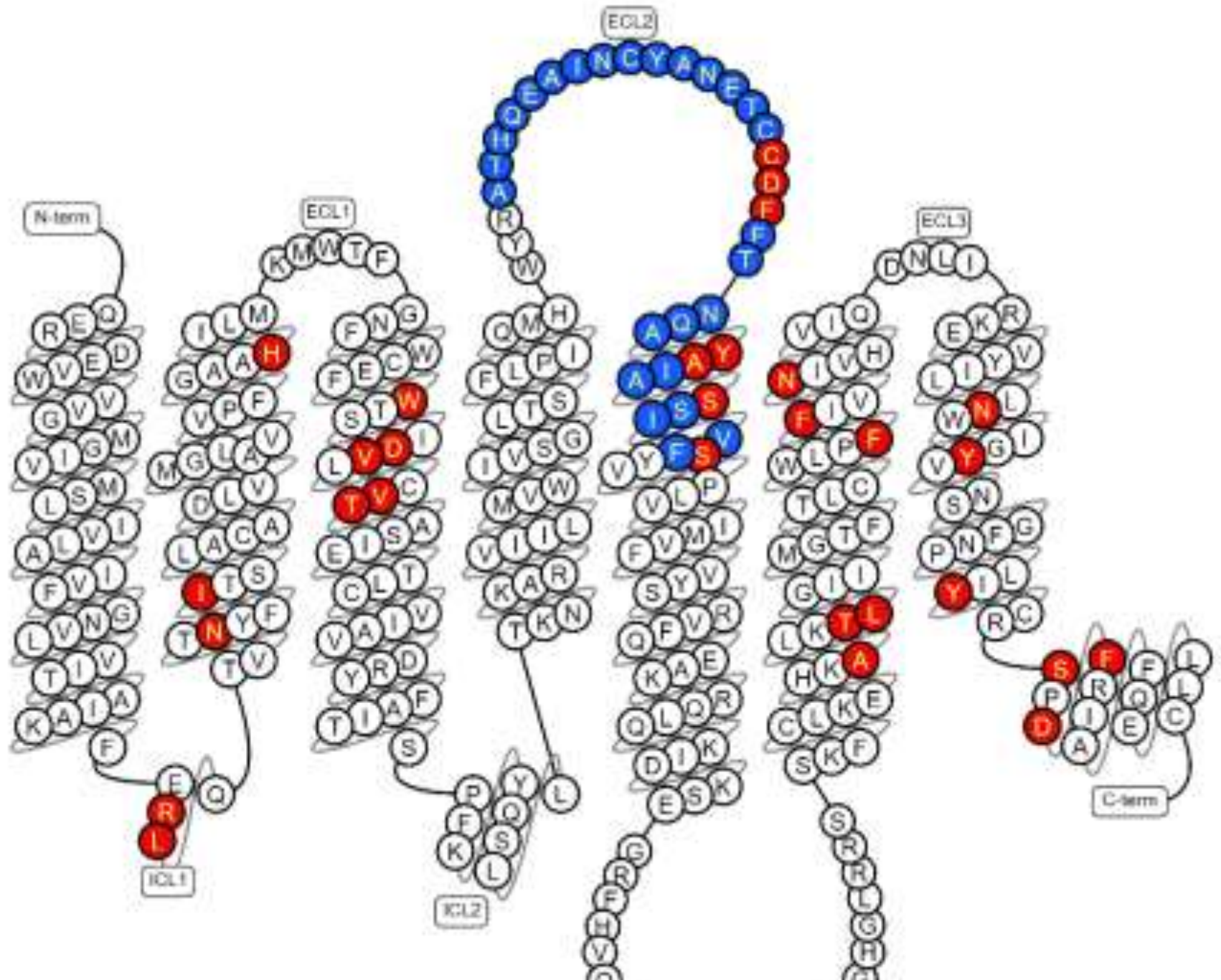

hsa:155 beta-3 adrenergic receptor

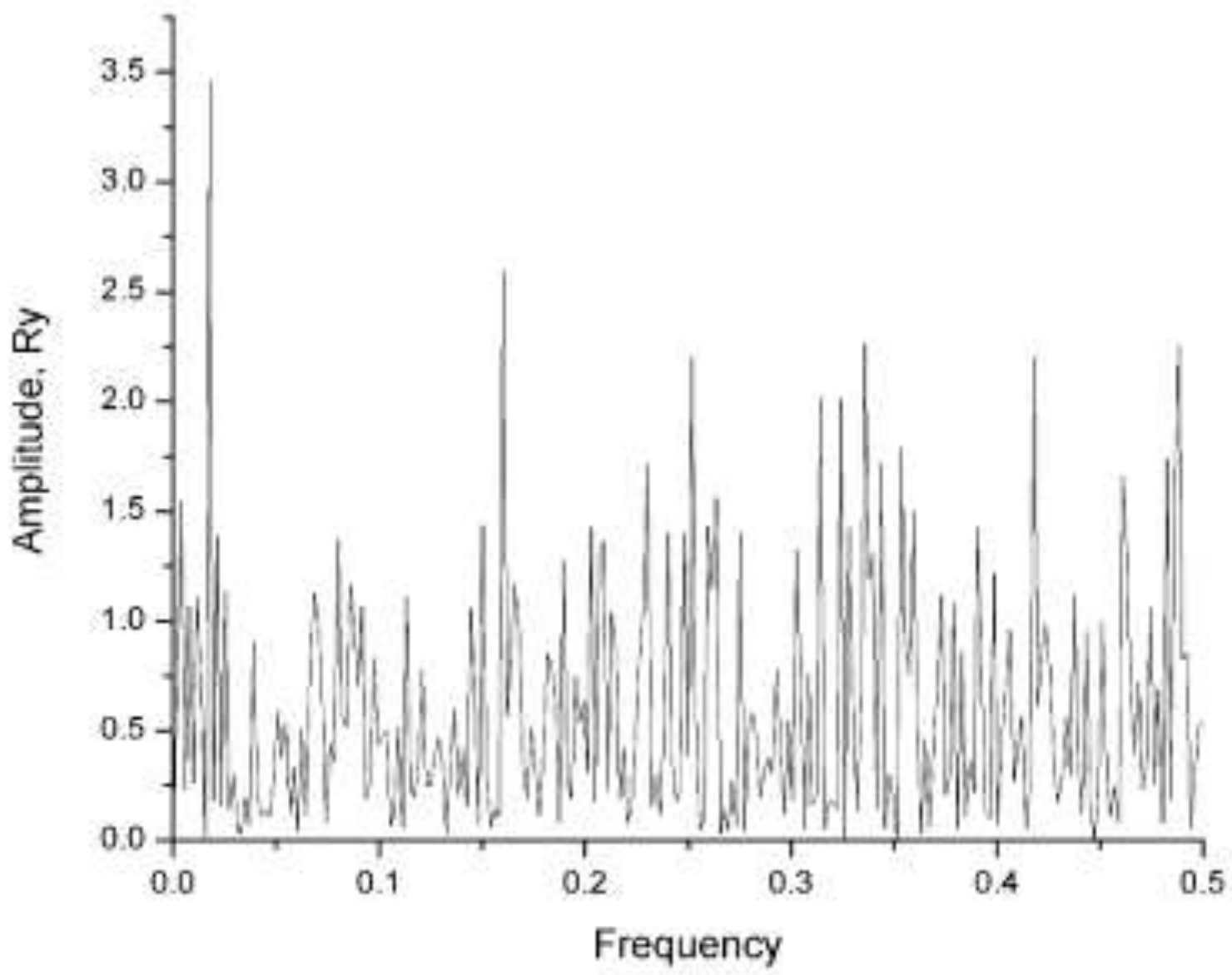

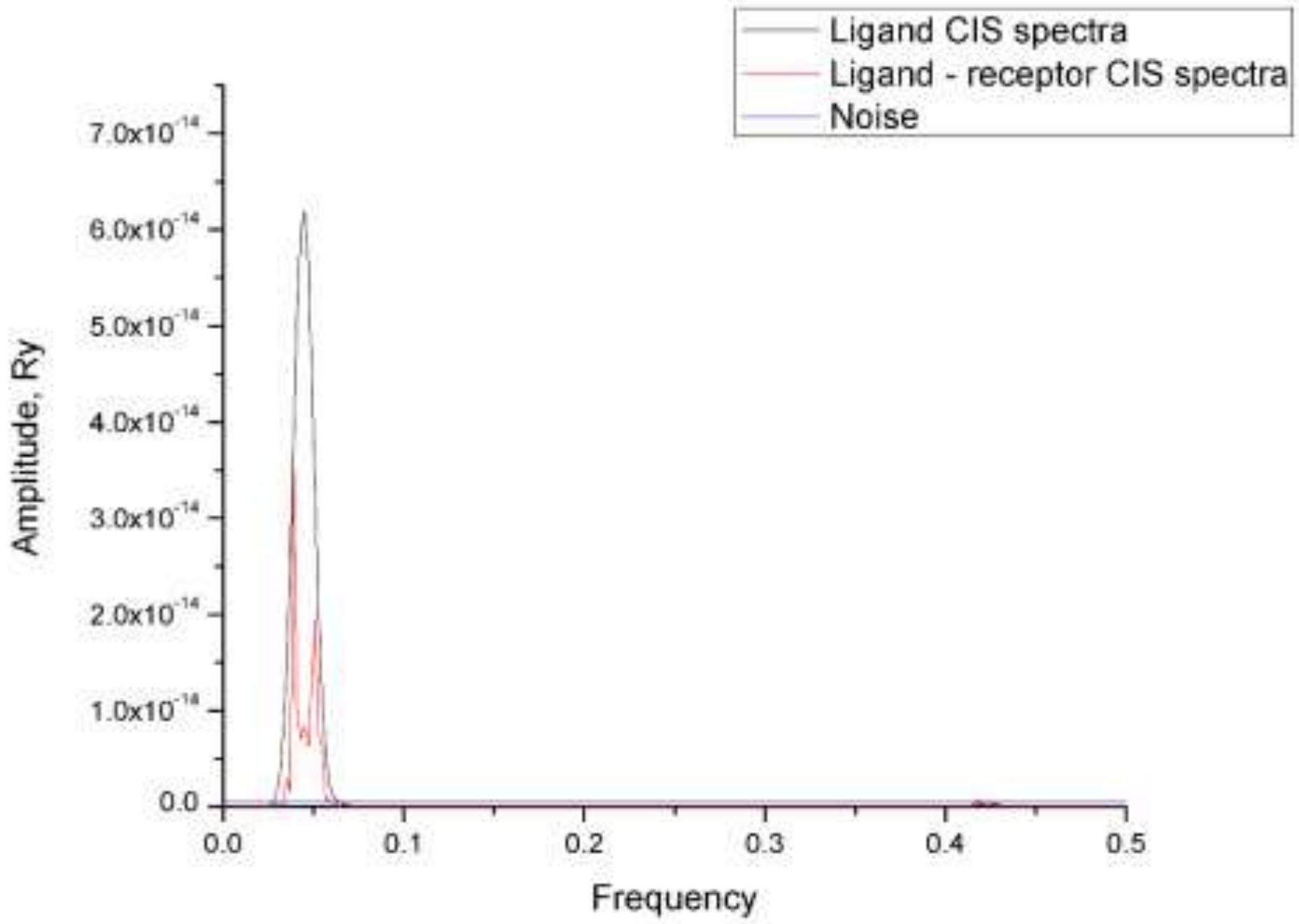

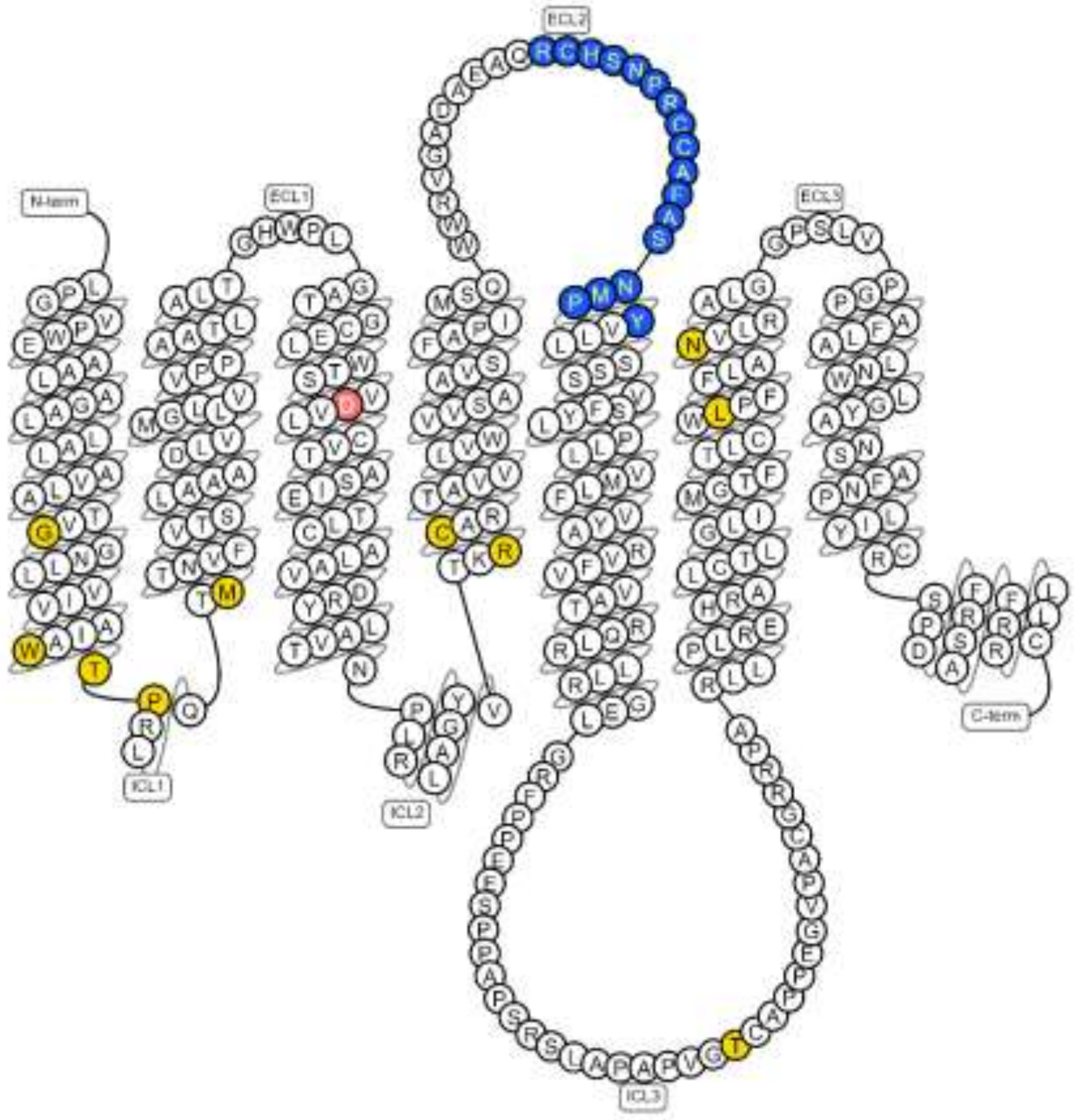

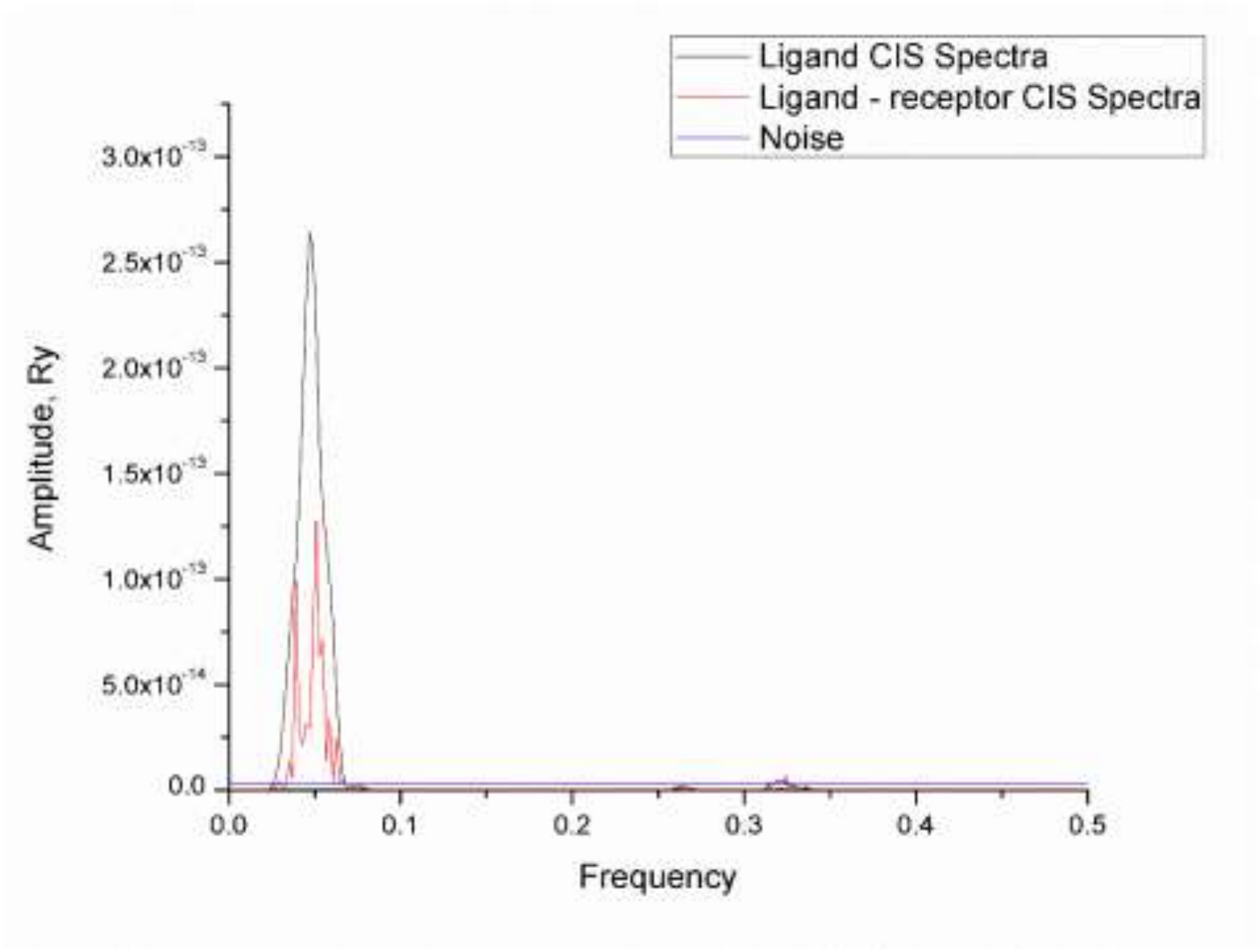

hsa 1128 muscarinic acetylcholine receptor m1

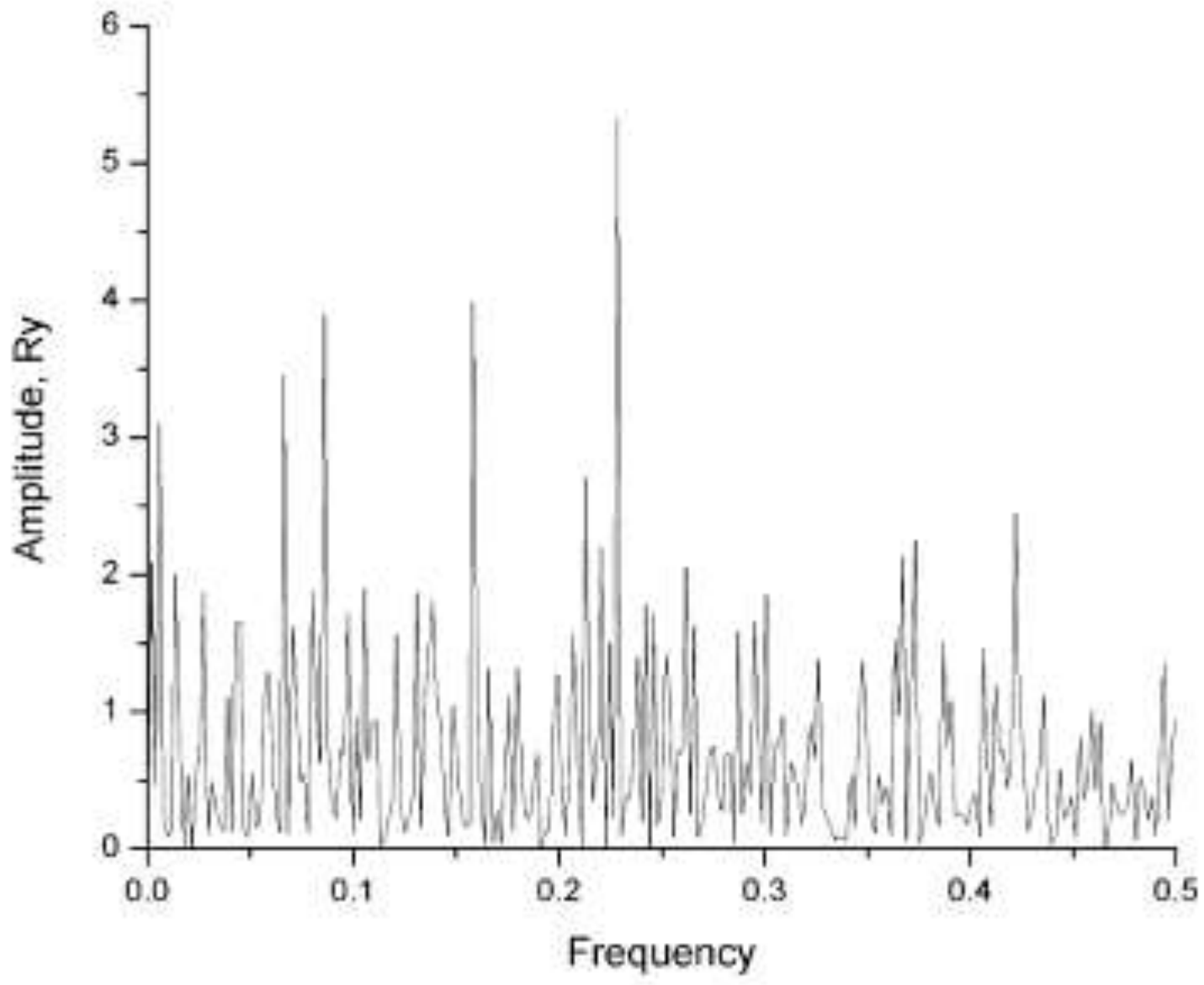

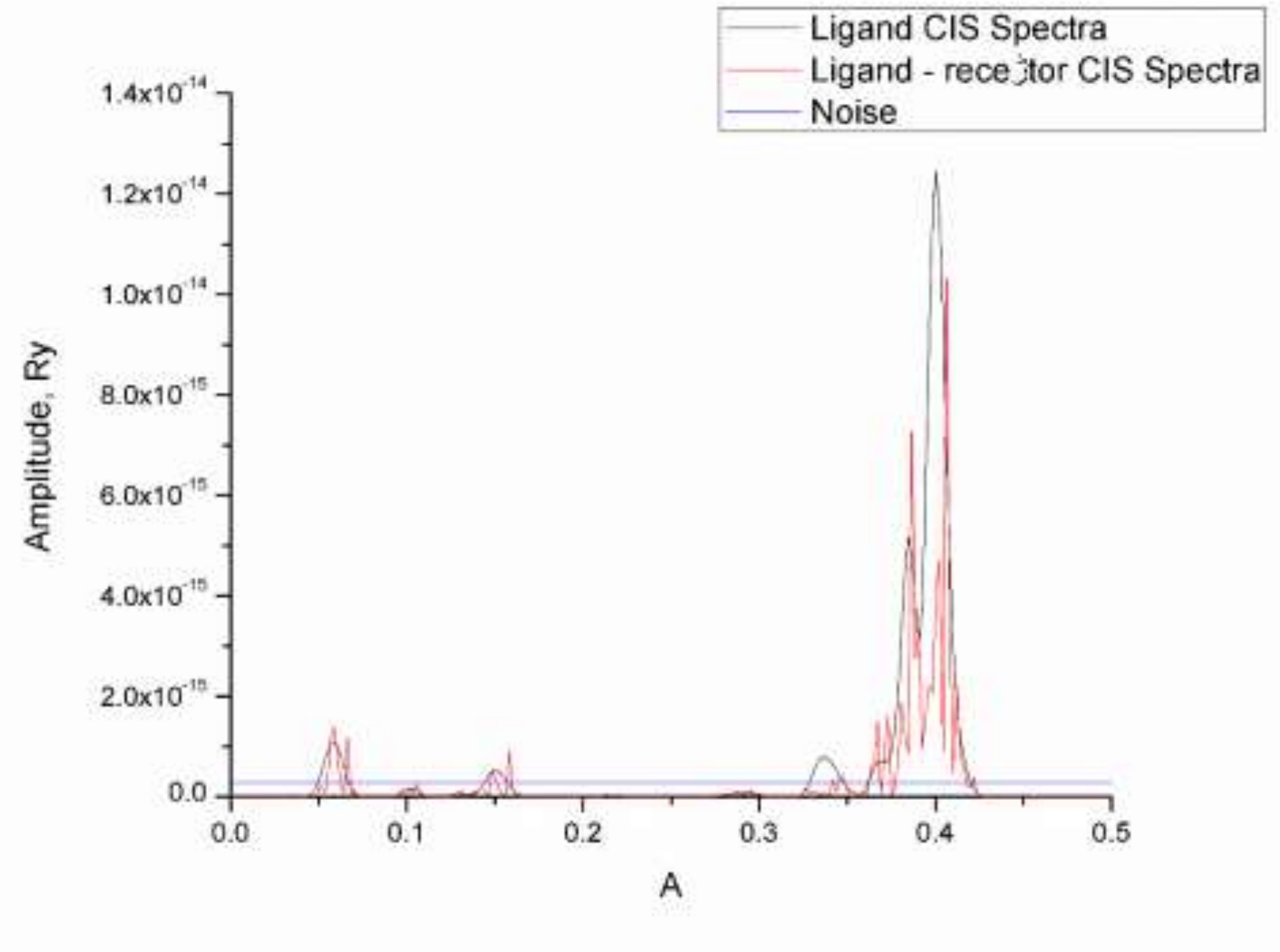

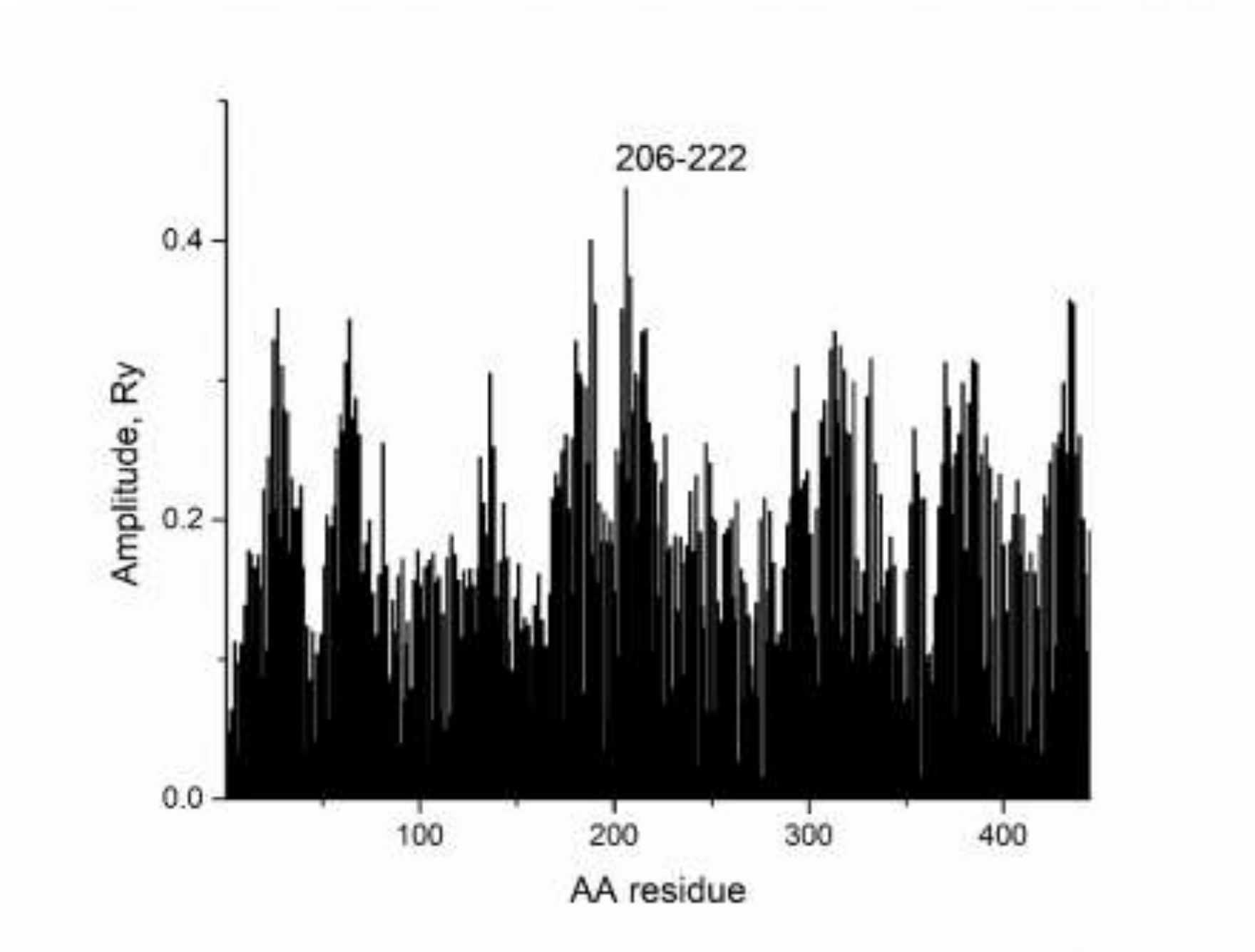

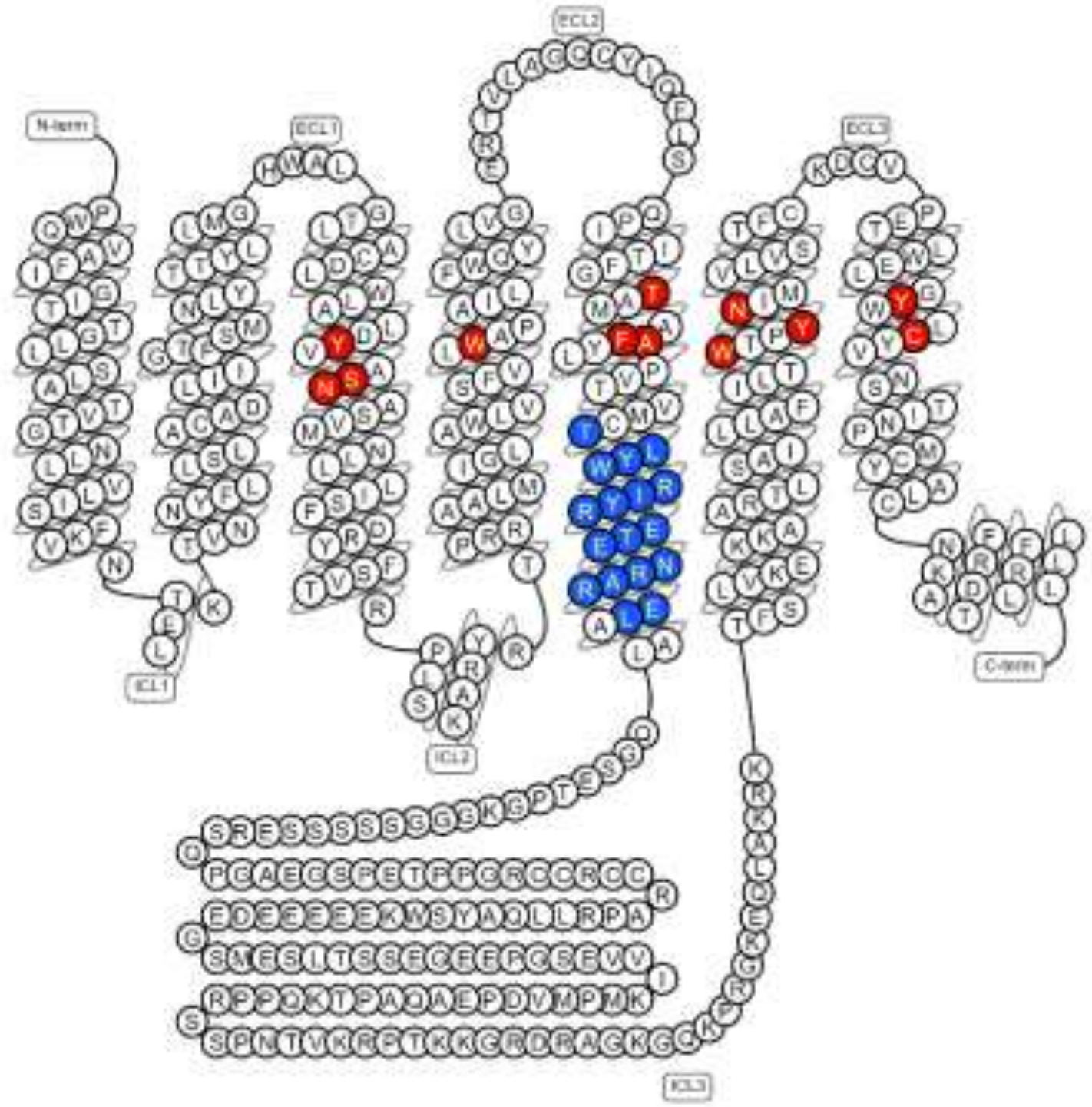

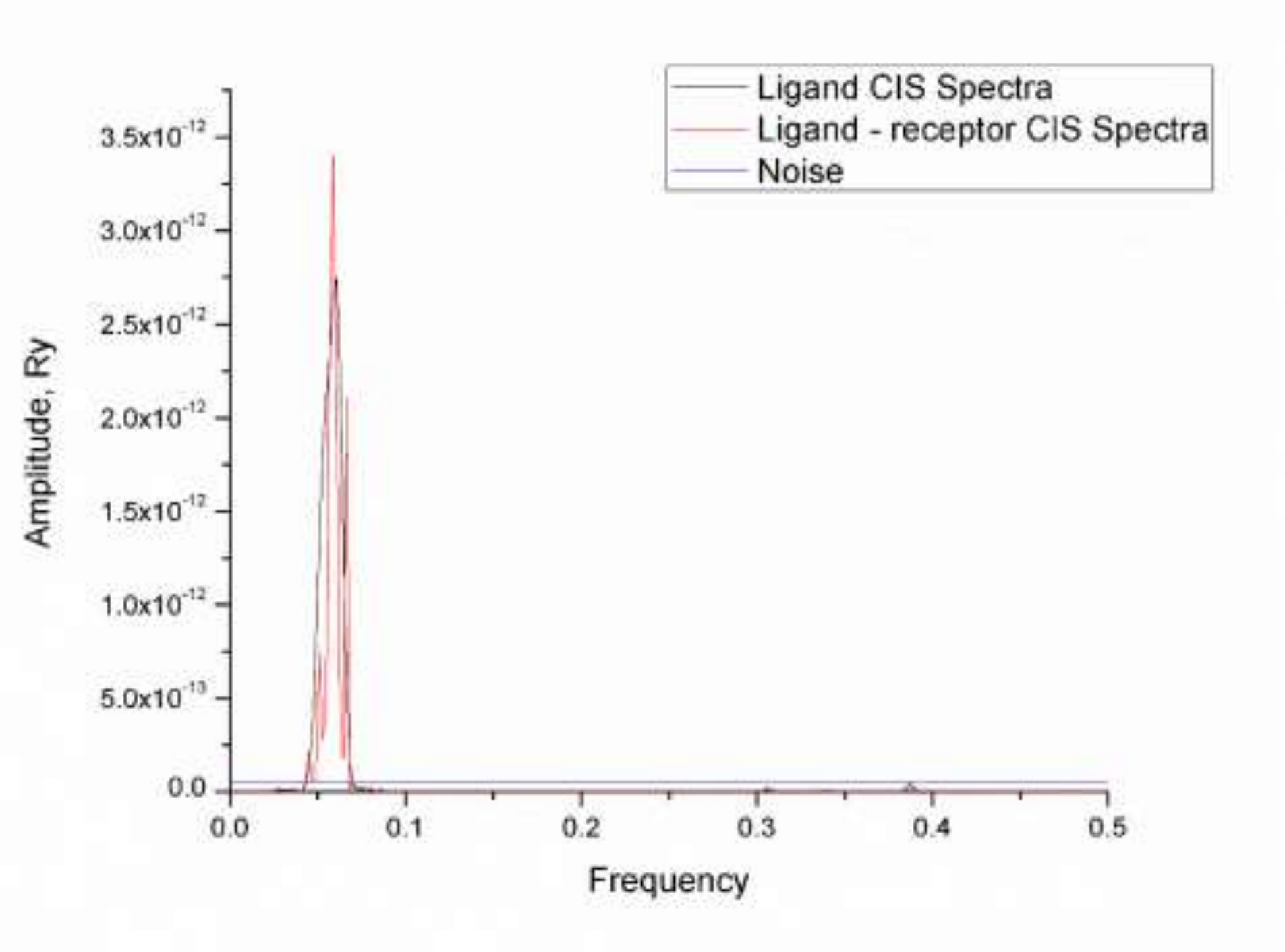

hsa 1129 muscarinic acetylcholine receptor m2

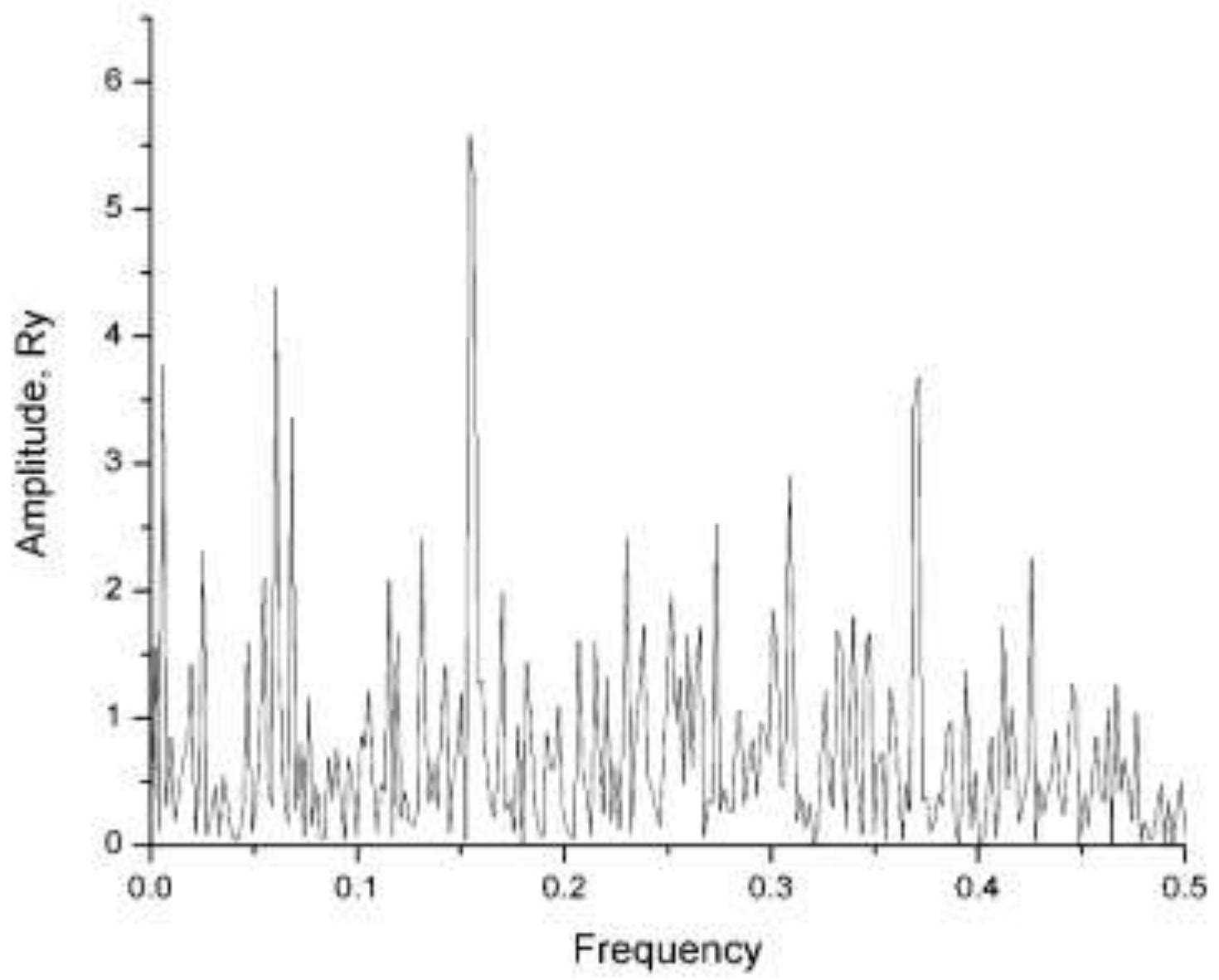

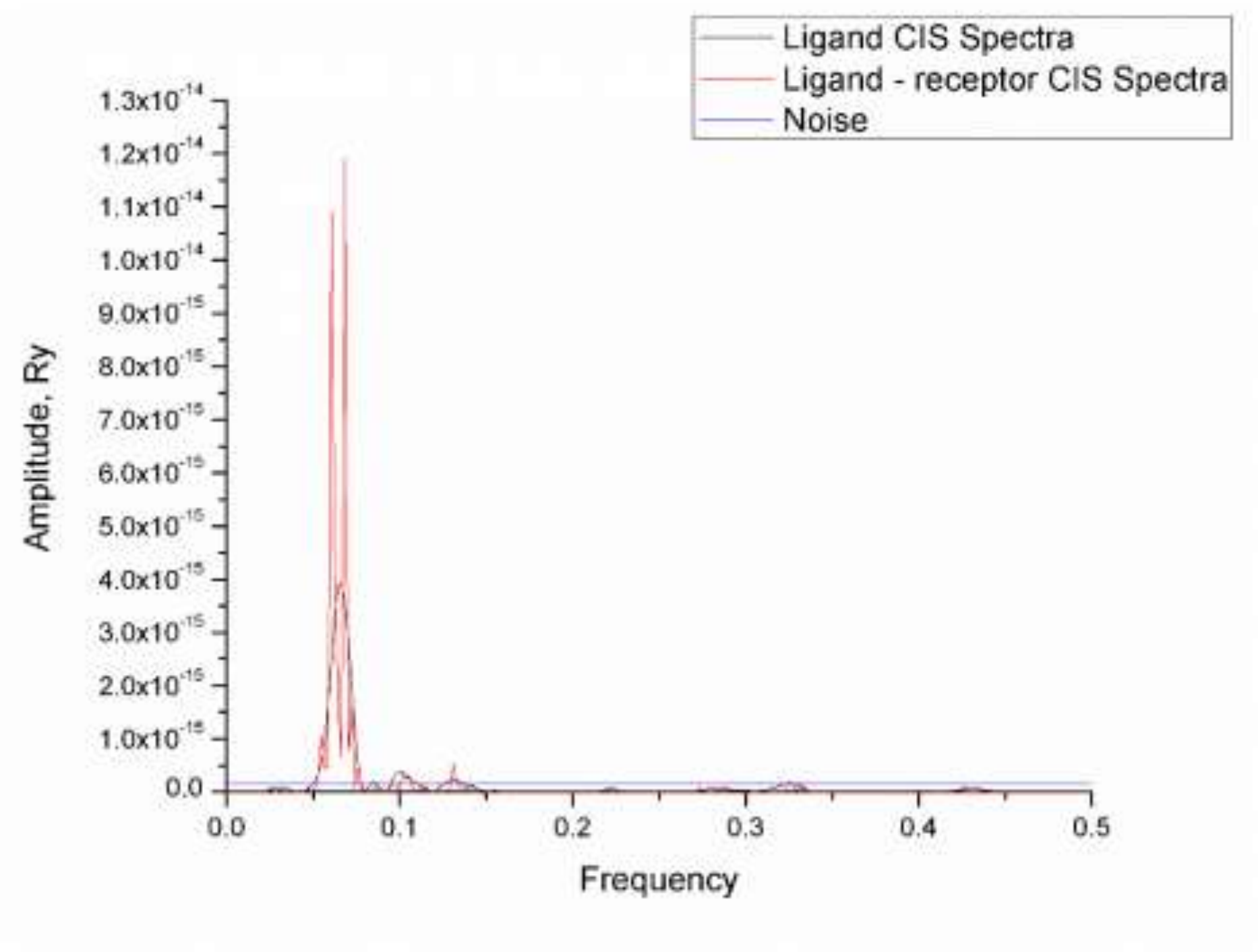

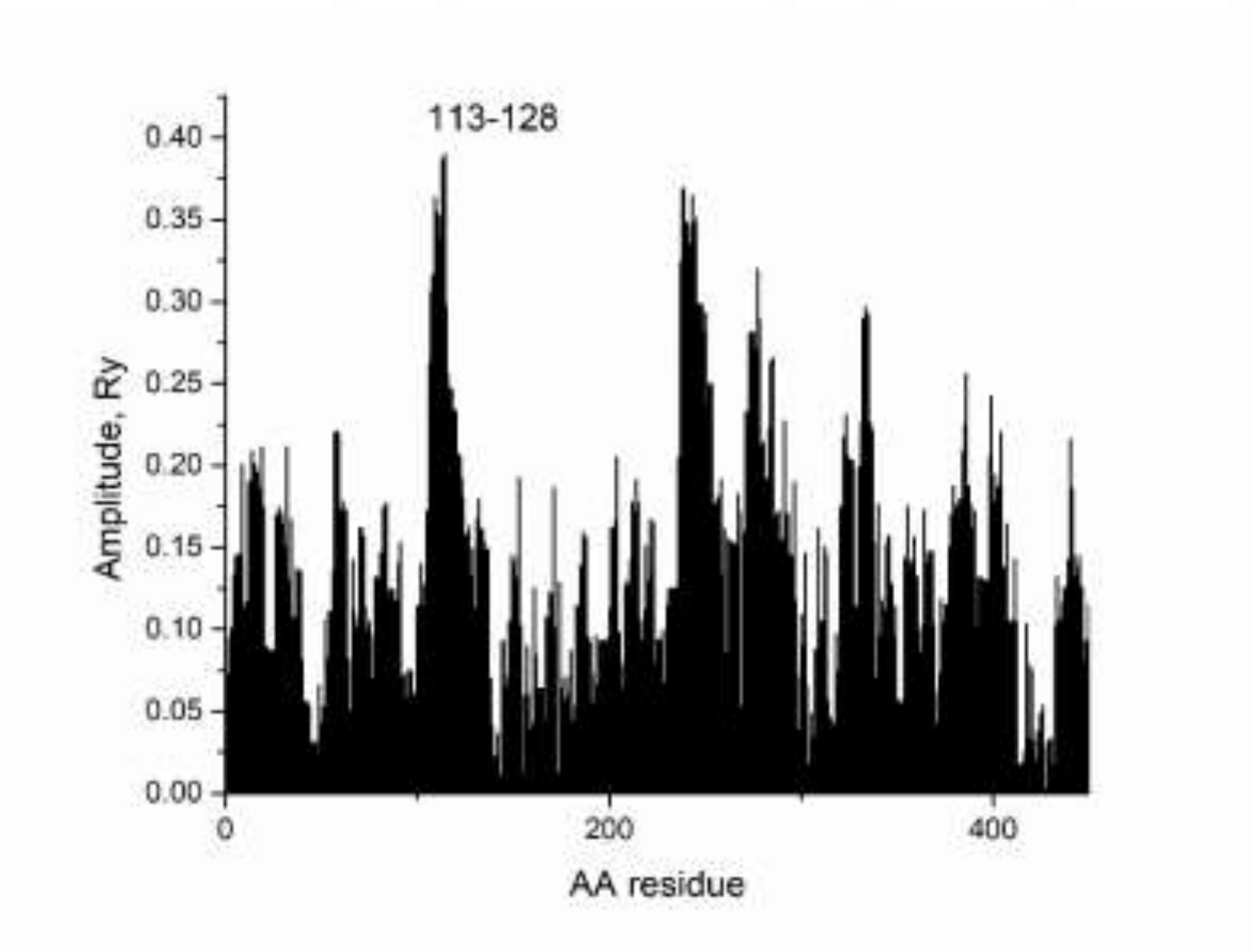

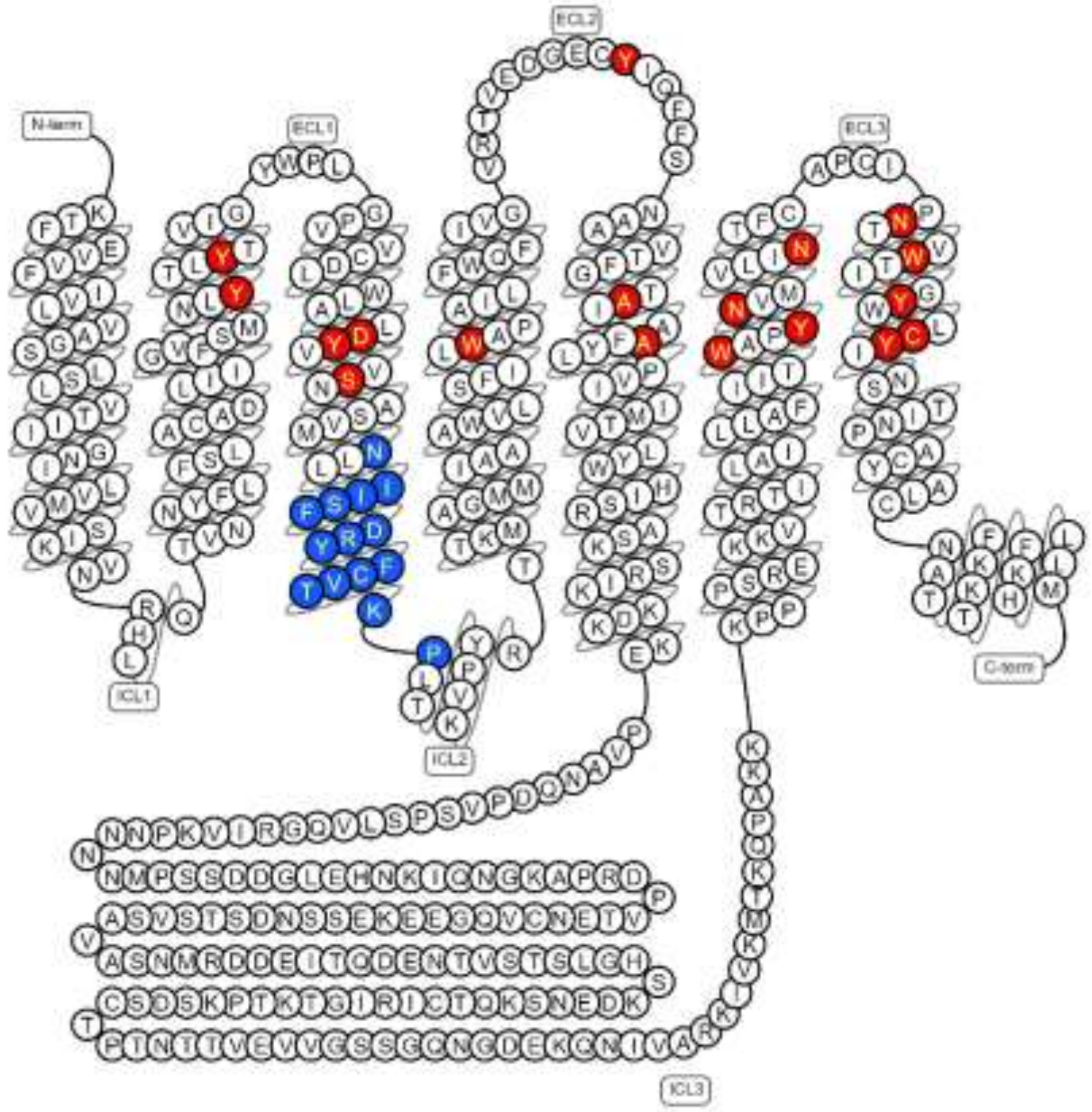

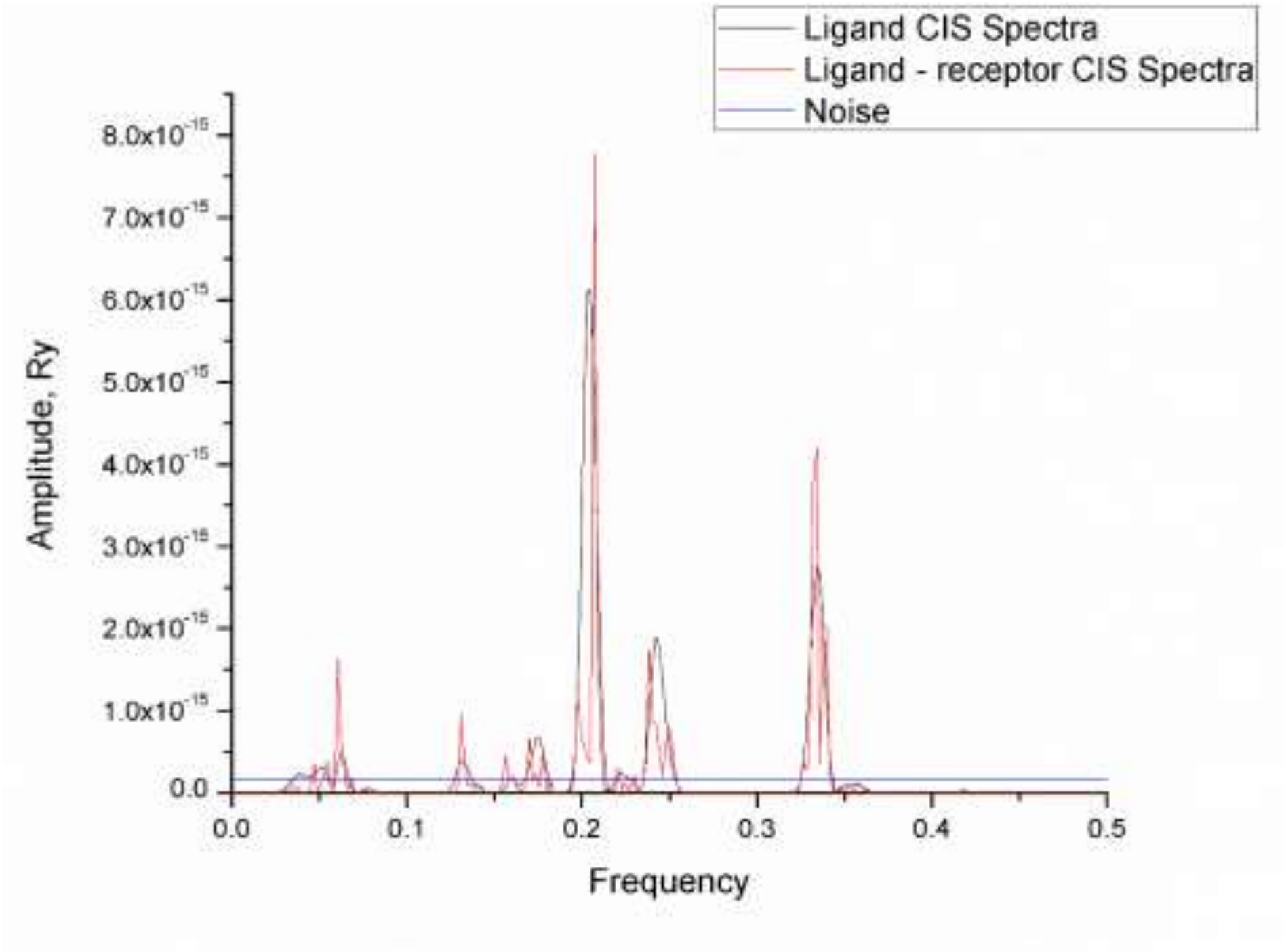

hsa 1131 muscarinic acetylcholine receptor m3

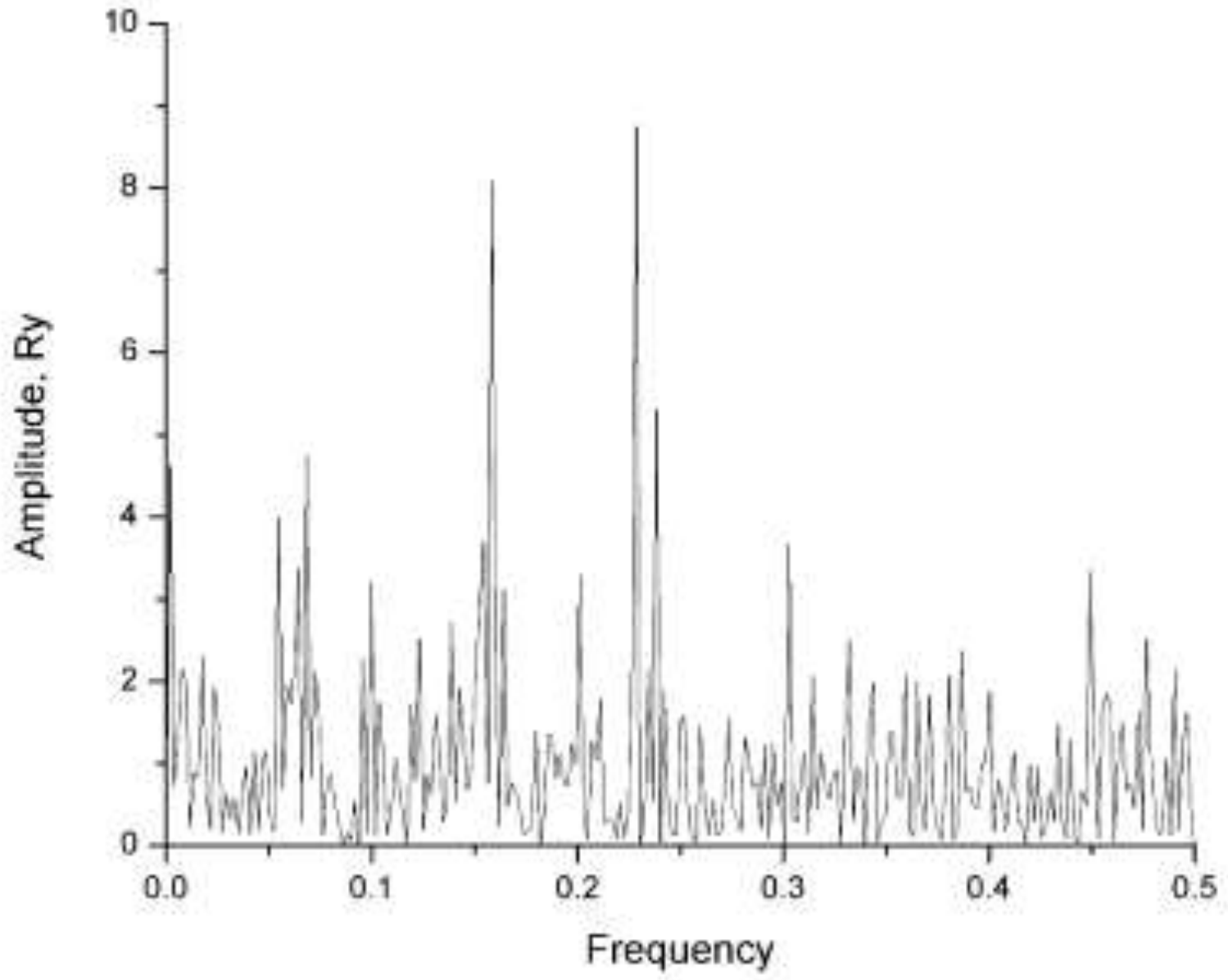

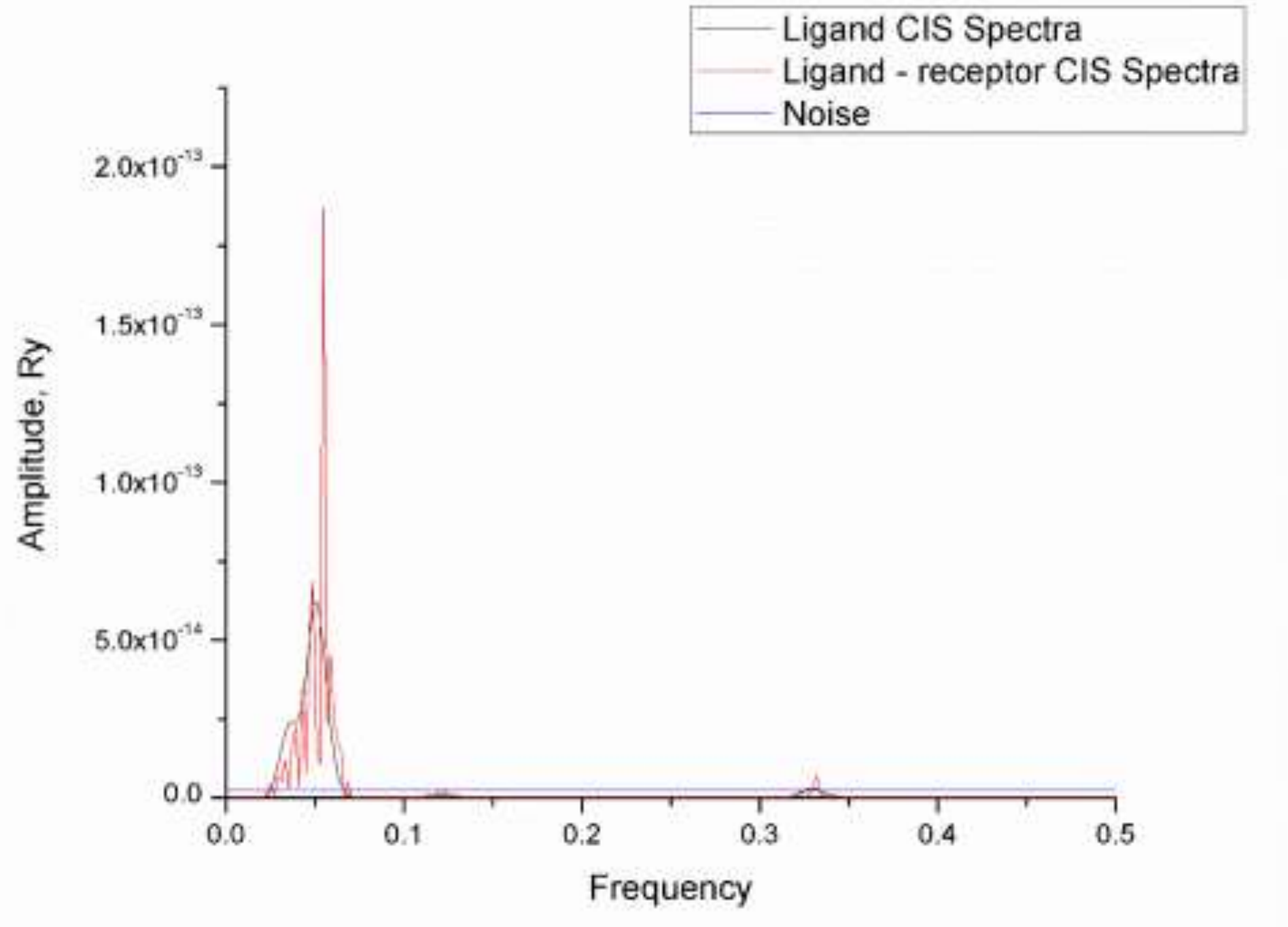
Ligand CIS Spectra
Ligand - receptor CIS Spectra
Noise
Amplitude, Ry
Frequency

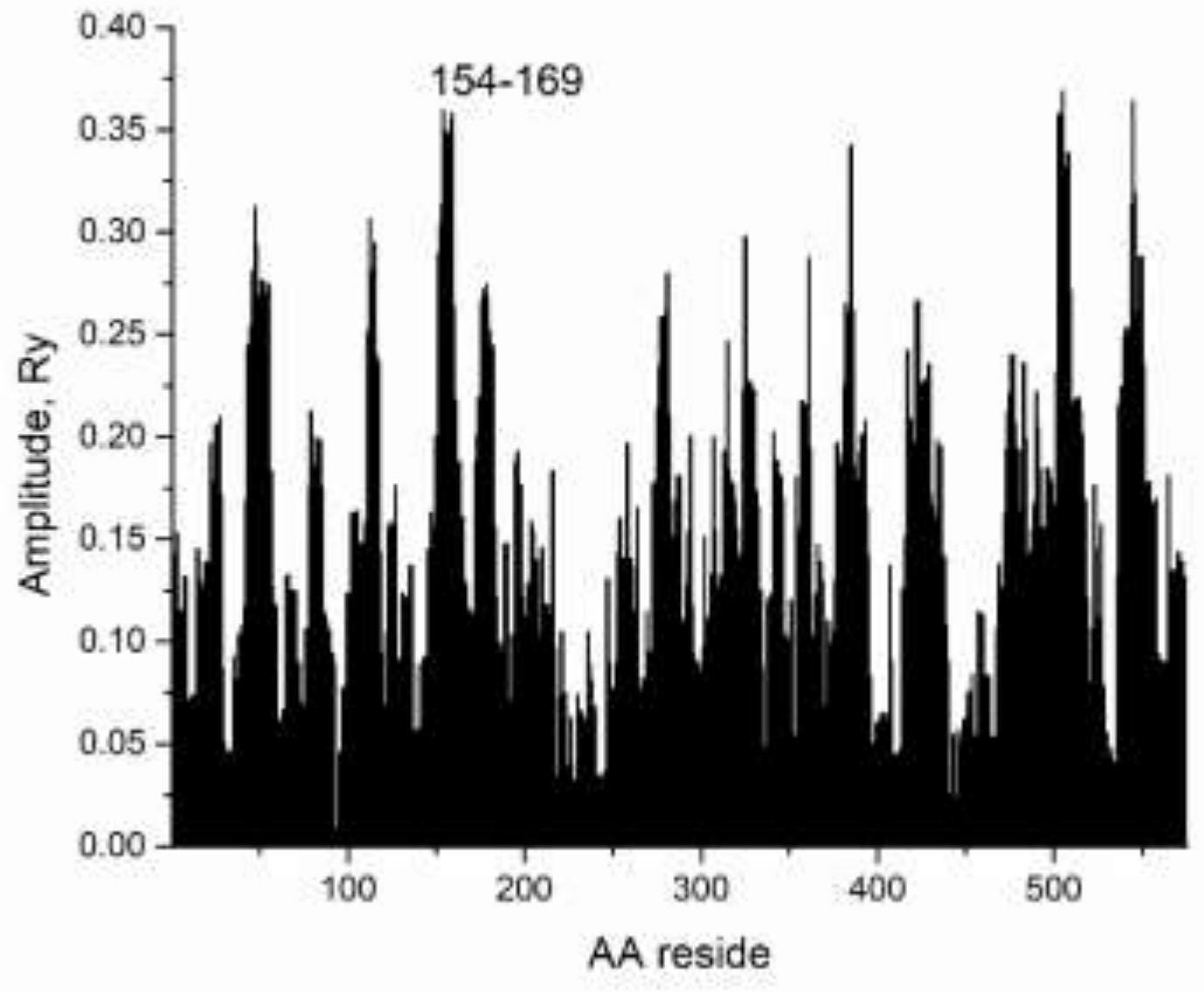
154-169
Amplitude, Ry
AA reside

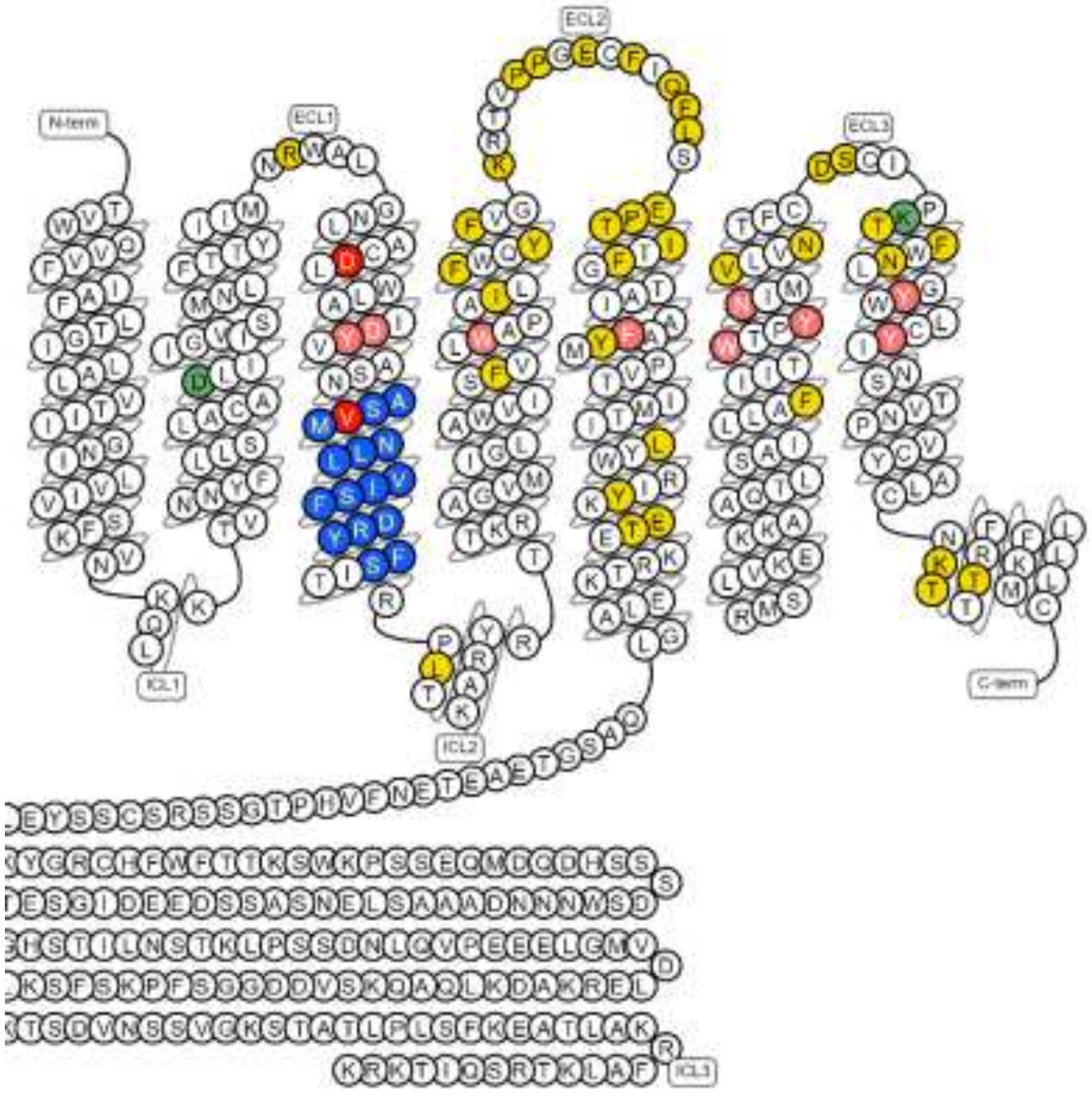

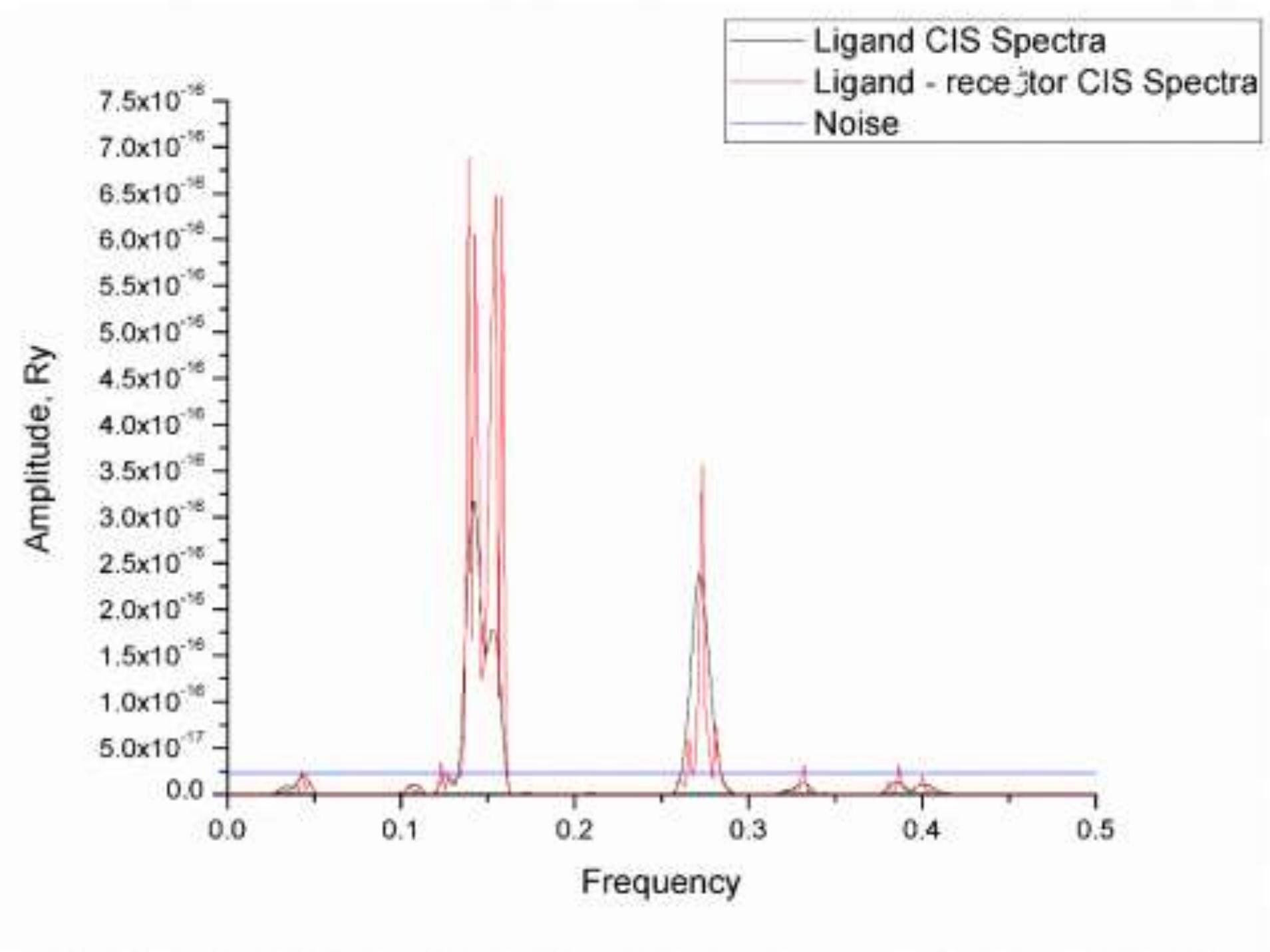

hsa 1812 d(1a) dopamine receptor

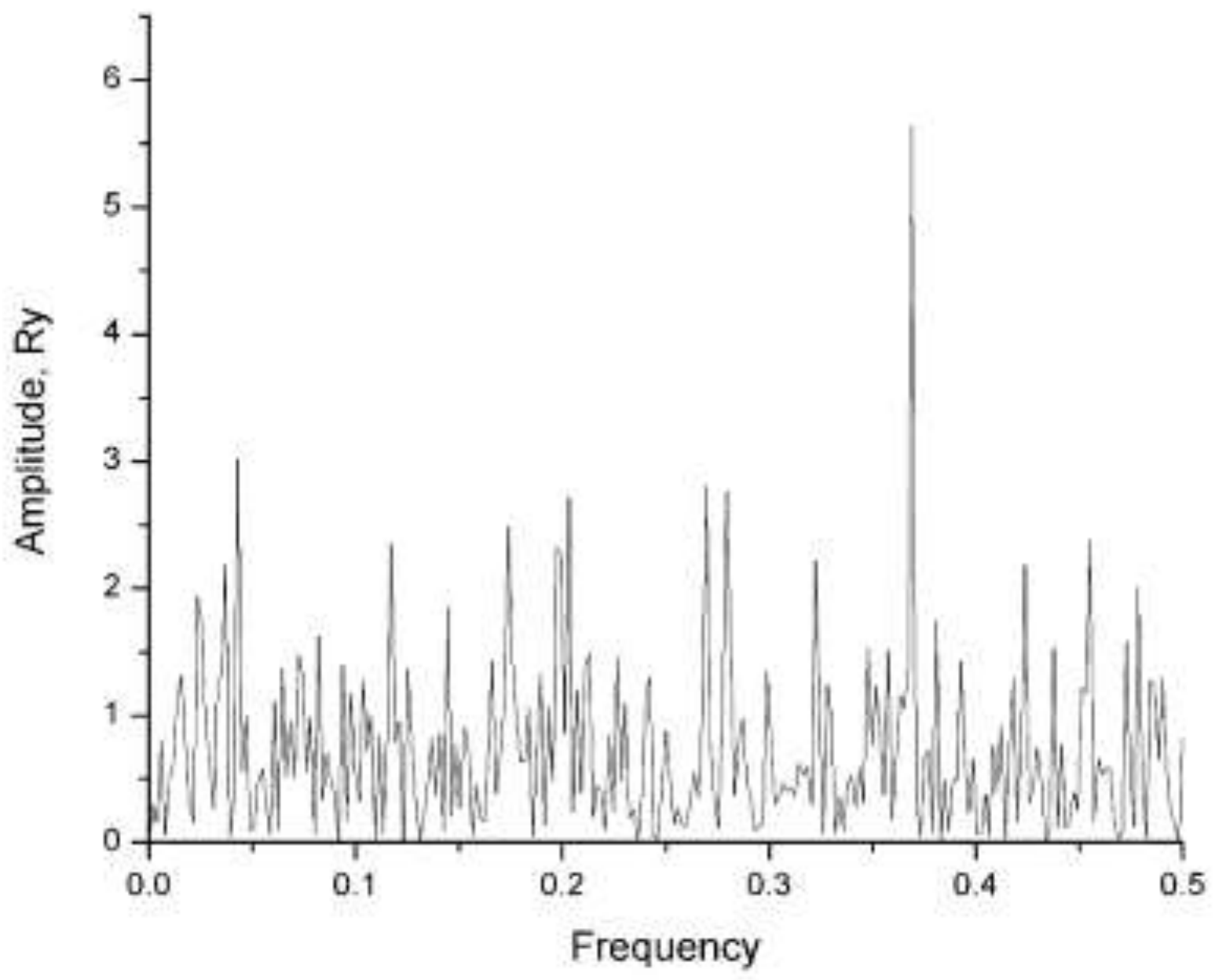

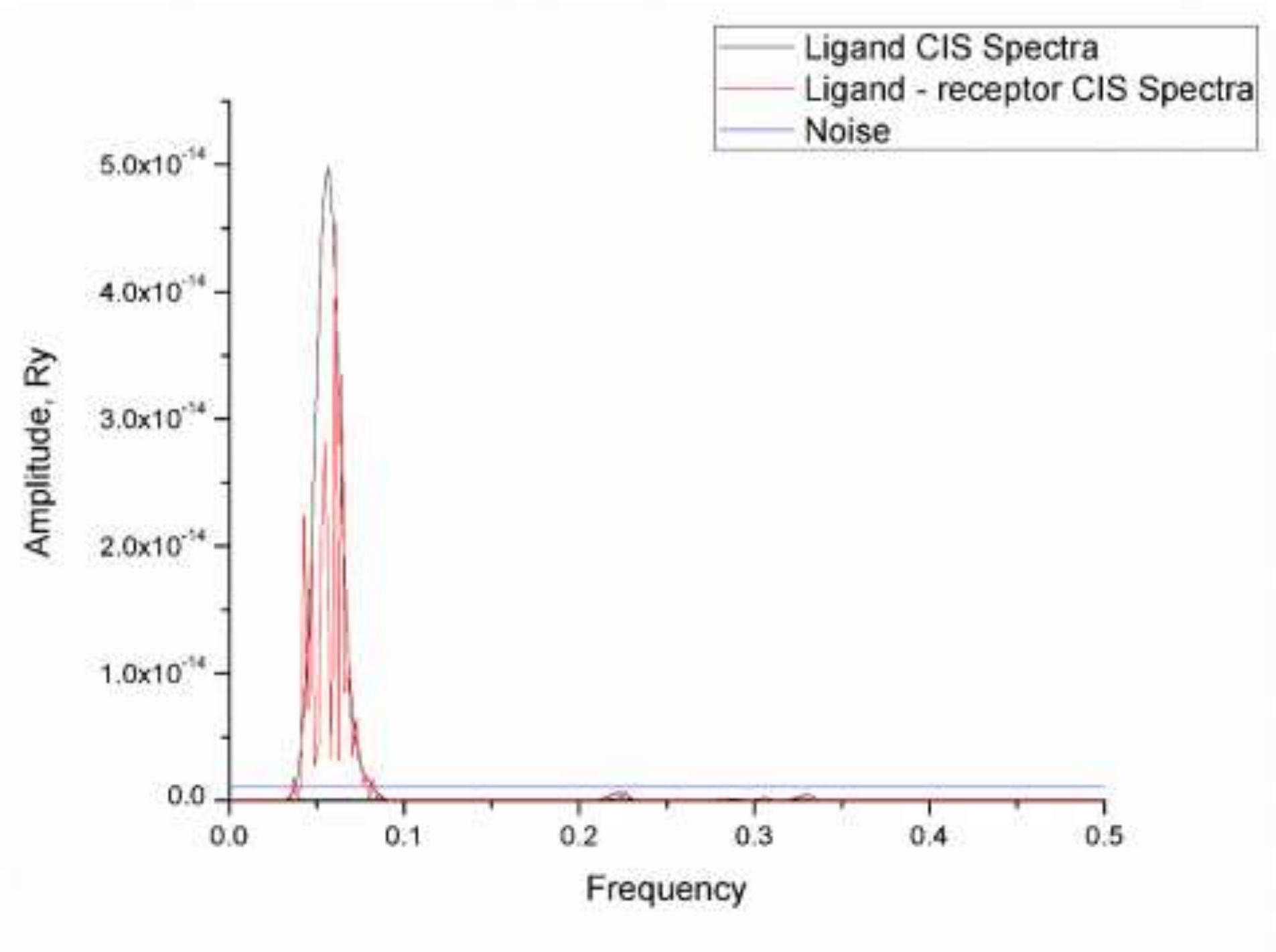

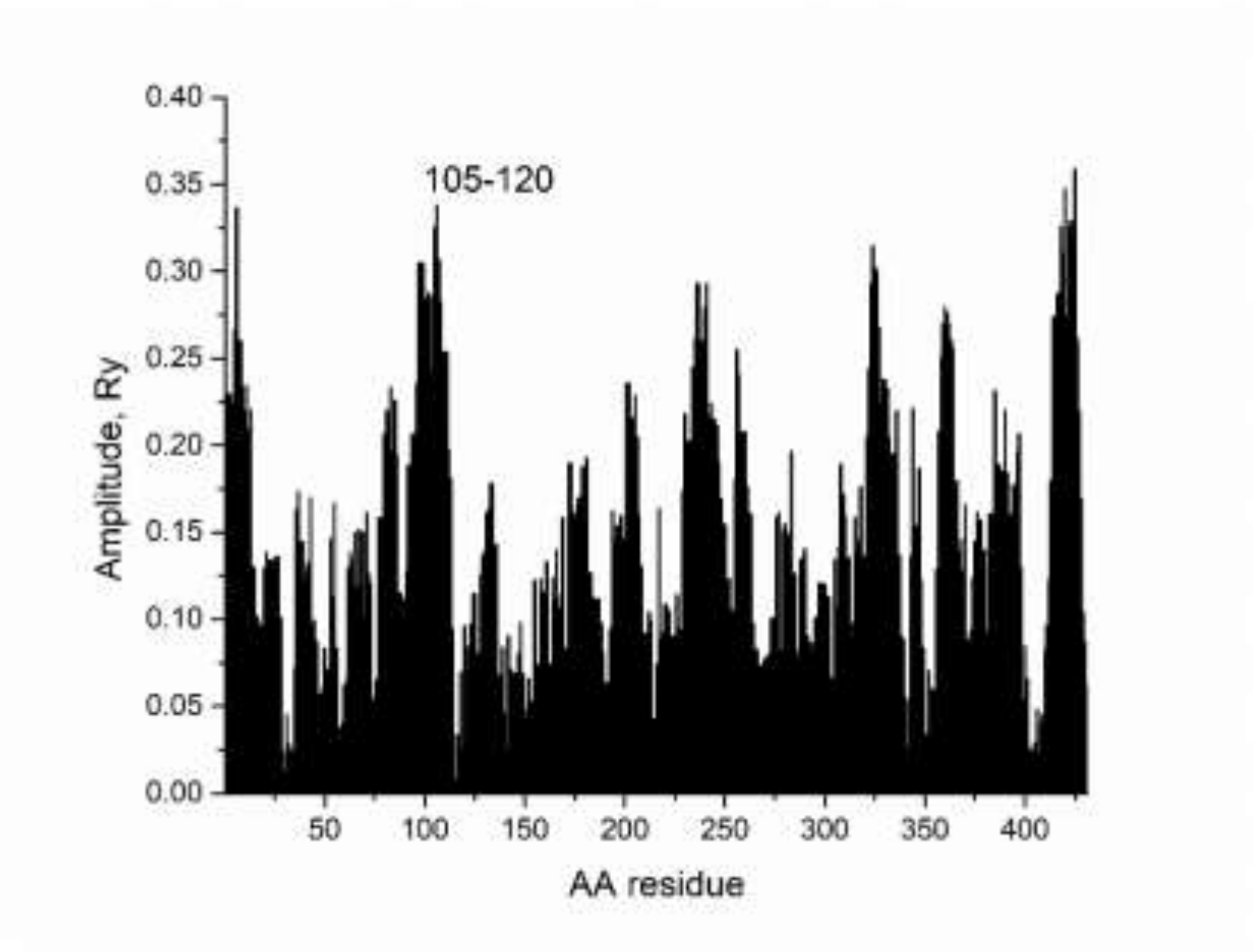

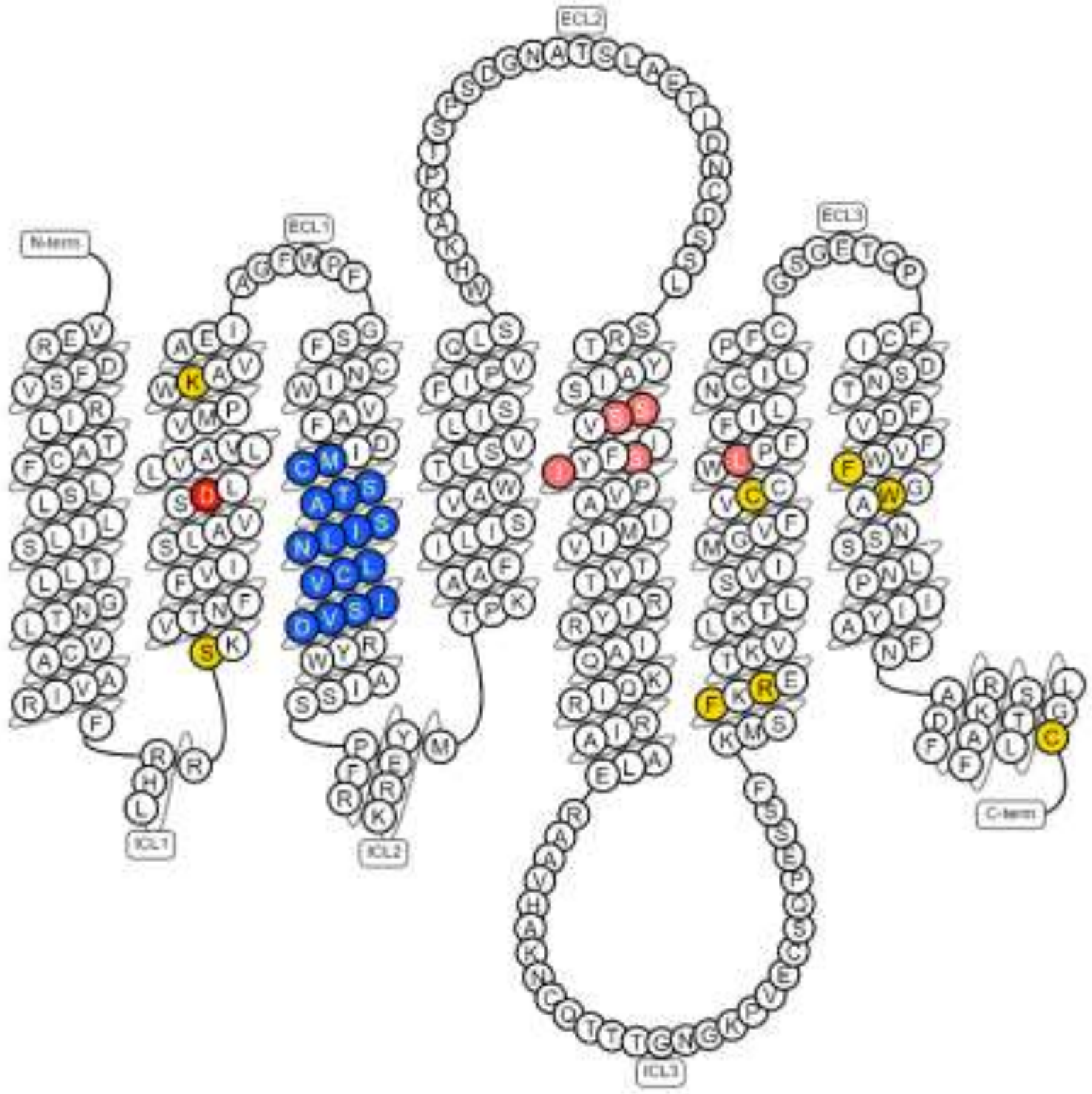

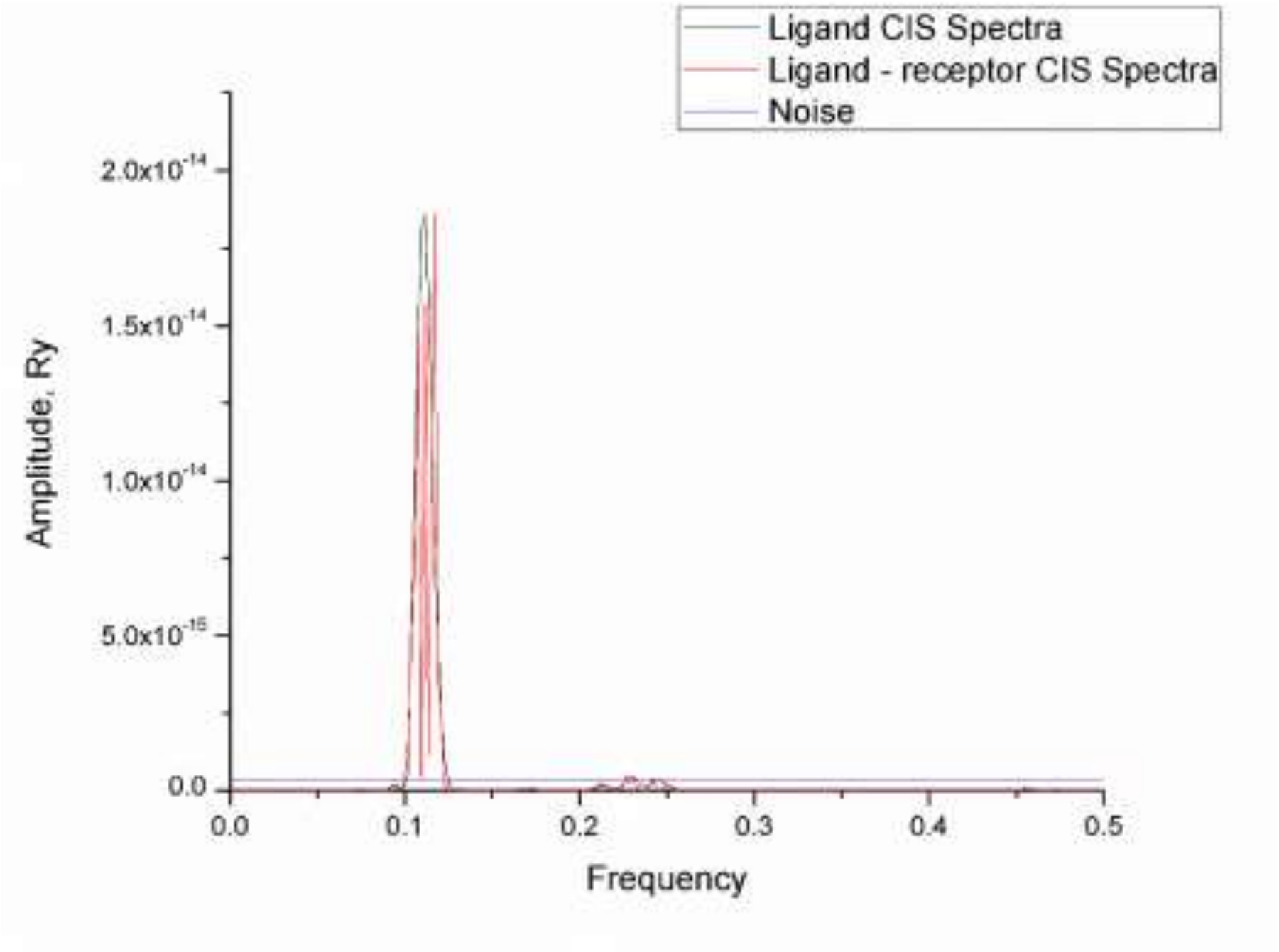

hsa 1813 d(2) dopamine receptor

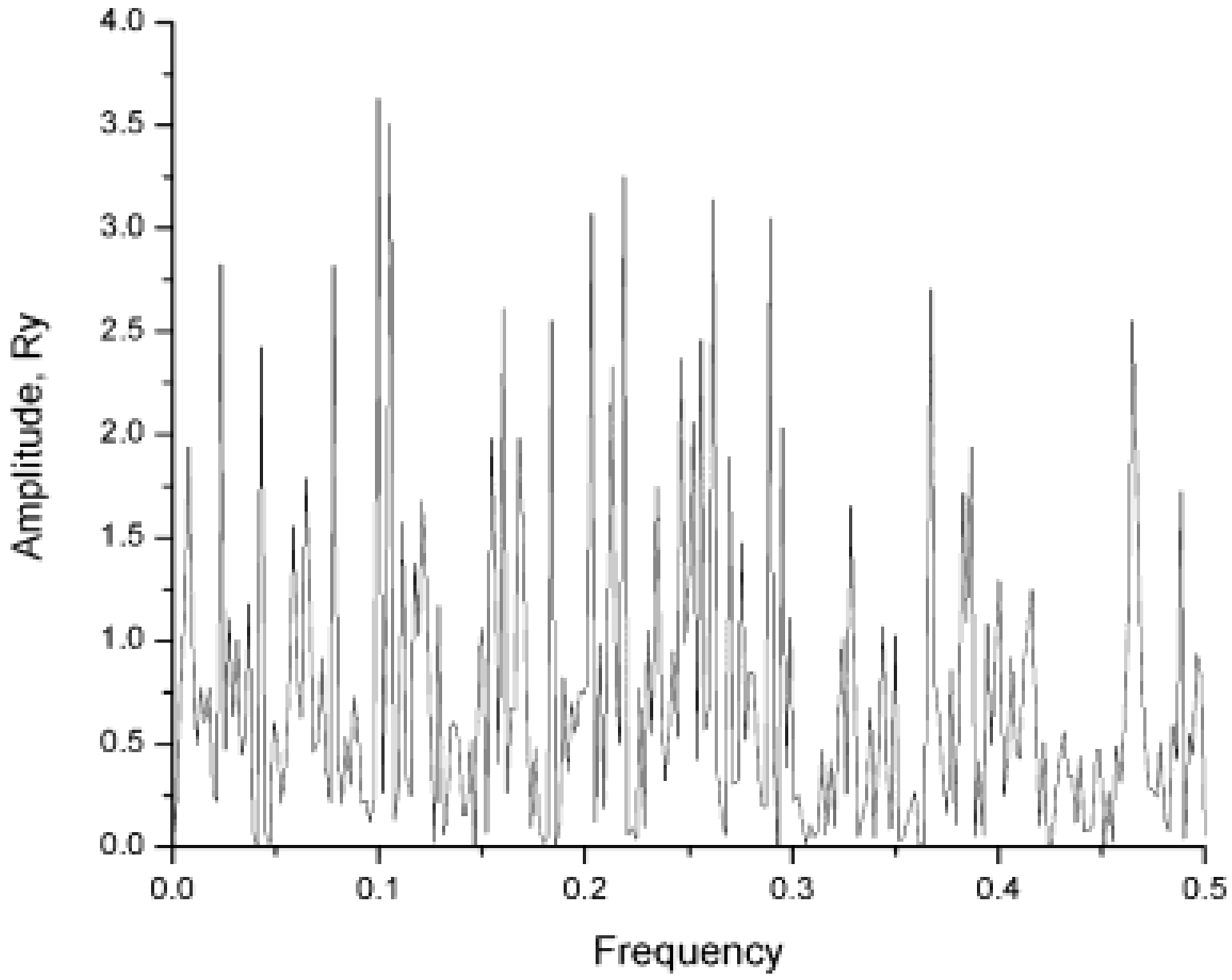

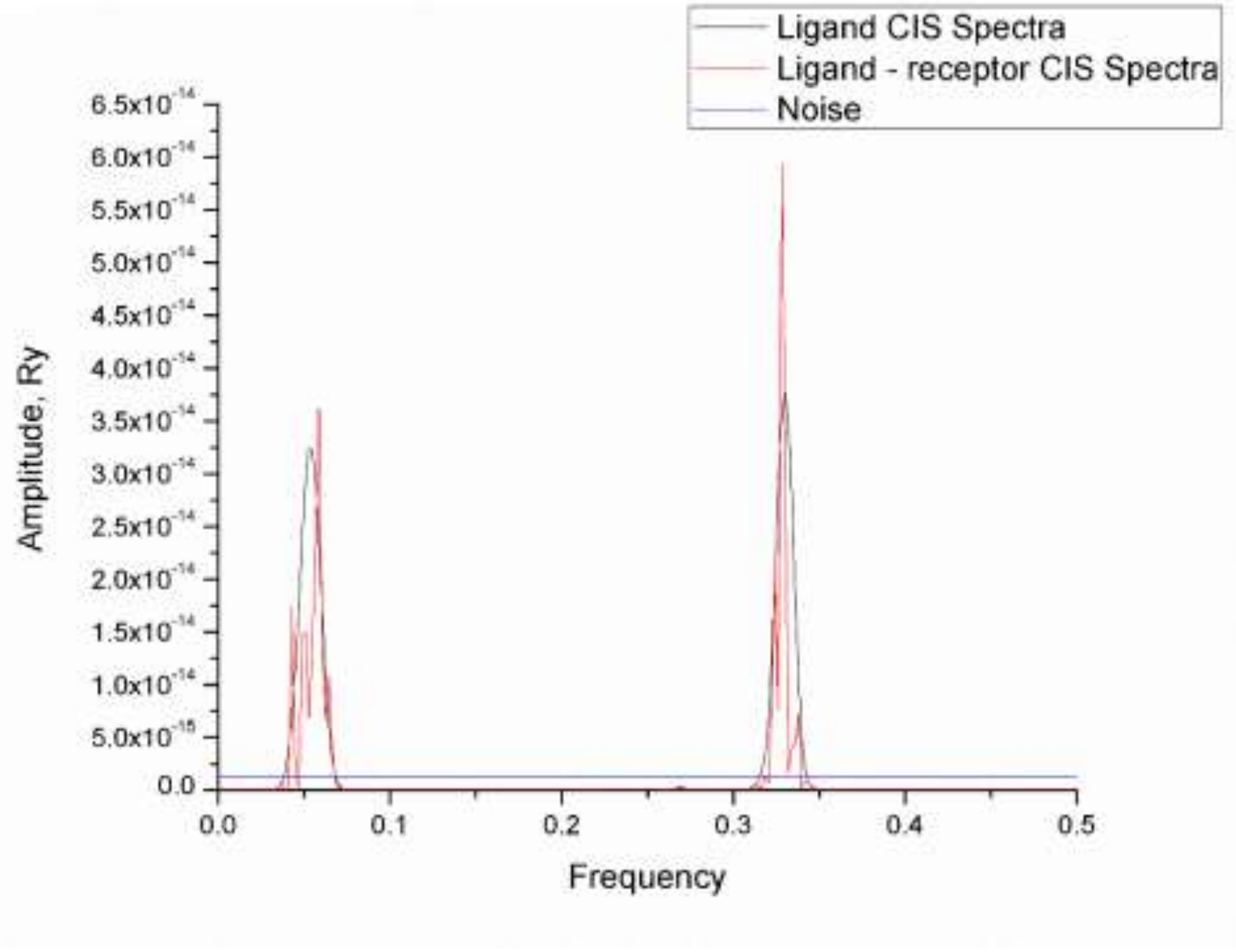
Ligand CIS Spectra
Ligand - receptor CIS Spectra
Noise
Amplitude, Ry
Frequency

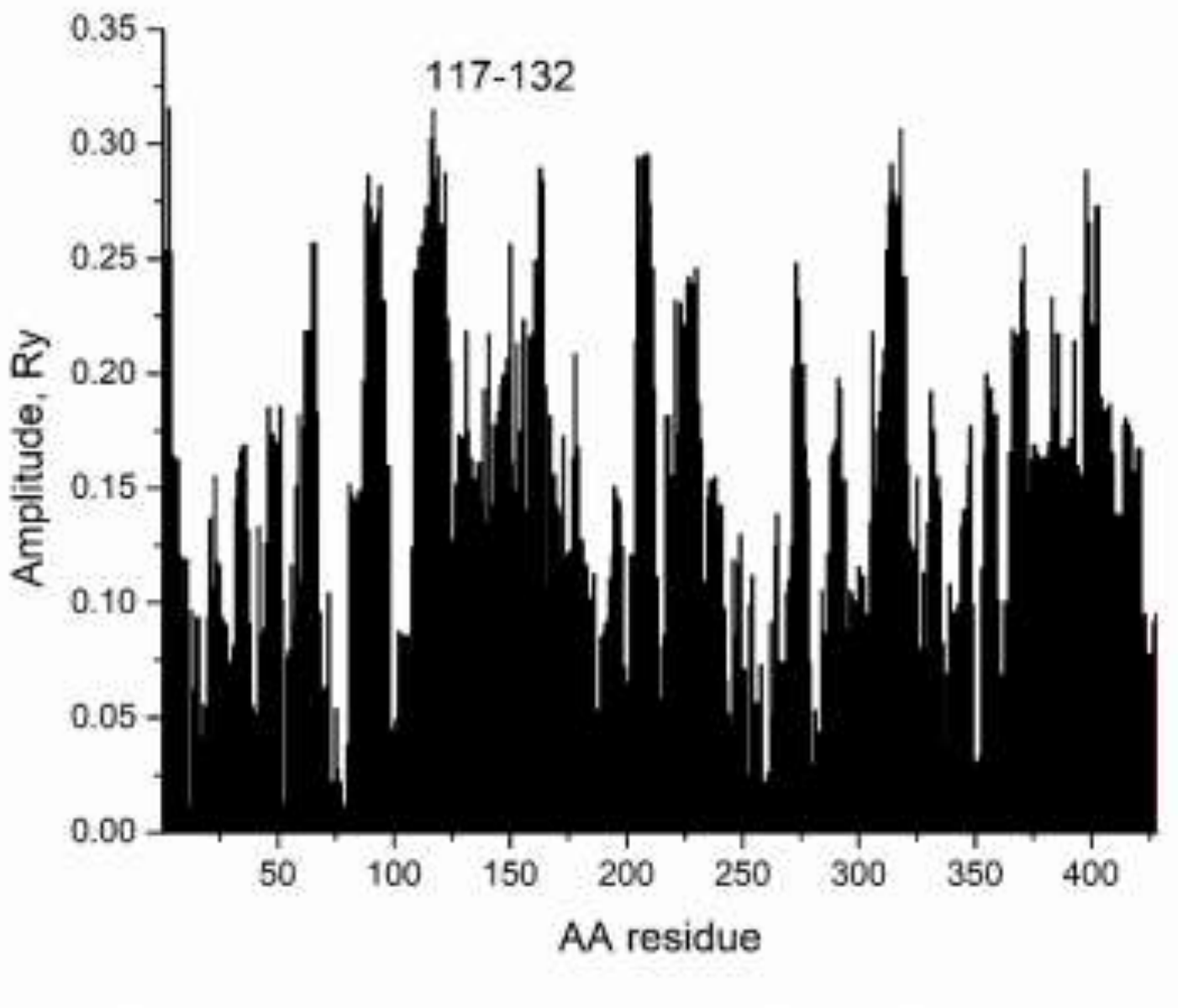
117-132
Amplitude, Ry
AA residue

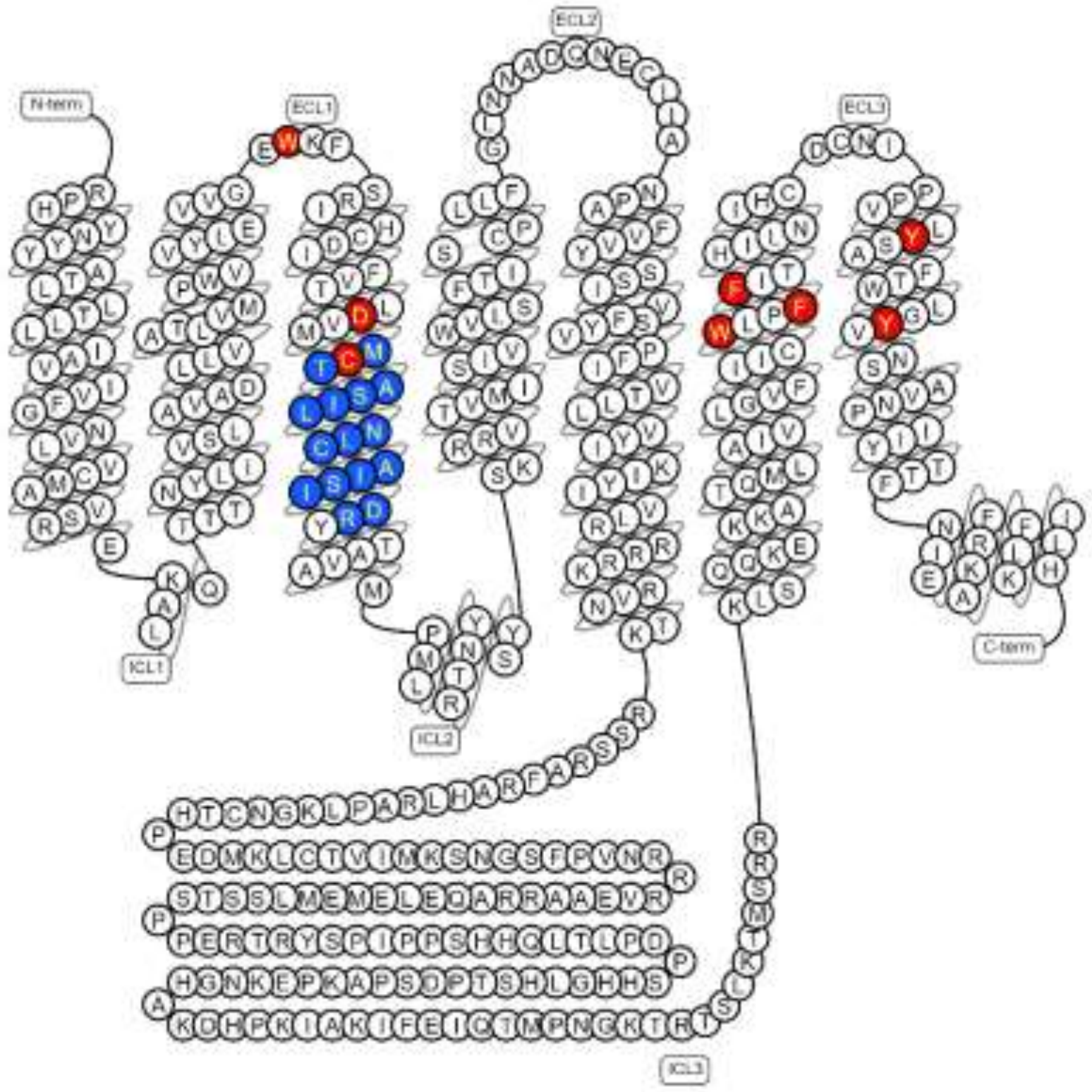

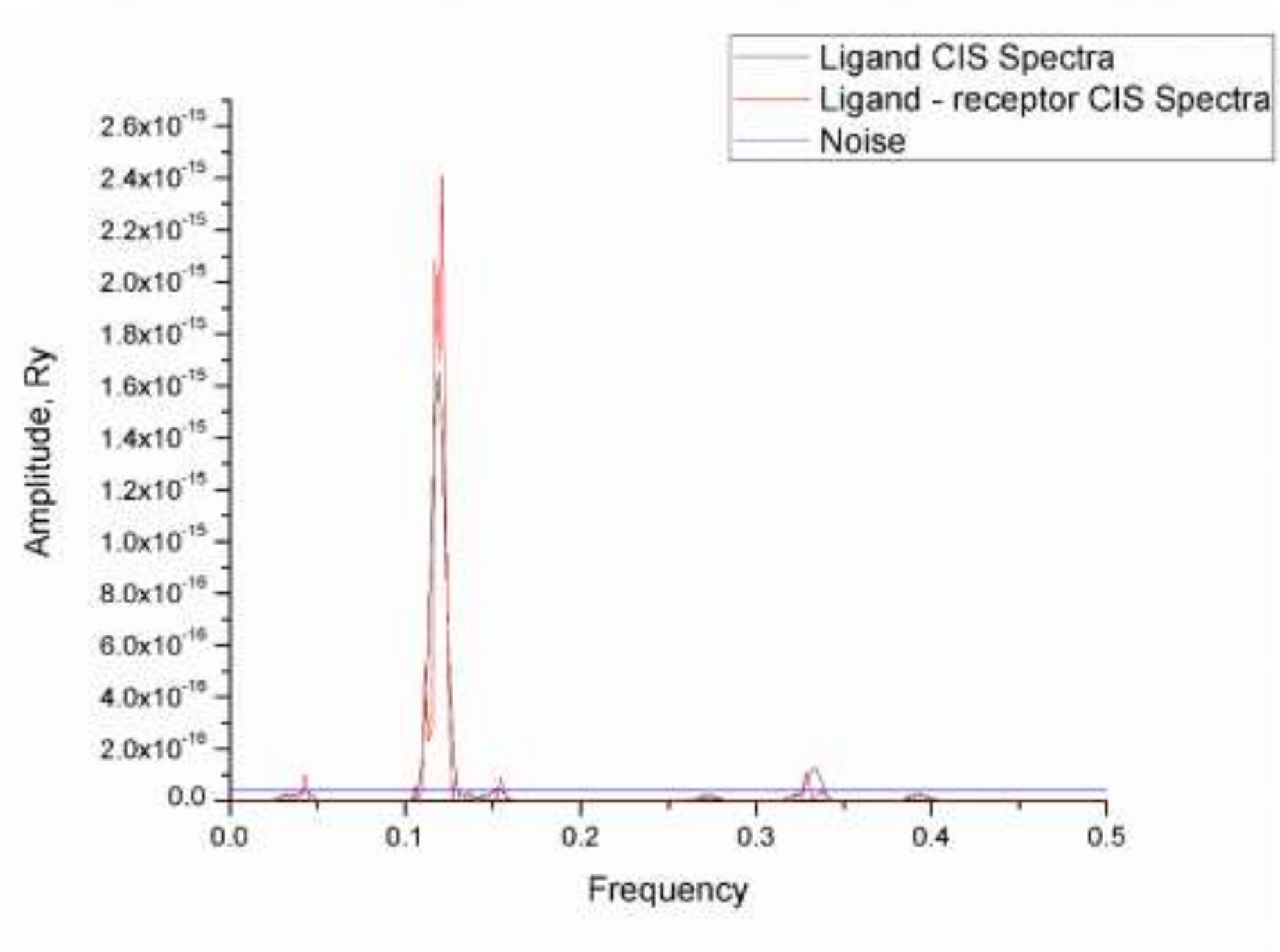
Ligand CIS Spectra
Ligand - receptor CIS Spectra
Noise
Amplitude, Ry
Frequency

hsa 1814 d(3) dopamine receptor

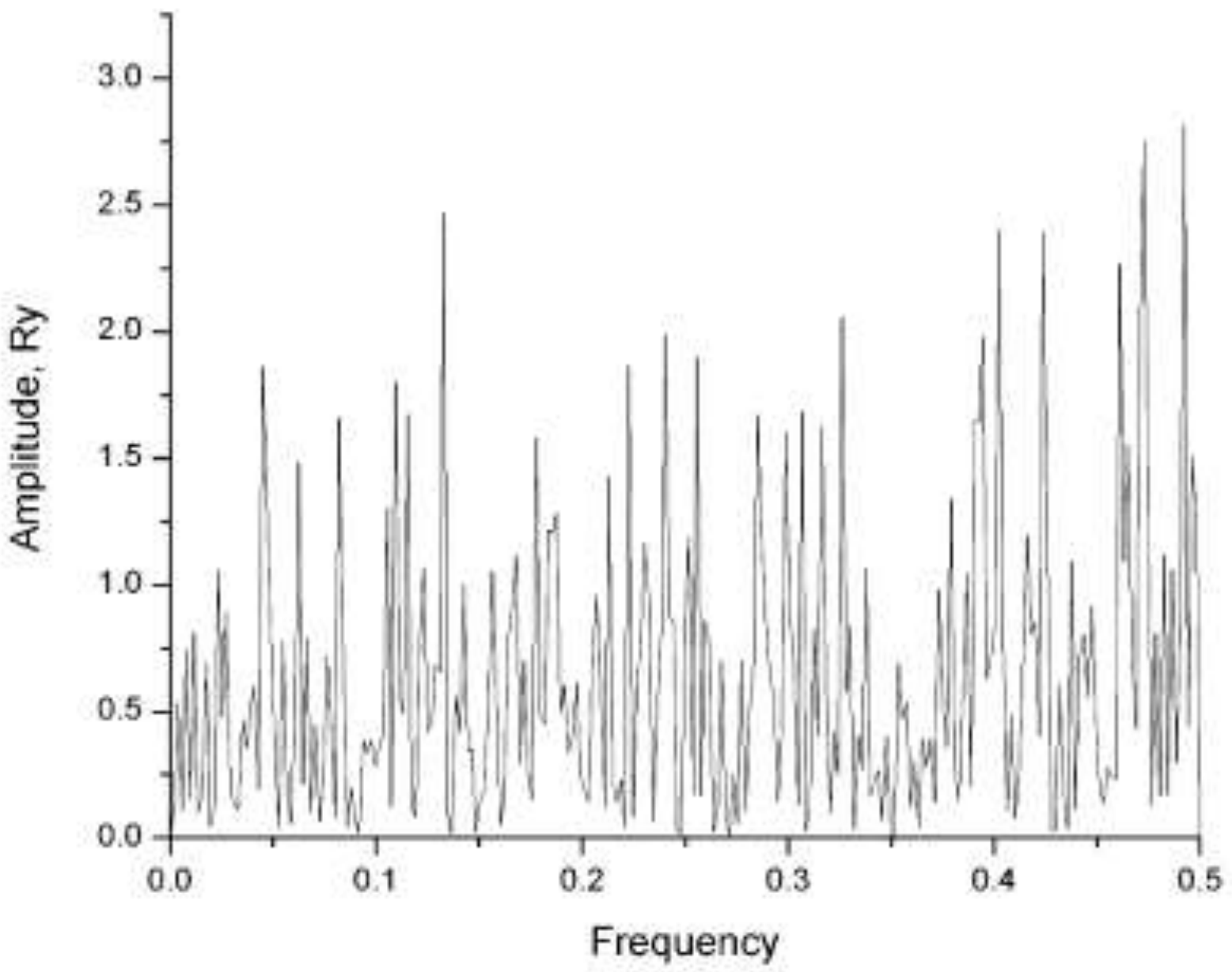

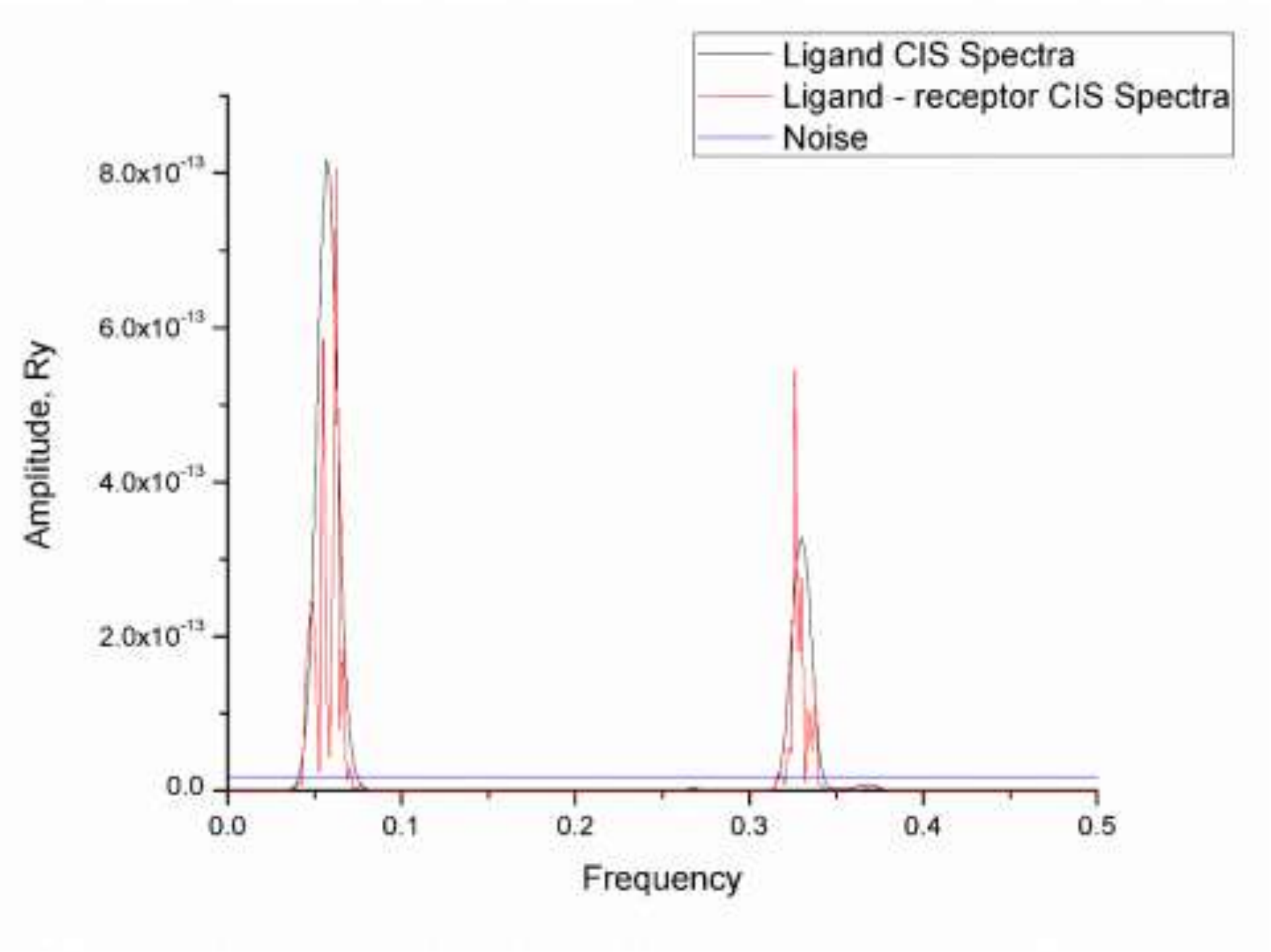

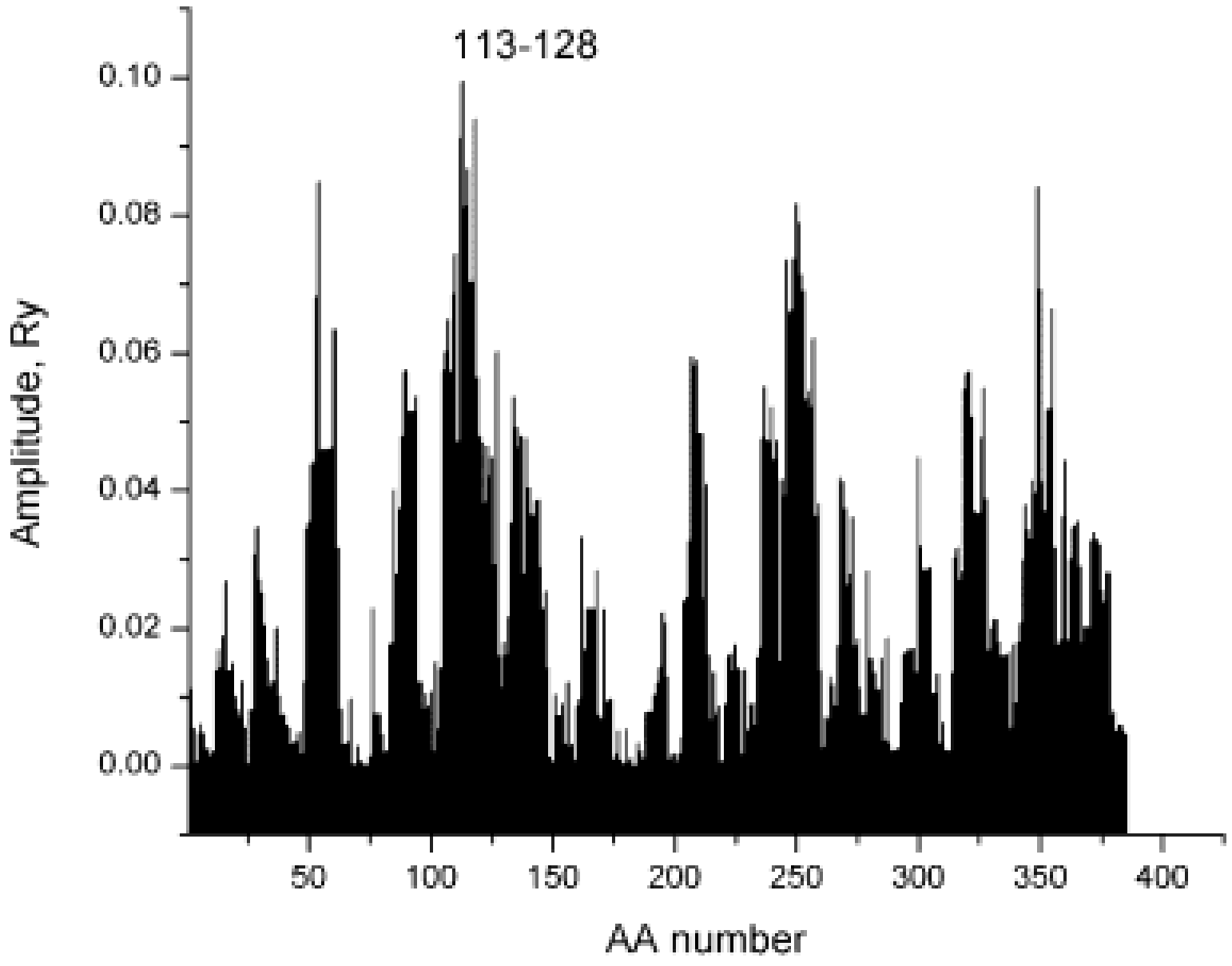

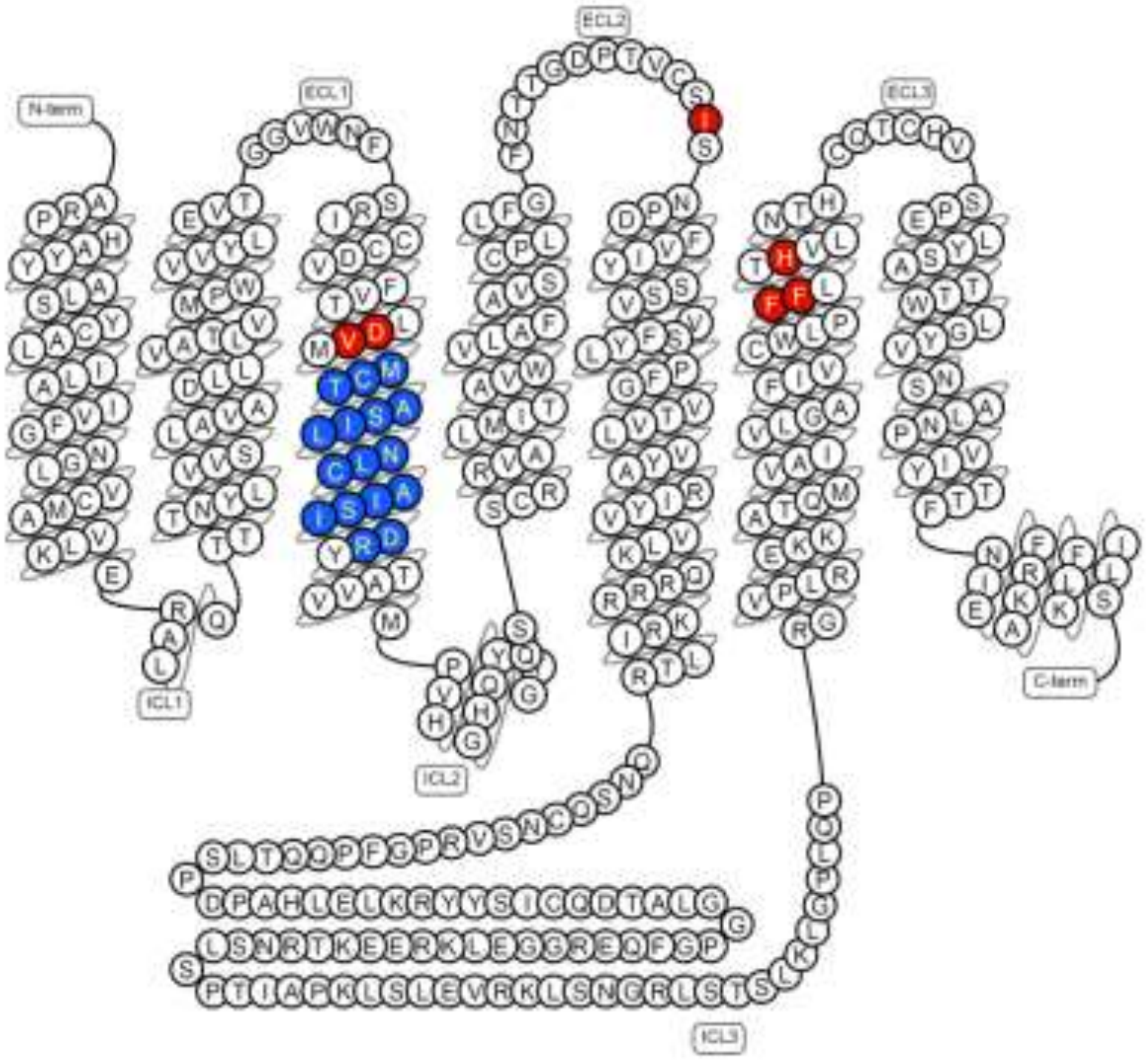

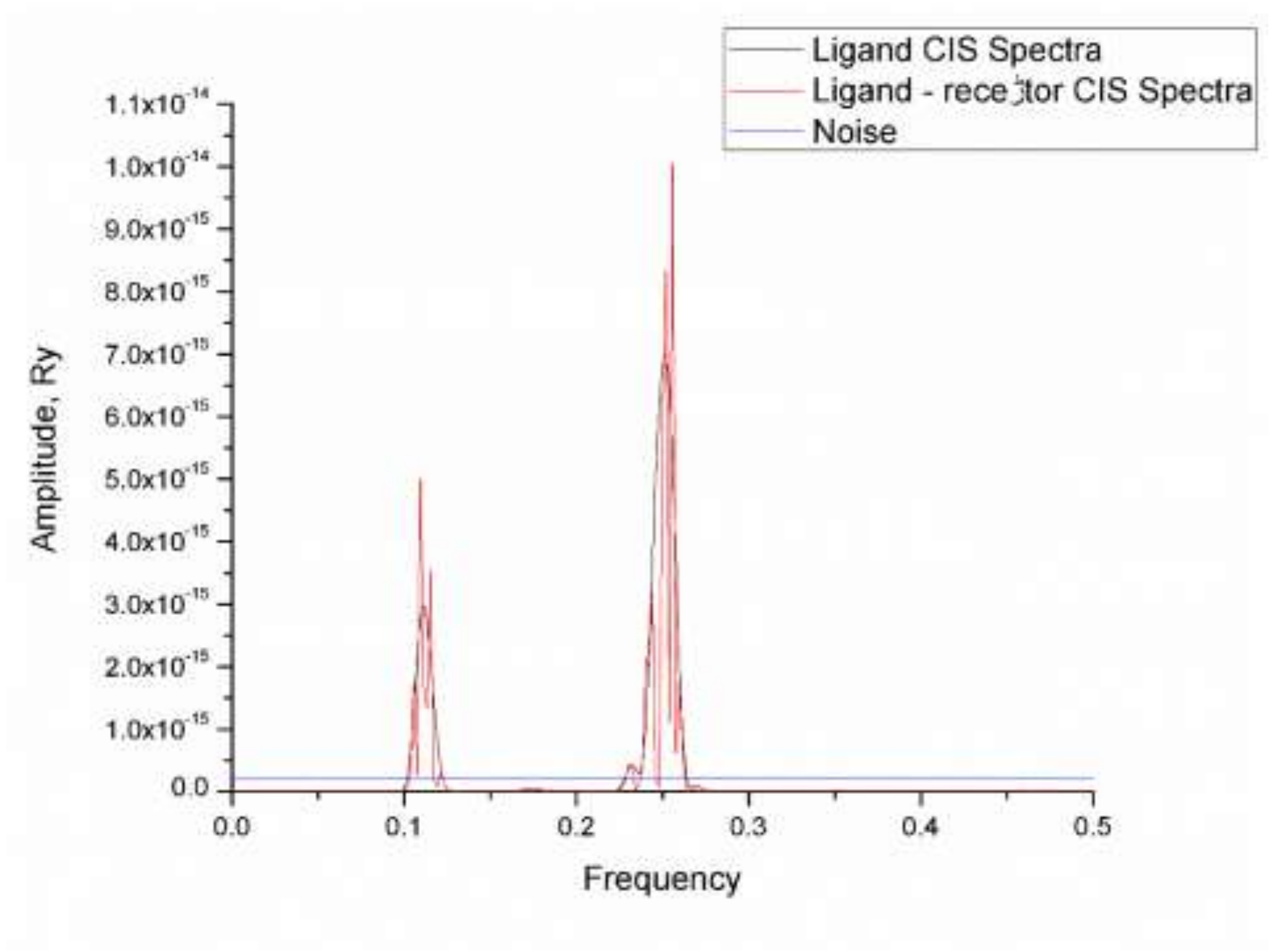

hsa:3269 histamine h1 receptor

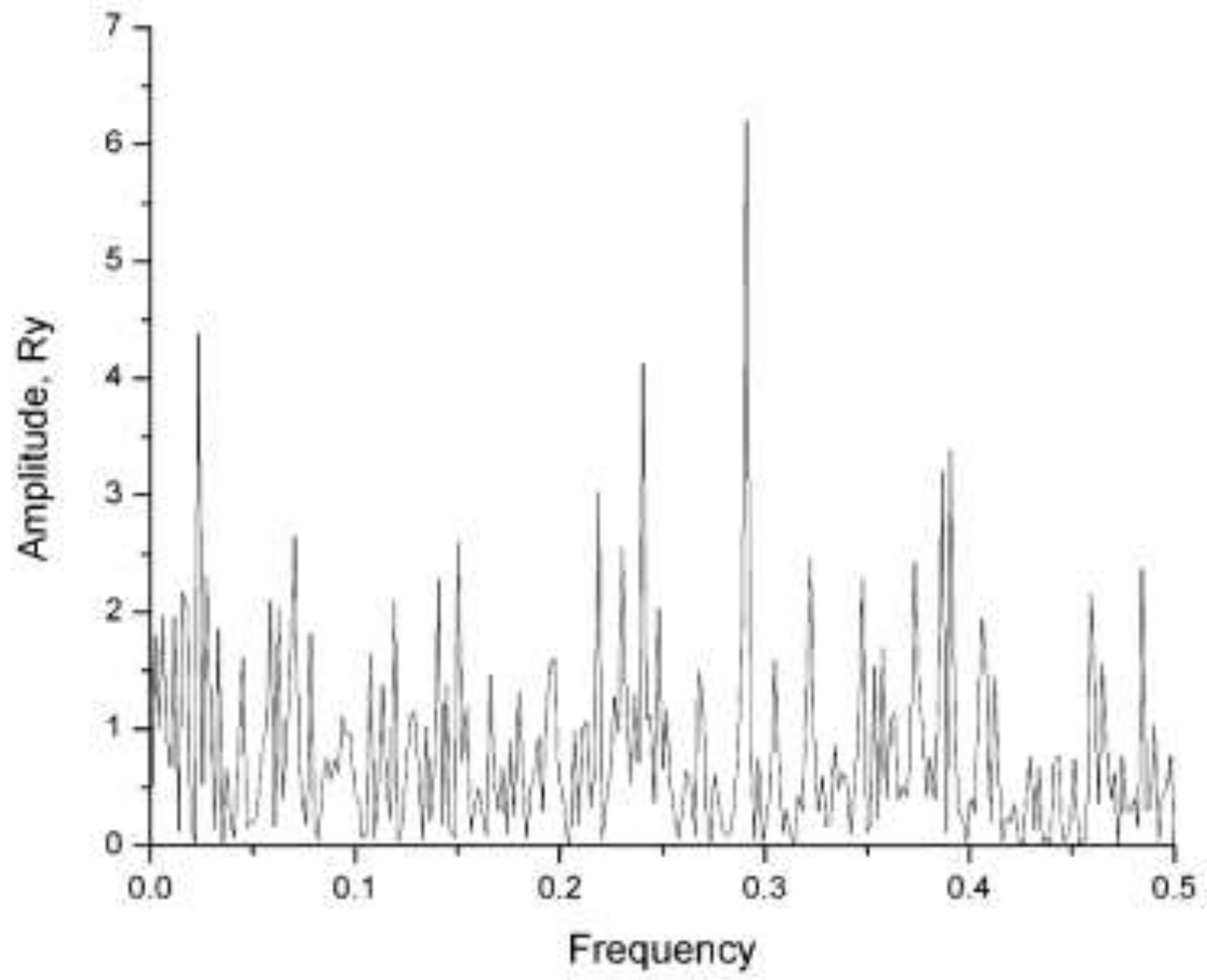

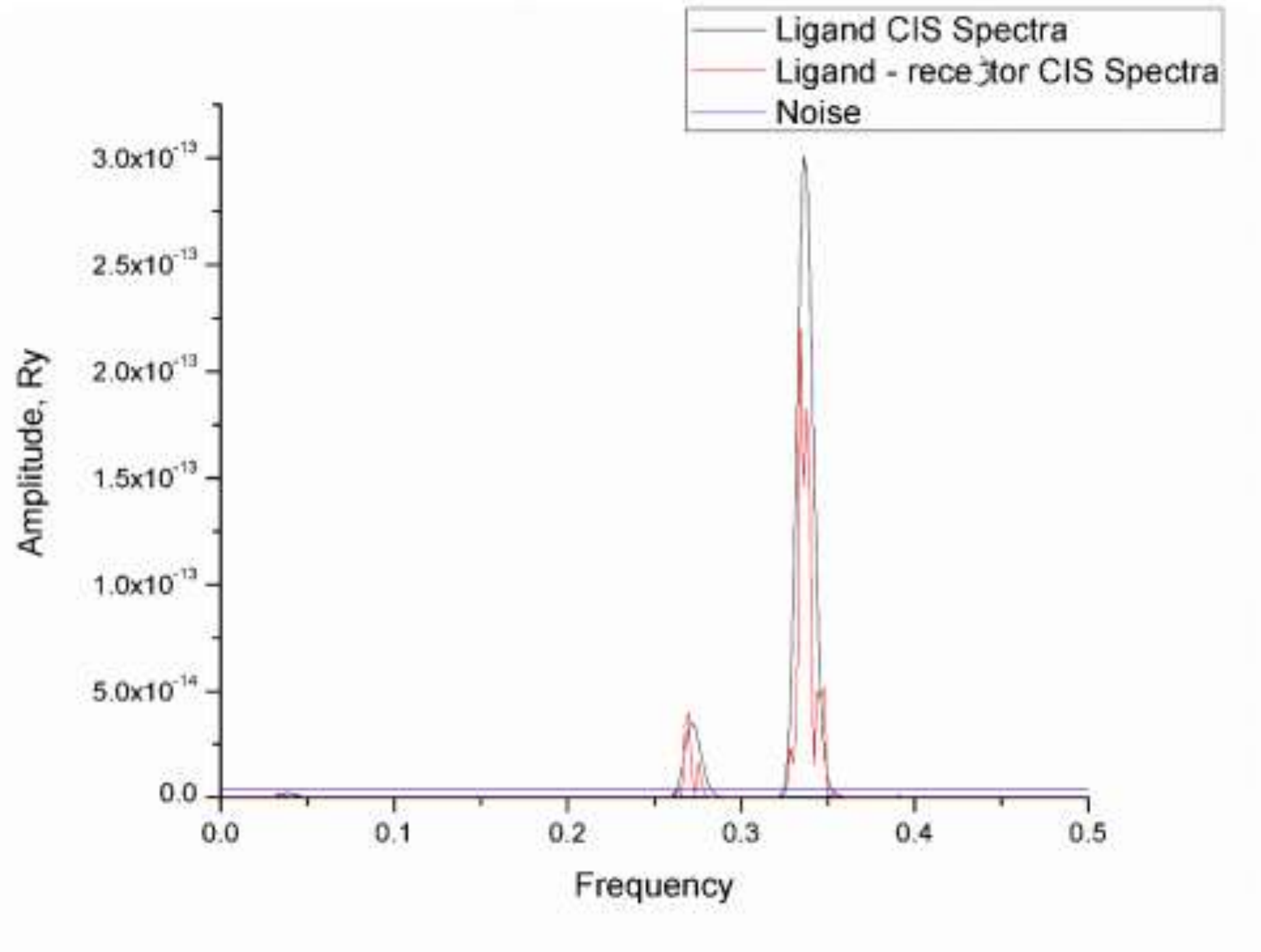
Ligand CIS Spectra
Ligand - receptor CIS Spectra
Noise
Amplitude, Ry
Frequency

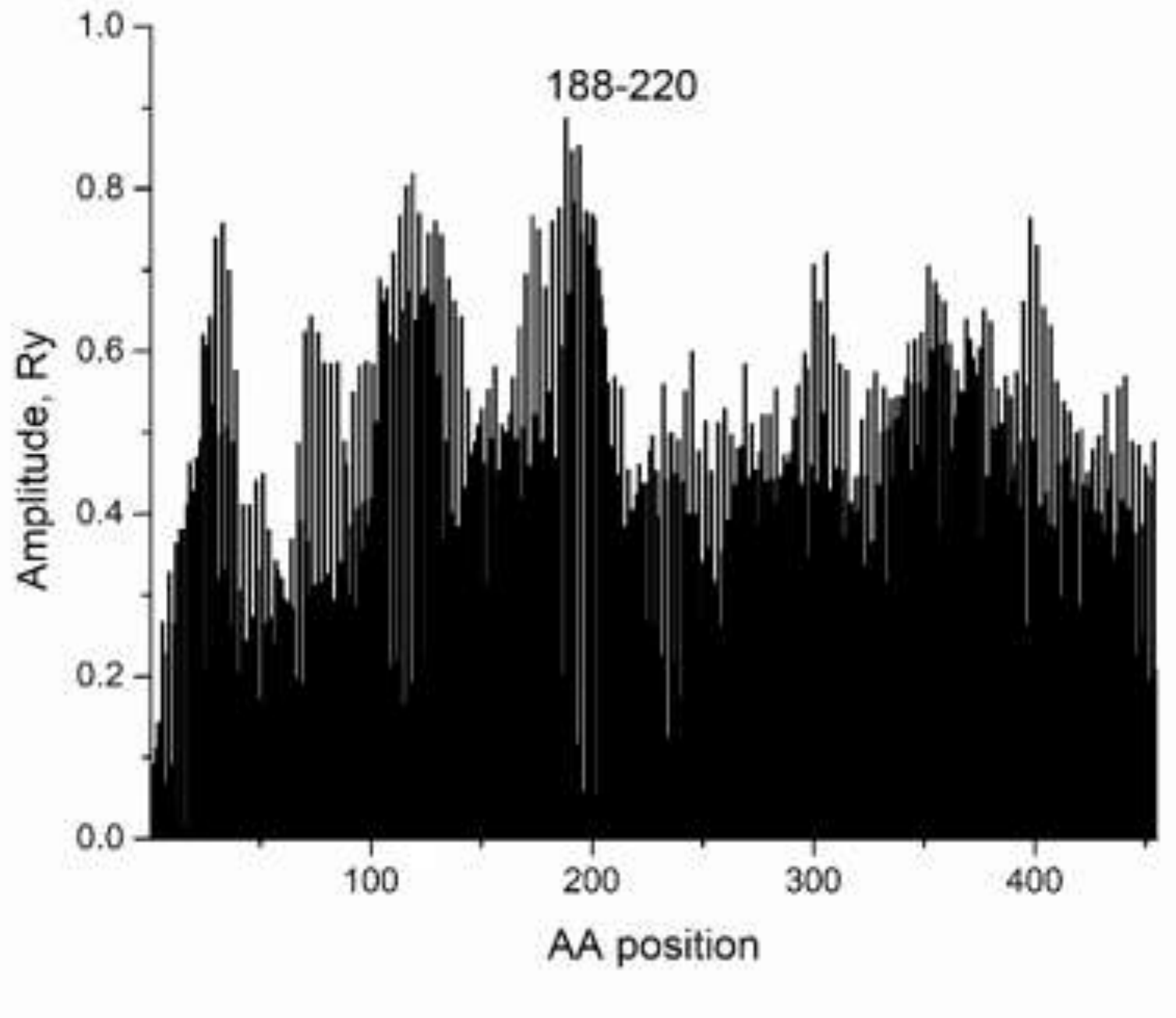
188-220
Amplitude, Ry
AA position

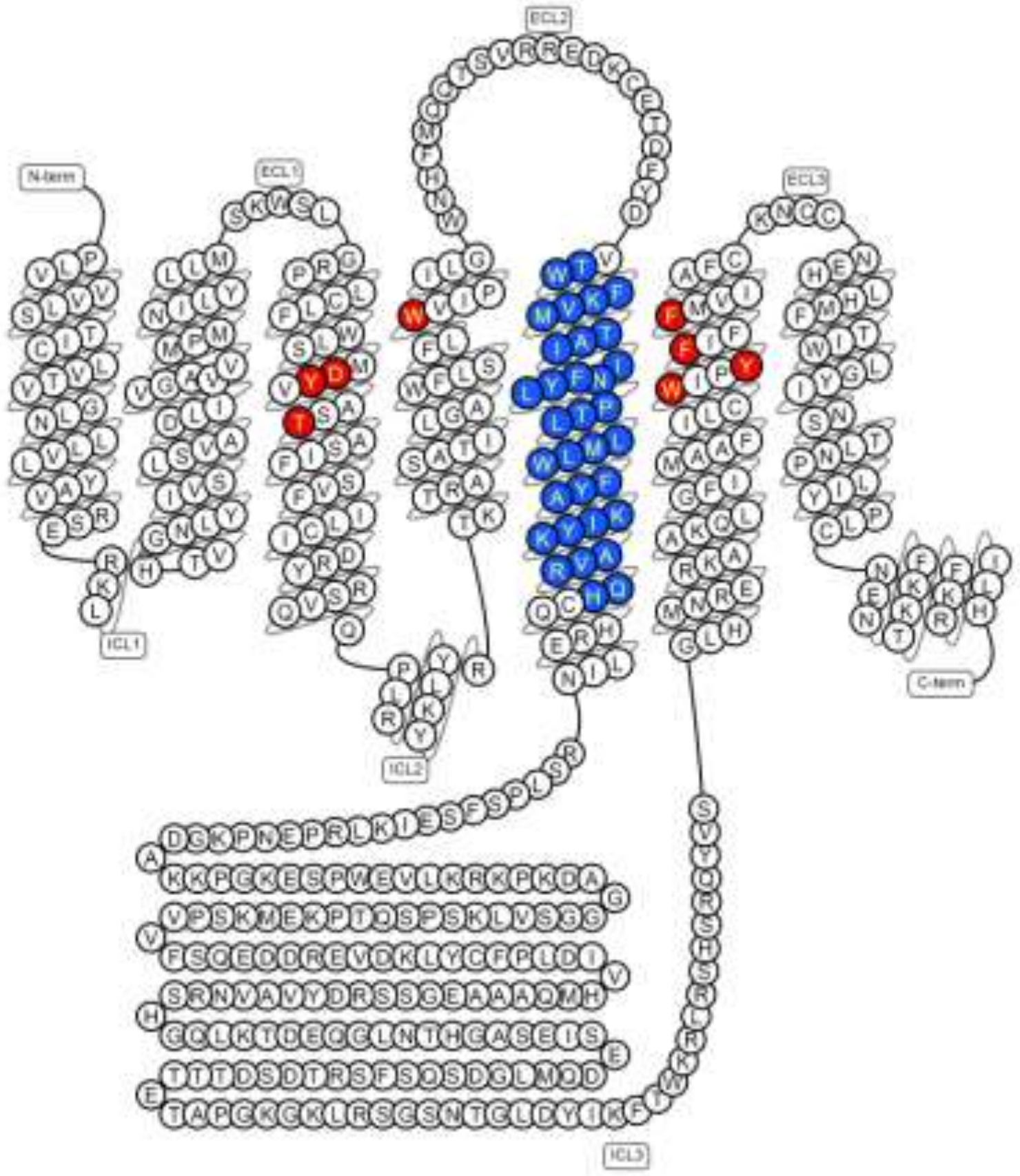

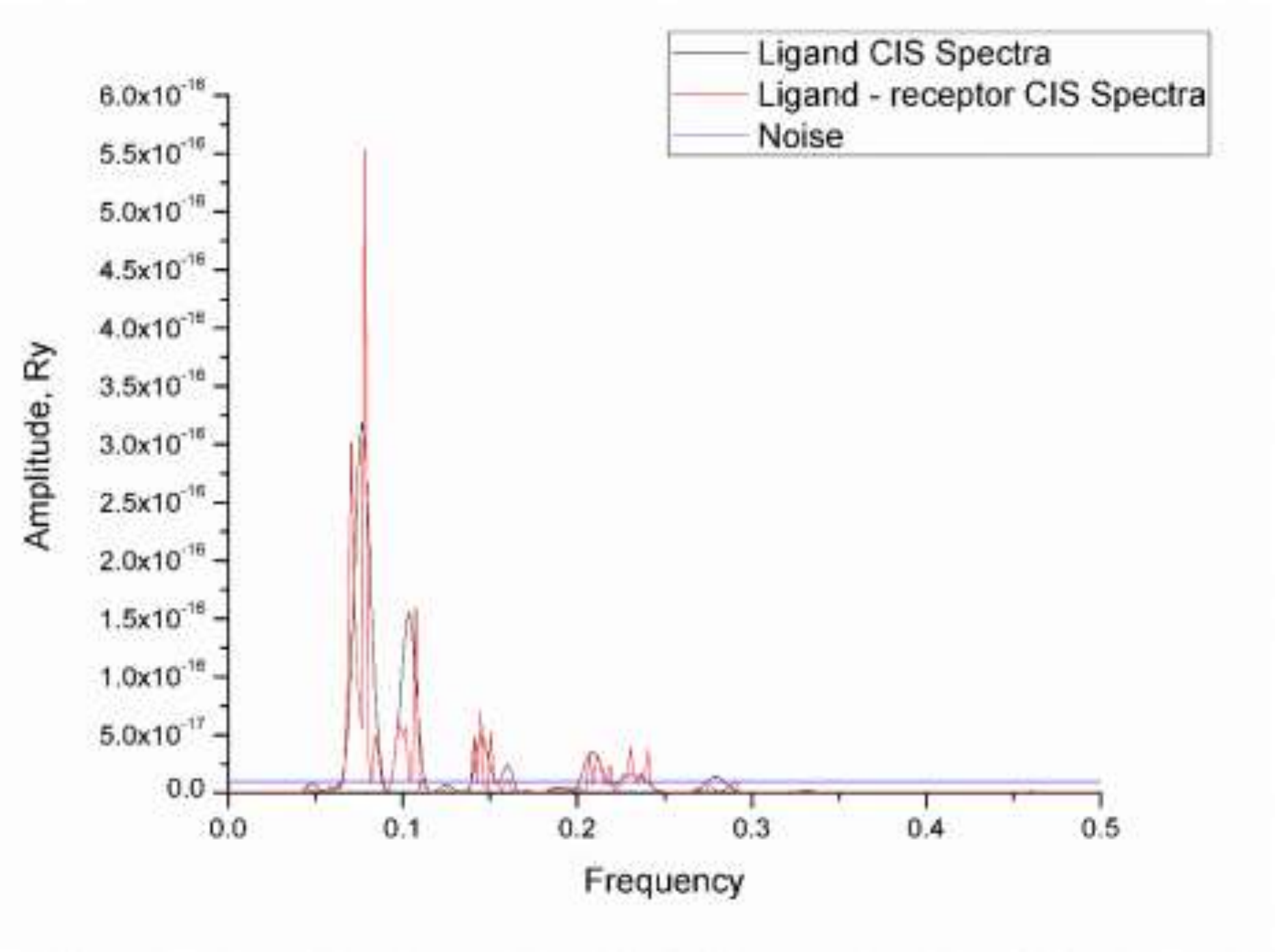

Ligand CIS Spectra
Ligand - receptor CIS Spectra
Noise
Amplitude, Ry
Frequency

hsa:3351 5-hydroxytryptamine receptor 1b

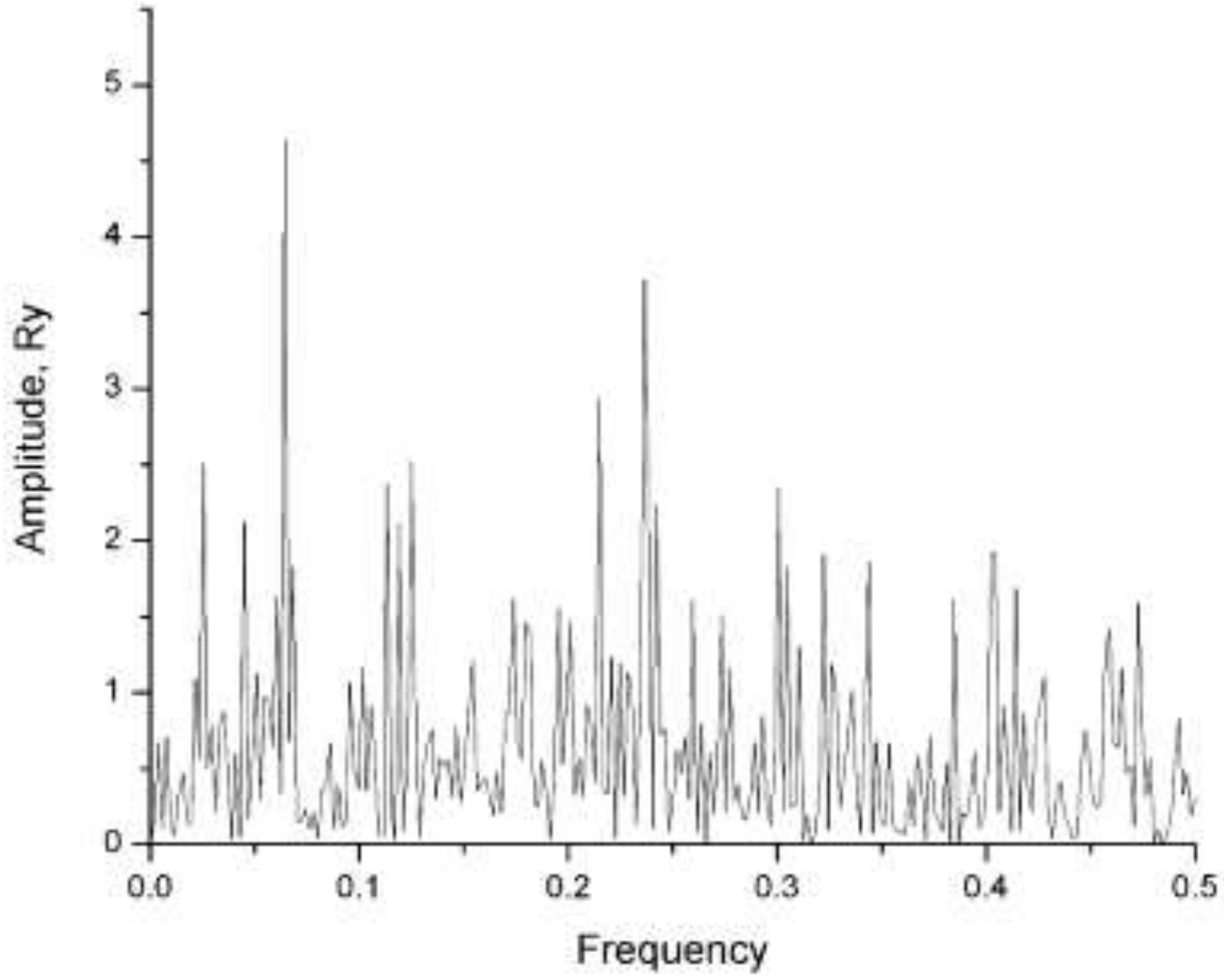

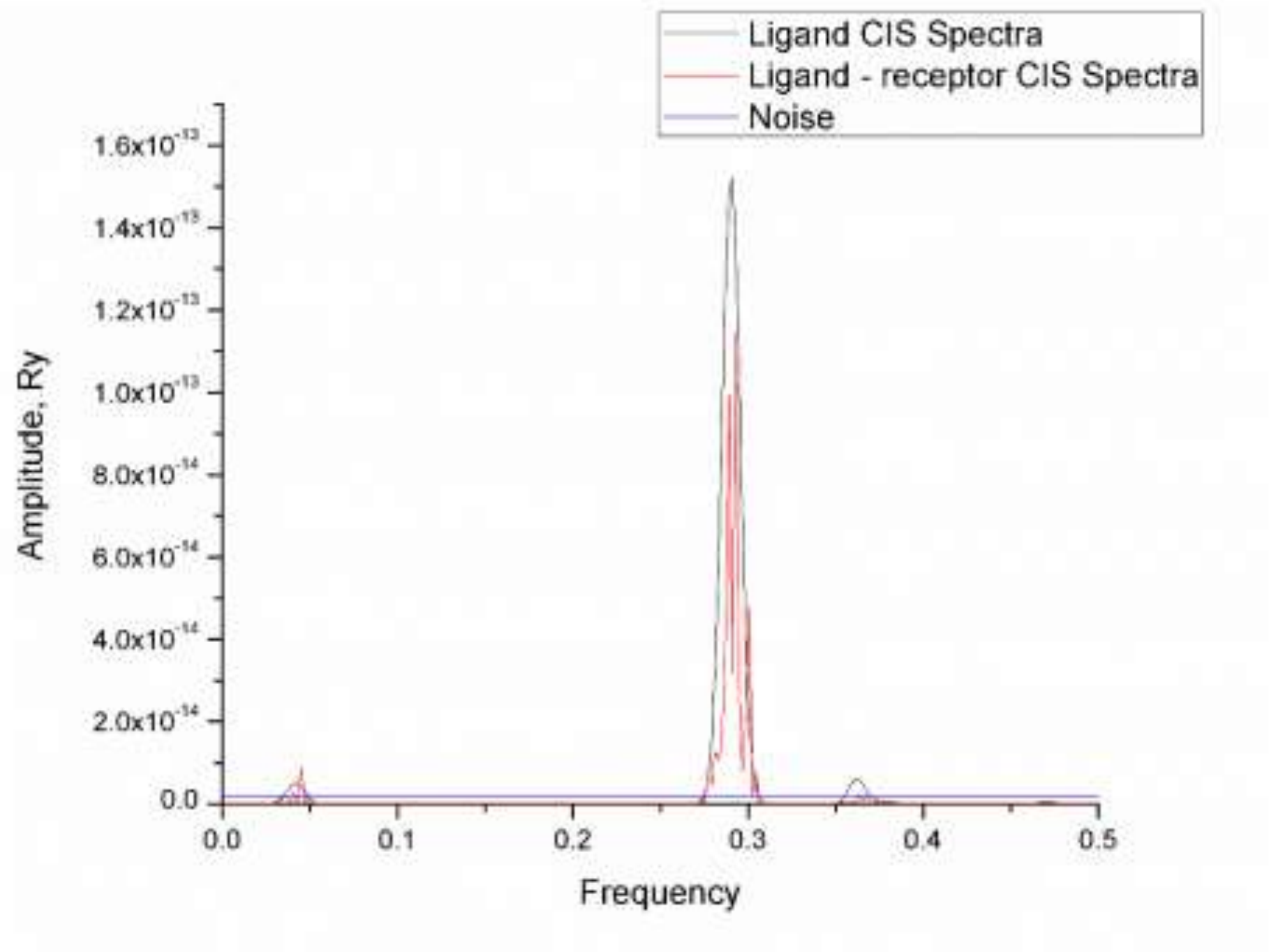

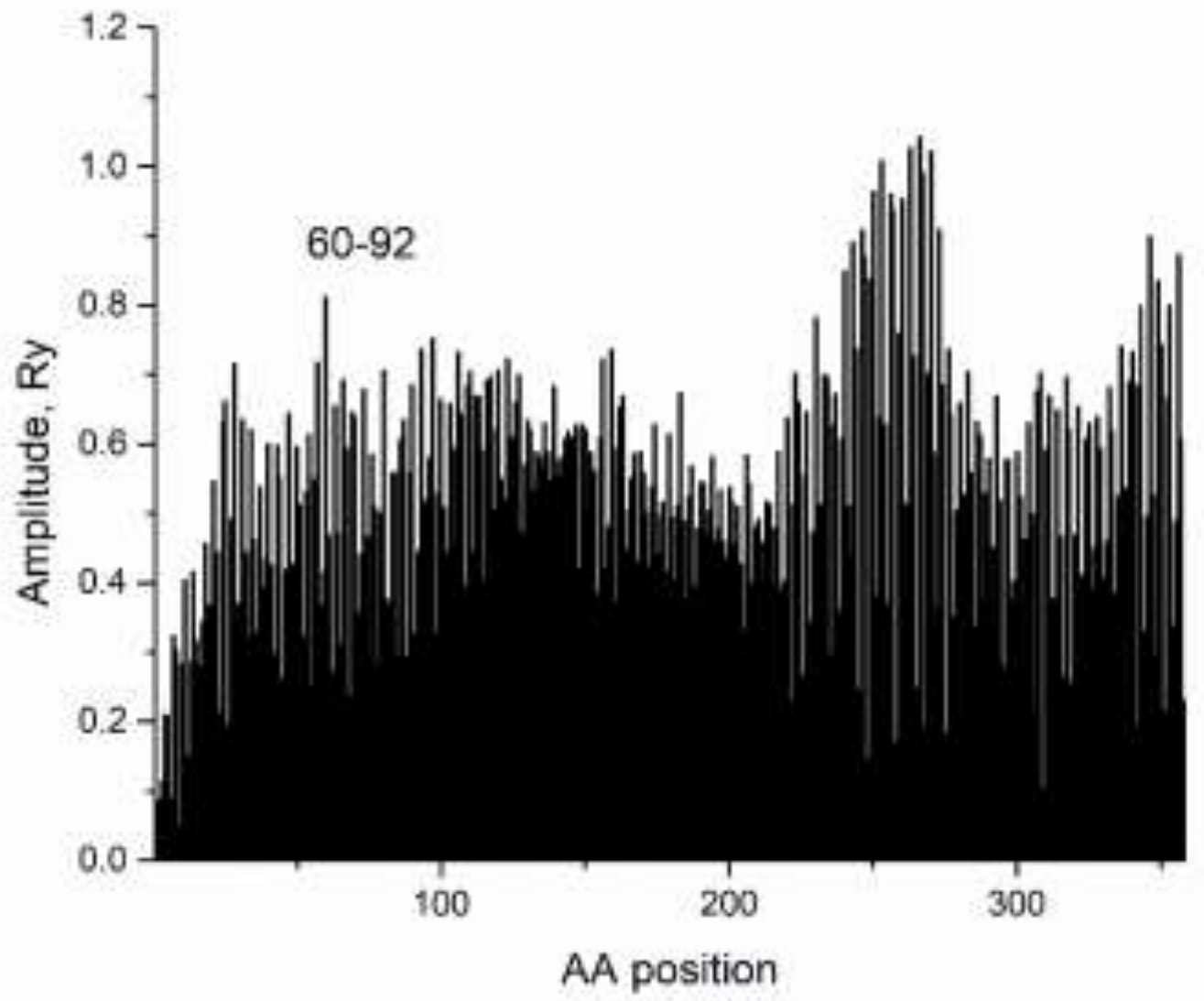

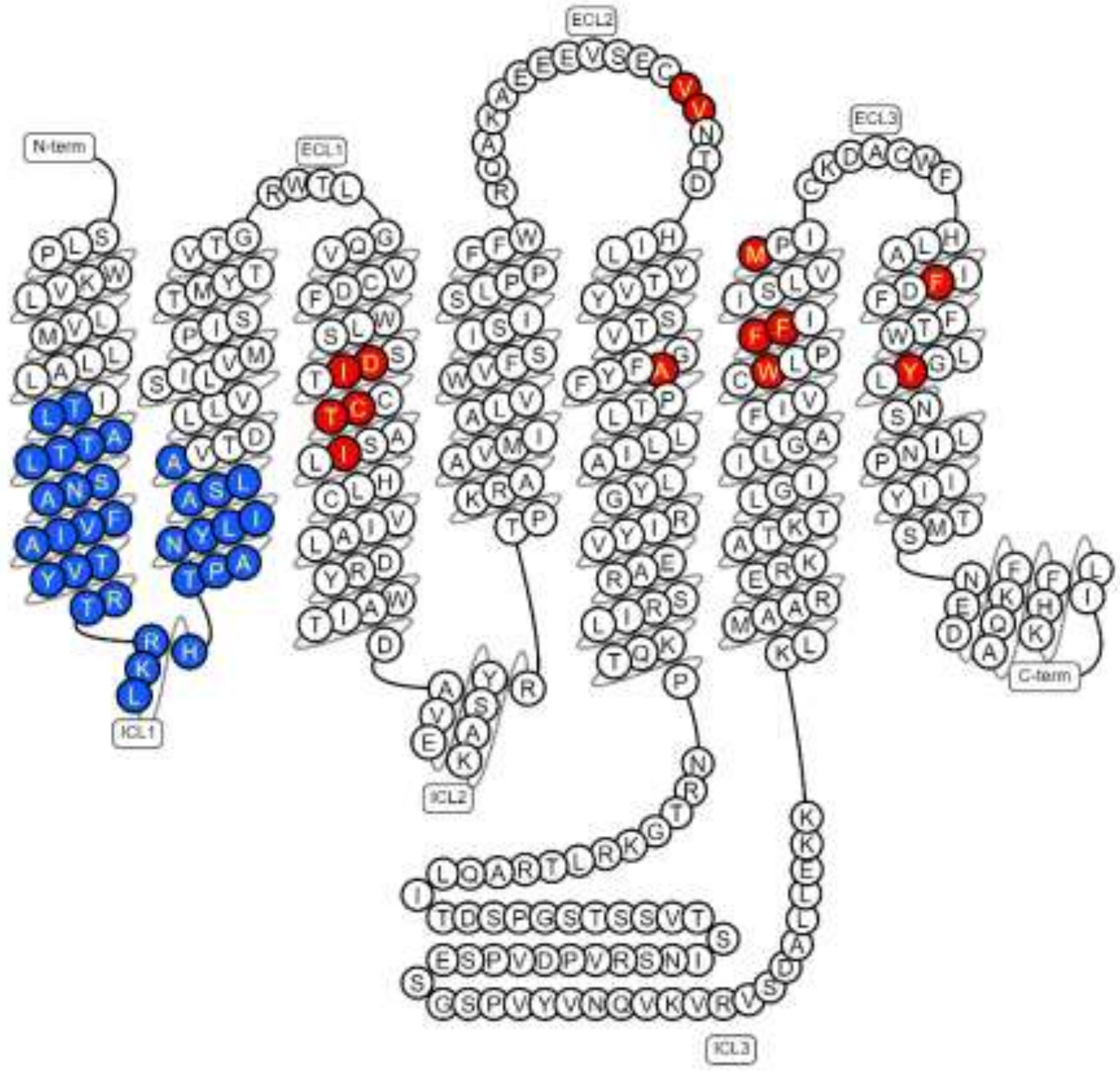

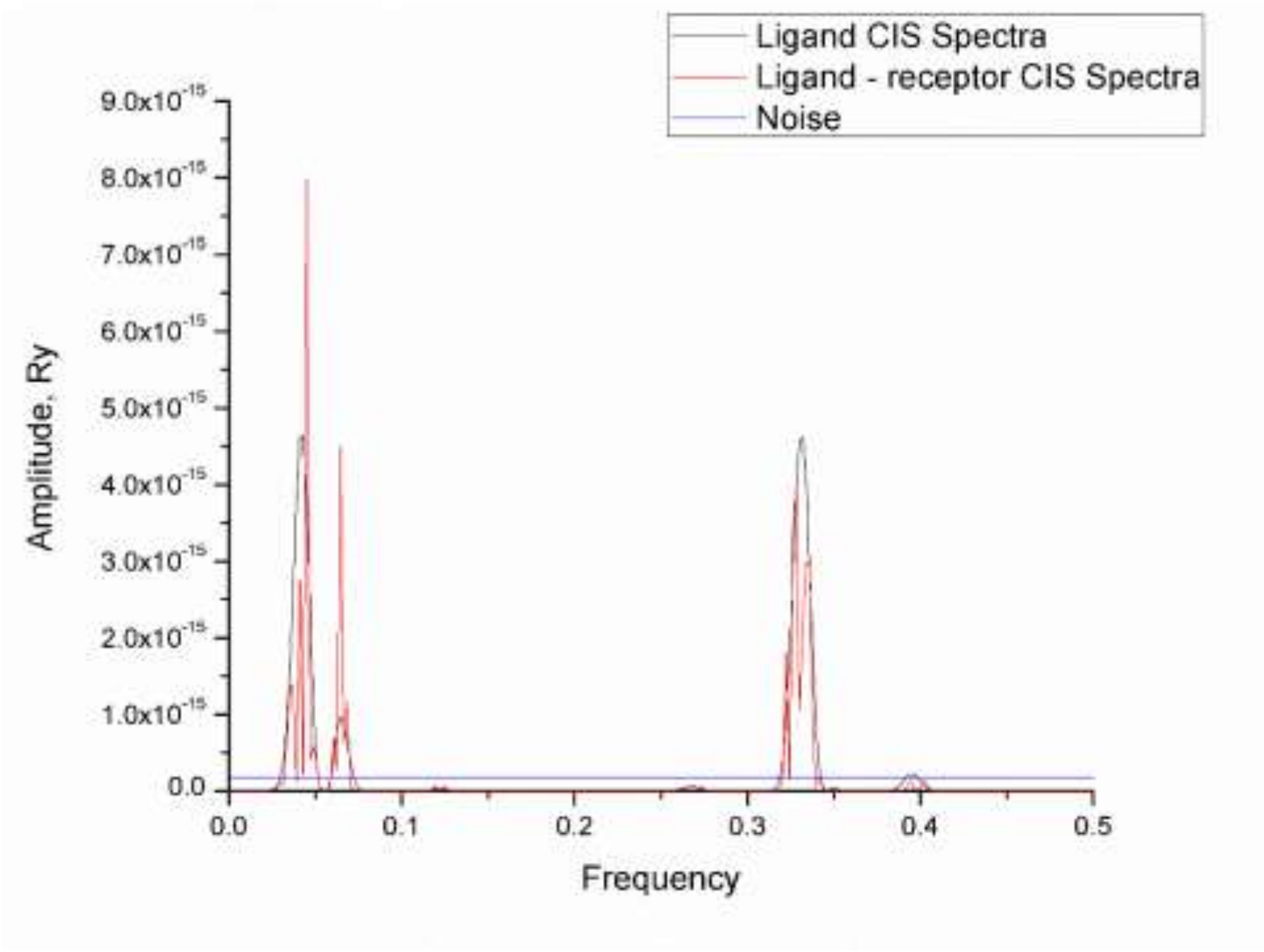

hsa:3352 5-hydroxytryptamine receptor 1d

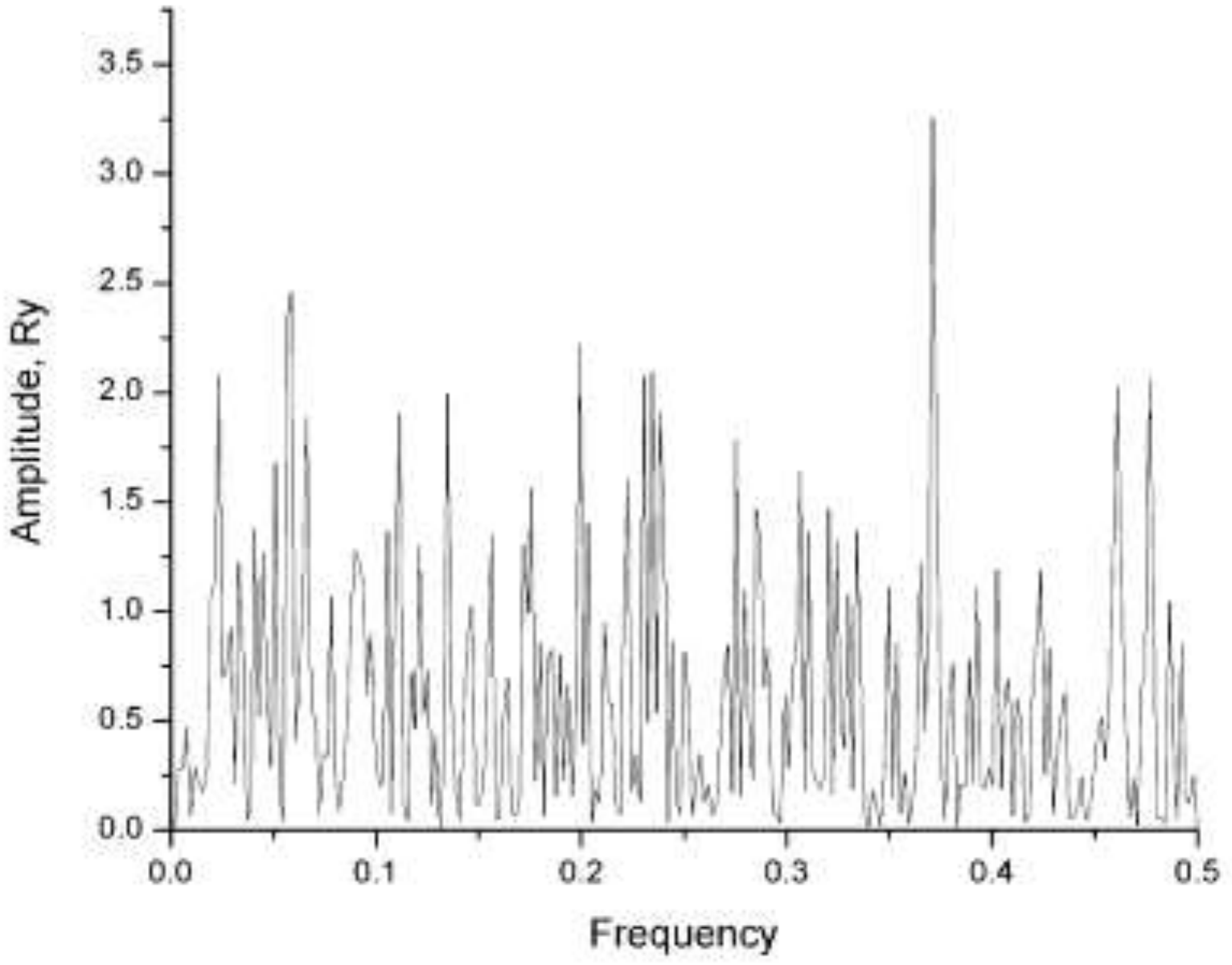

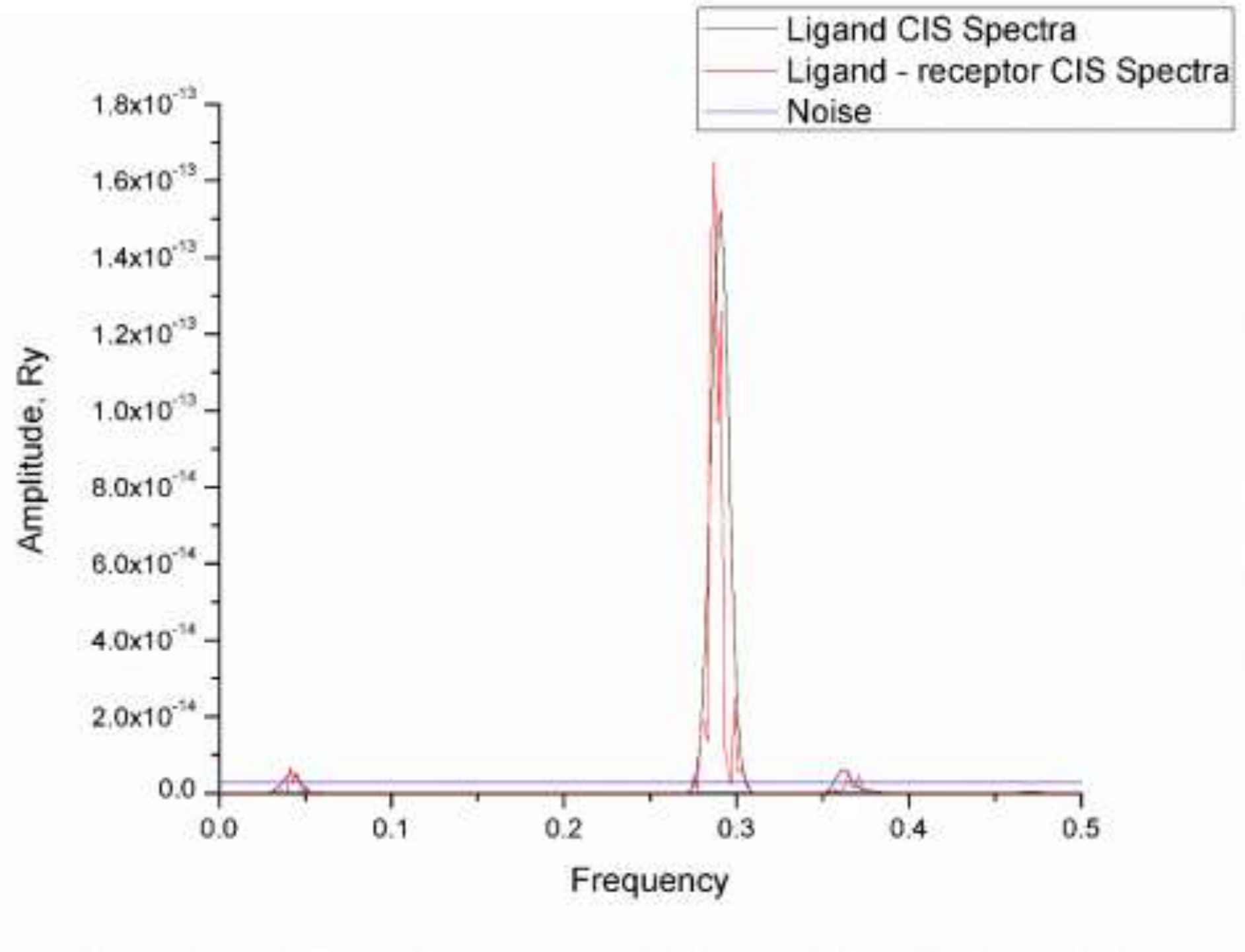

Ligand CIS Spectra
Ligand - receptor CIS Spectra
Noise
Amplitude, Ry
Frequency

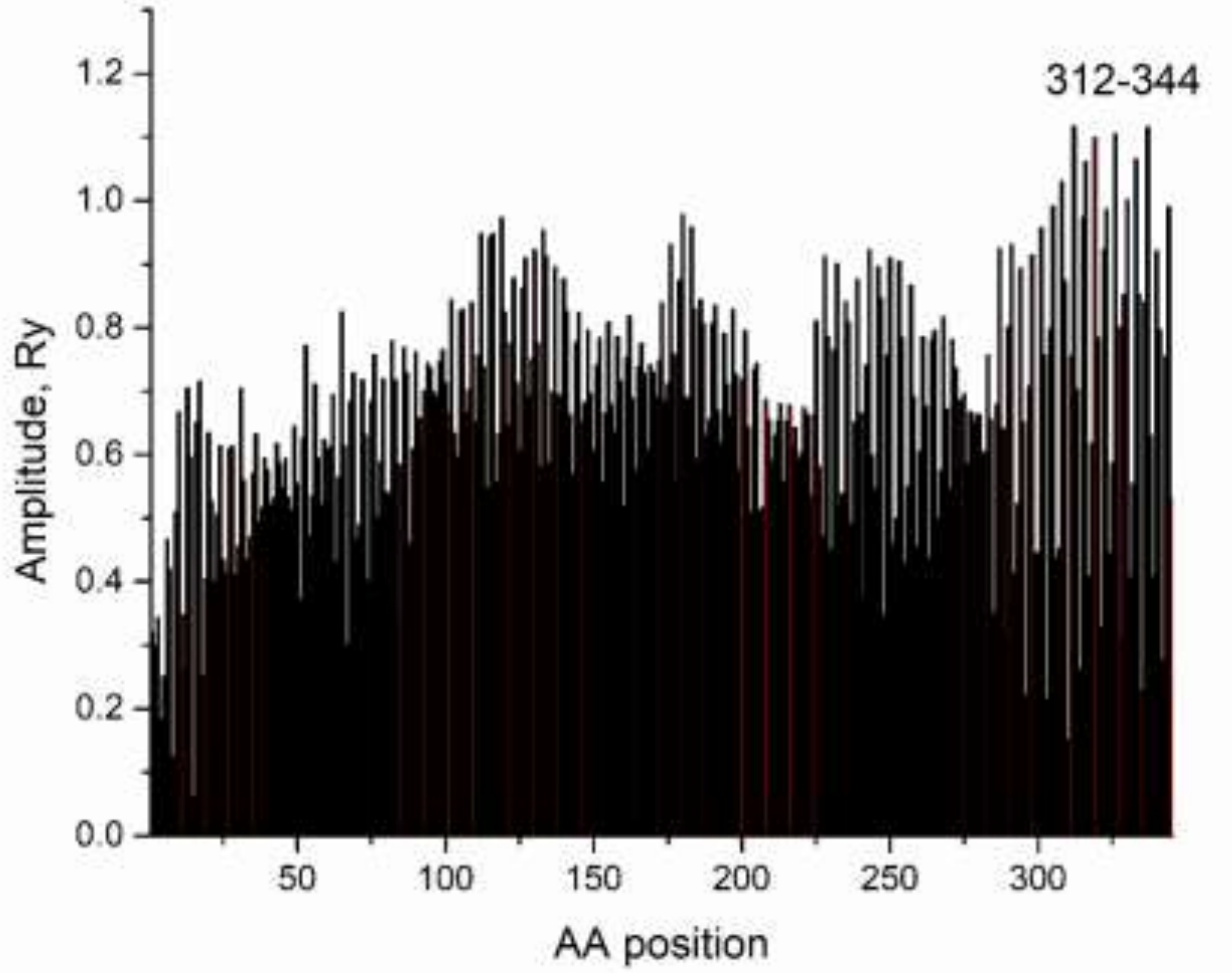

312-344
Amplitude, Ry
AA position

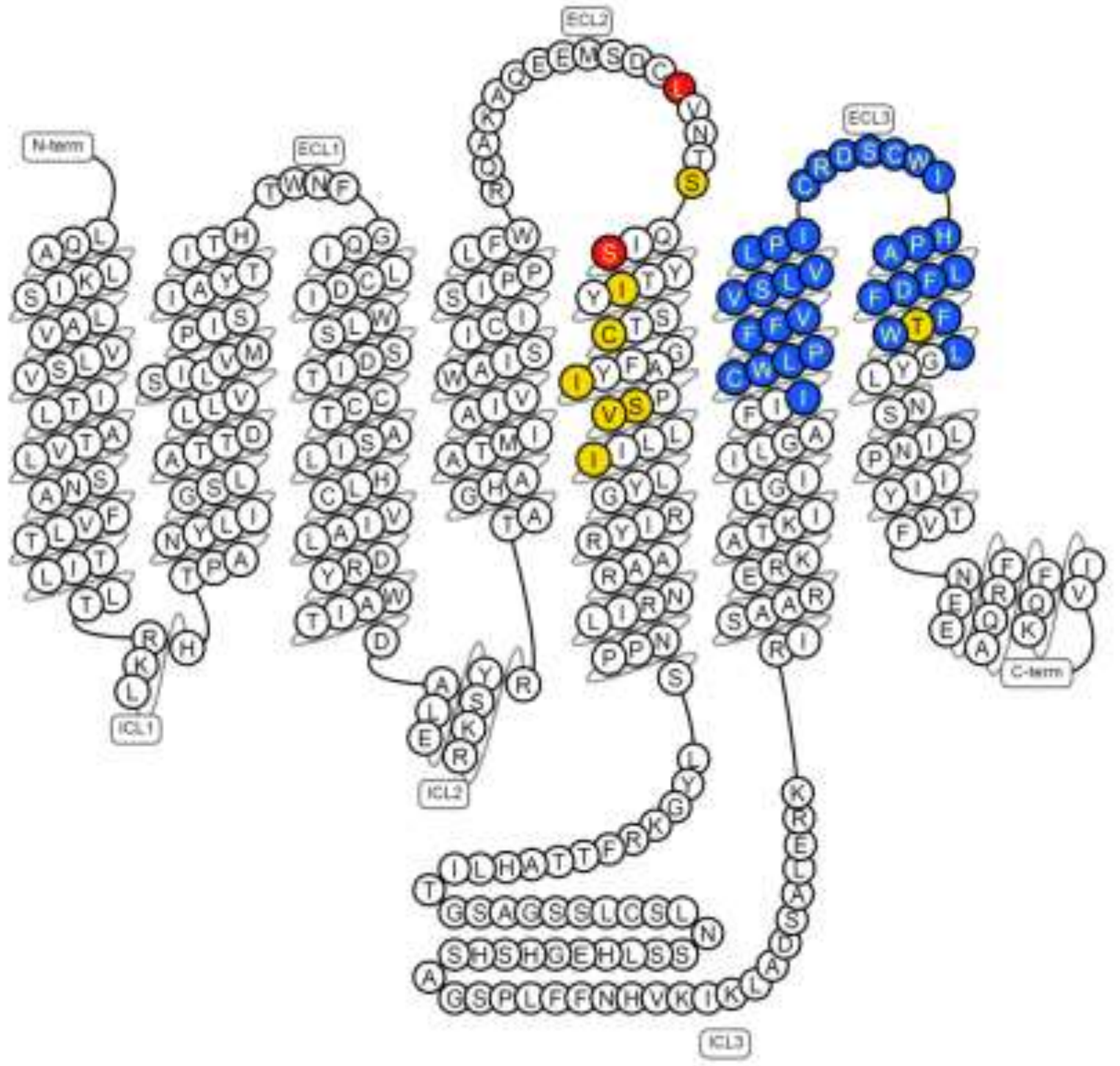

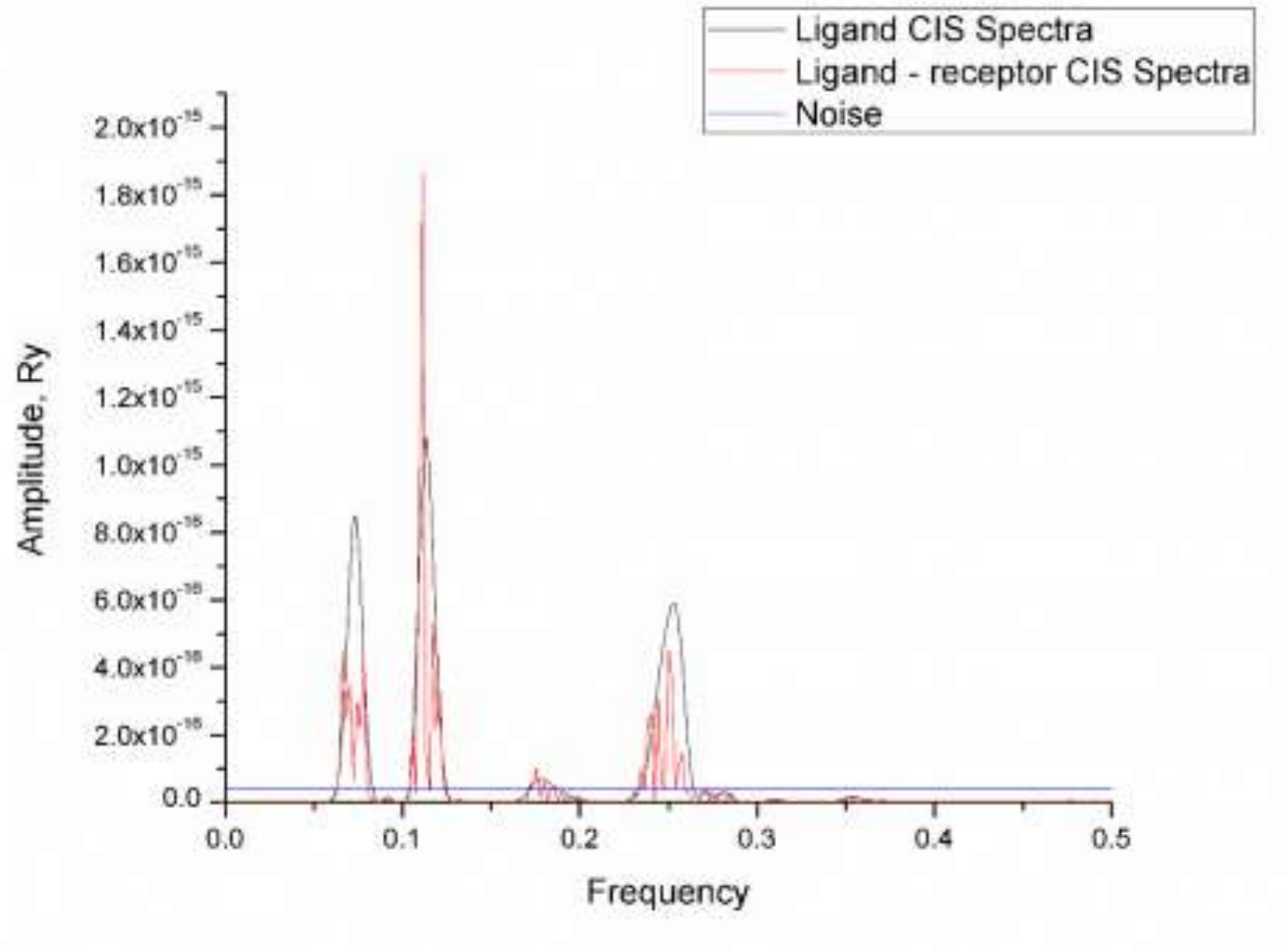

hsa:3356 5-hydroxytryptamine receptor 2a

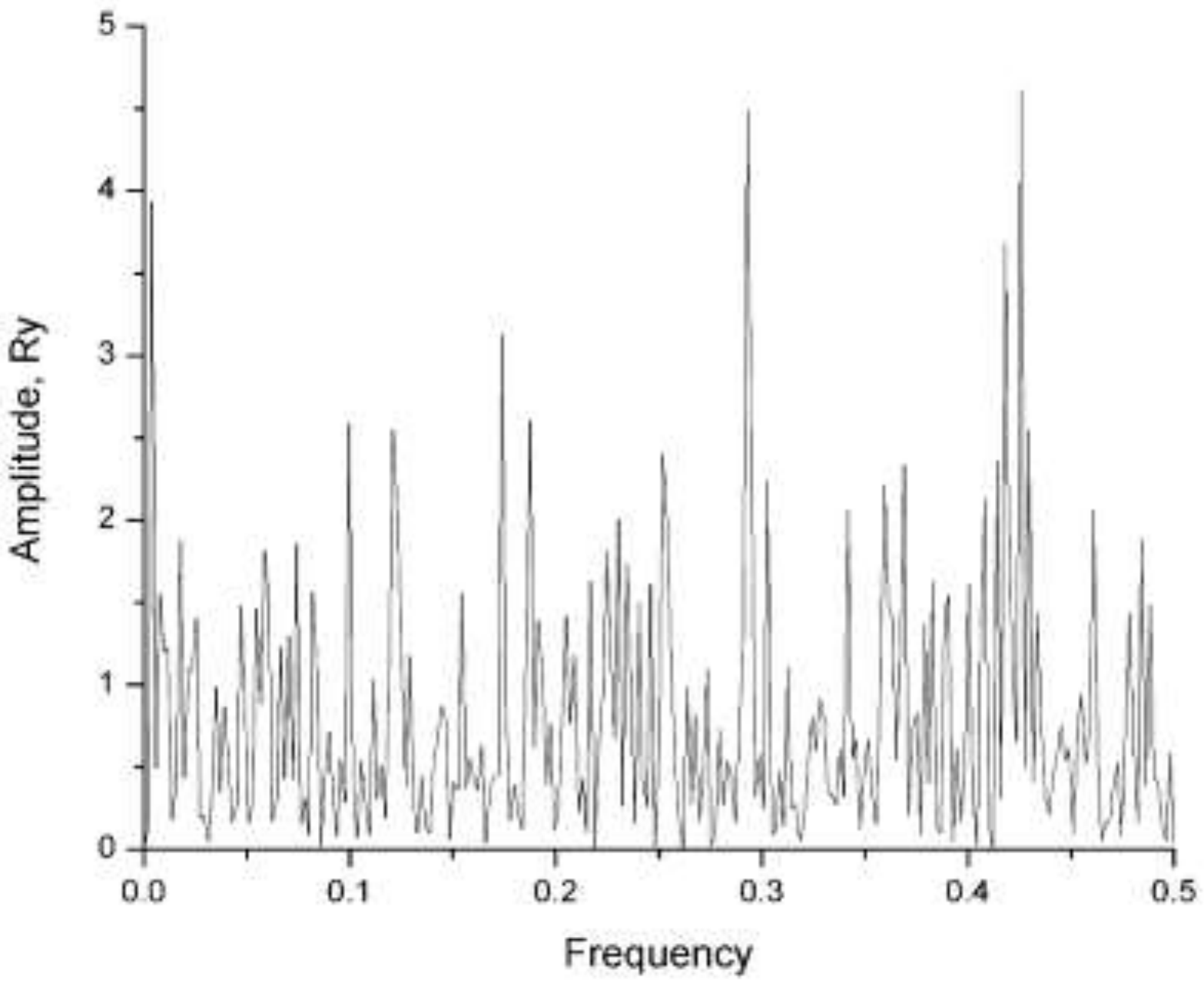

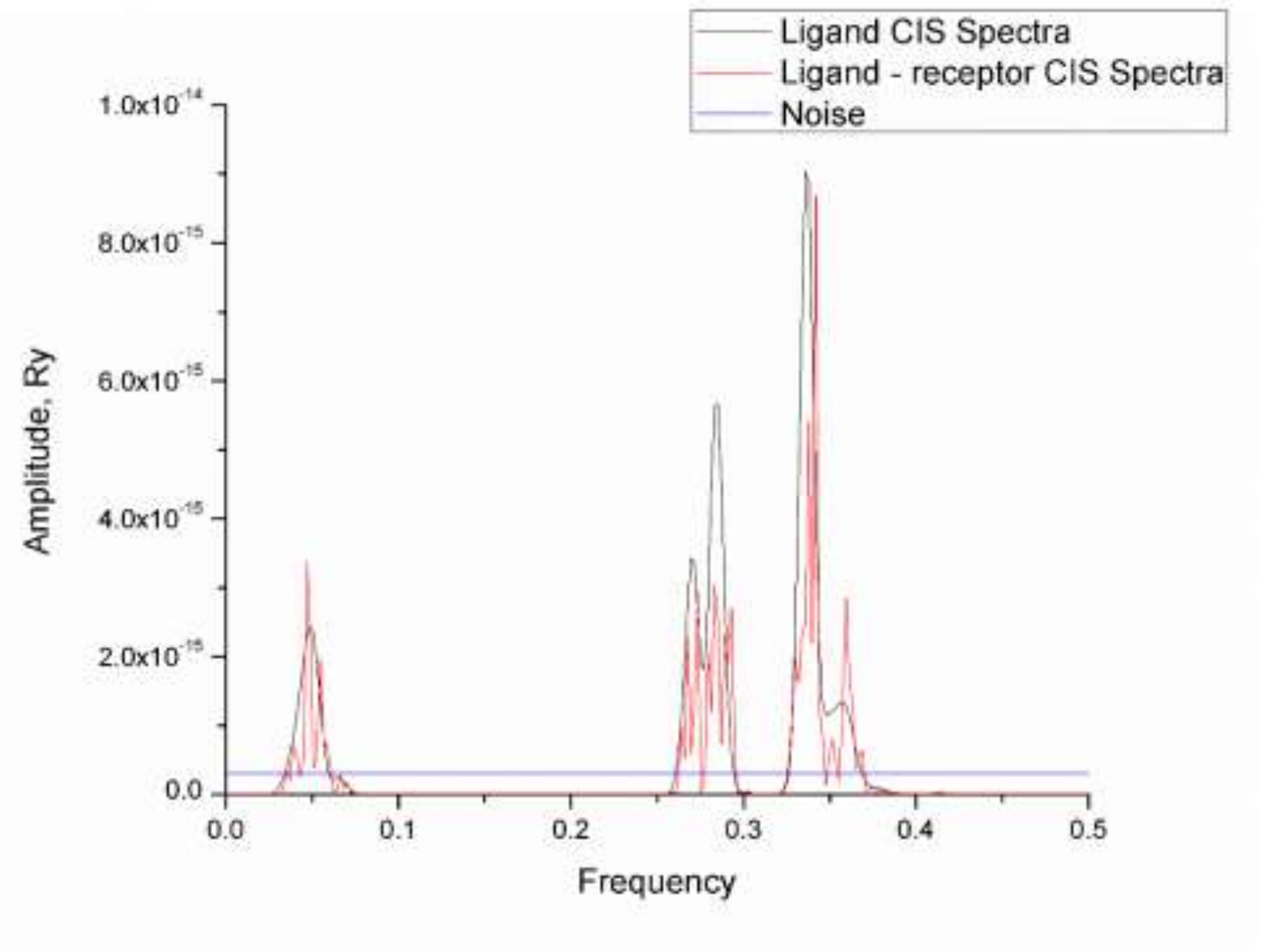

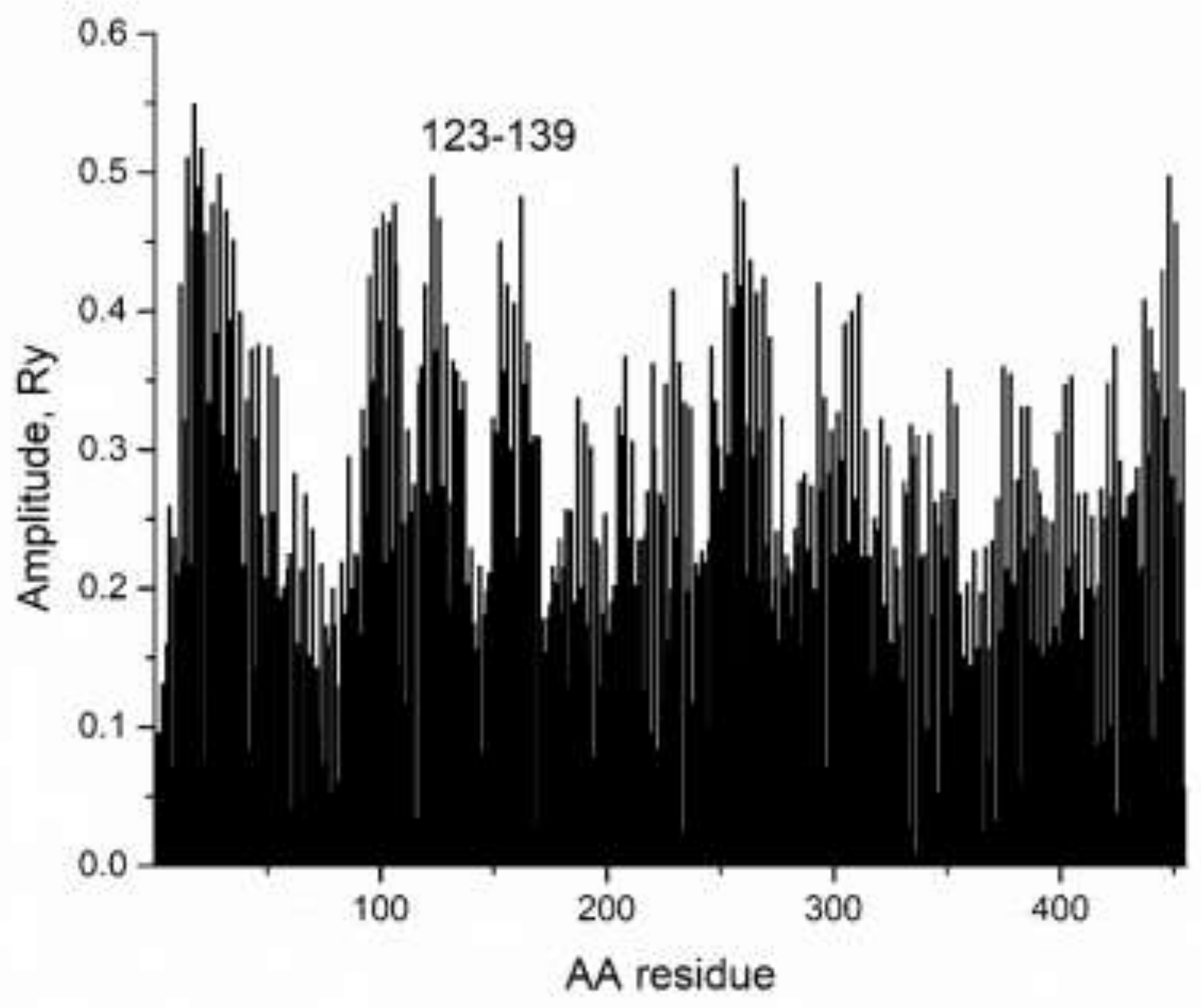

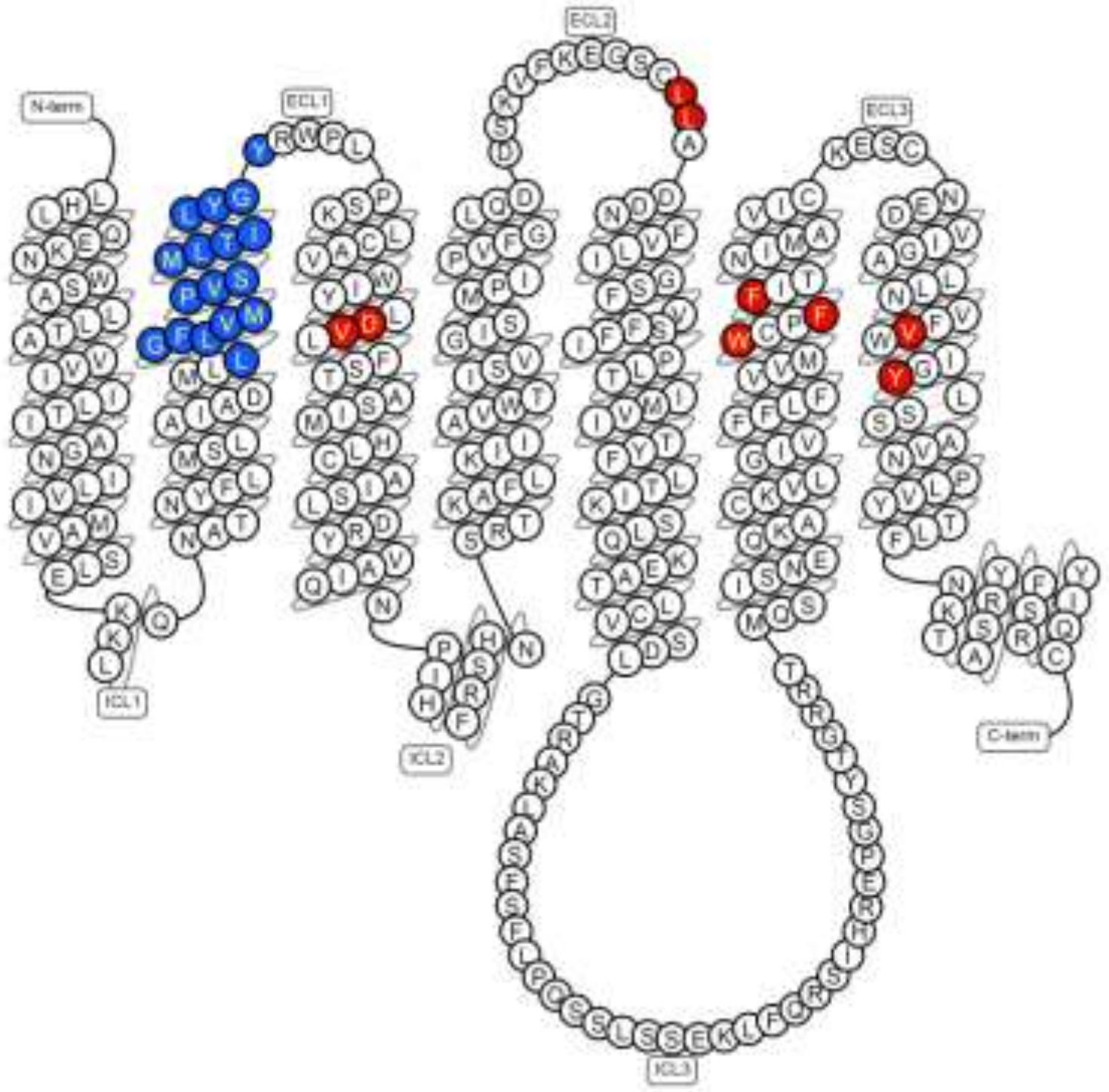

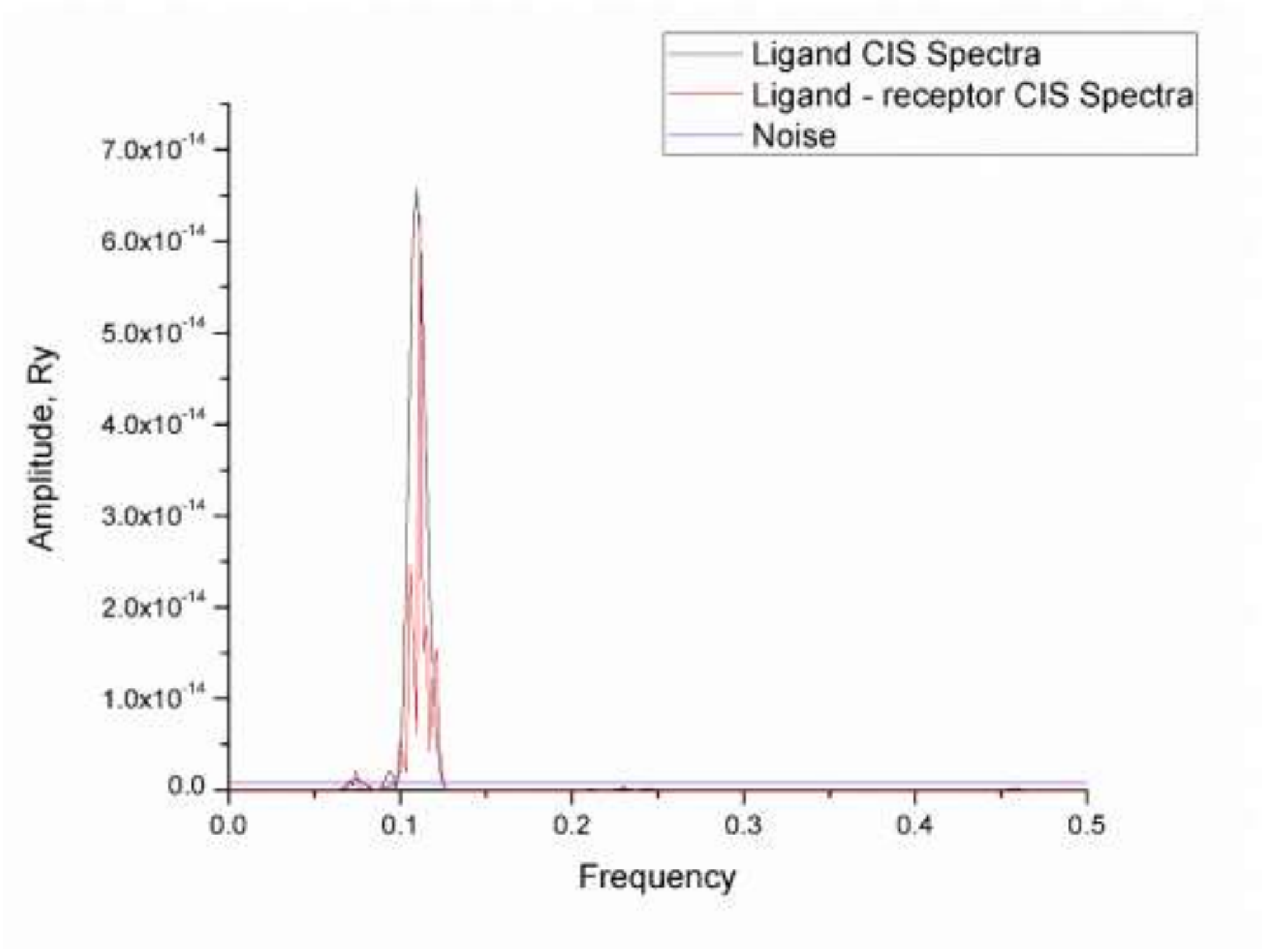

hsa:3358 5-hydroxytryptamine receptor 2c

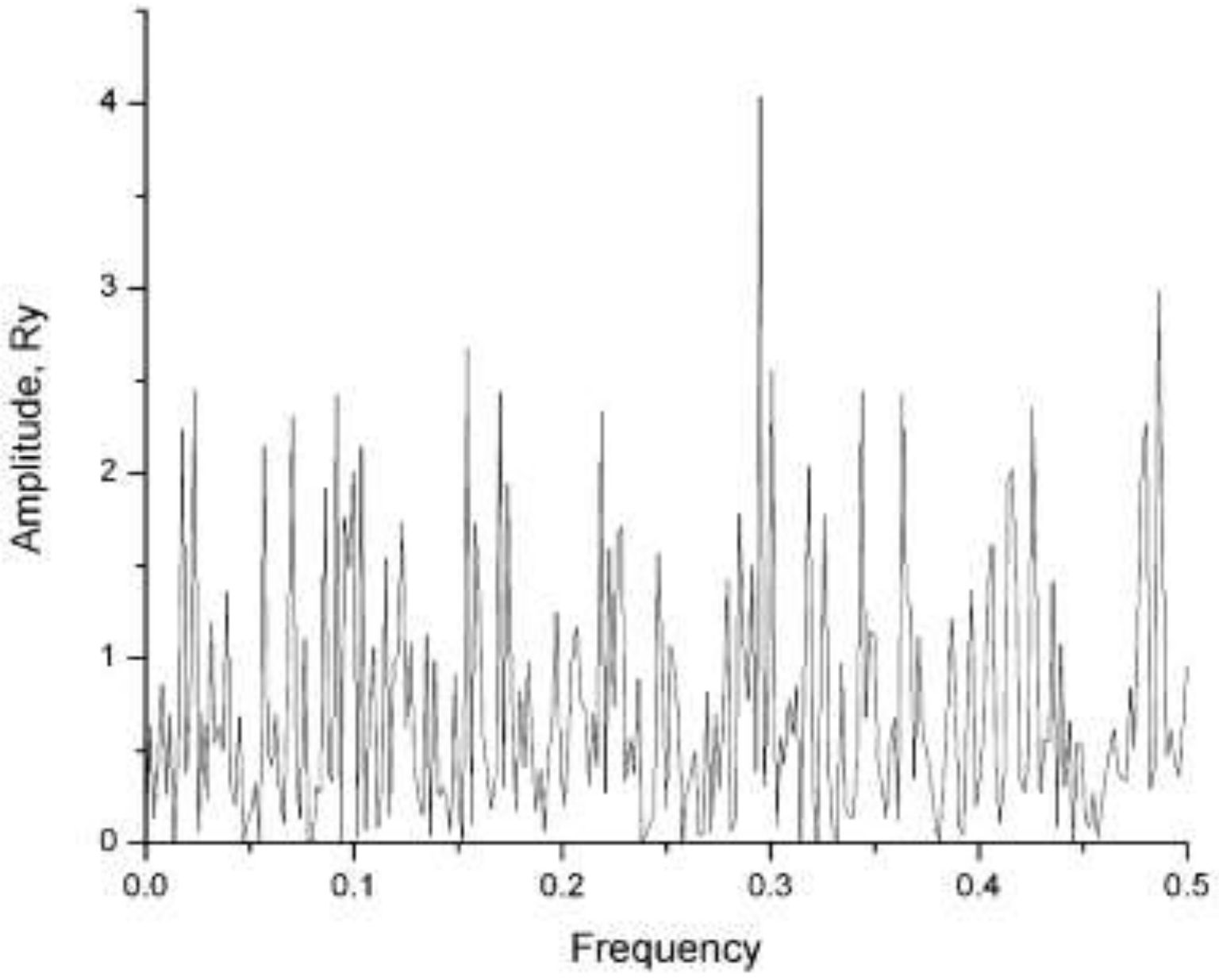

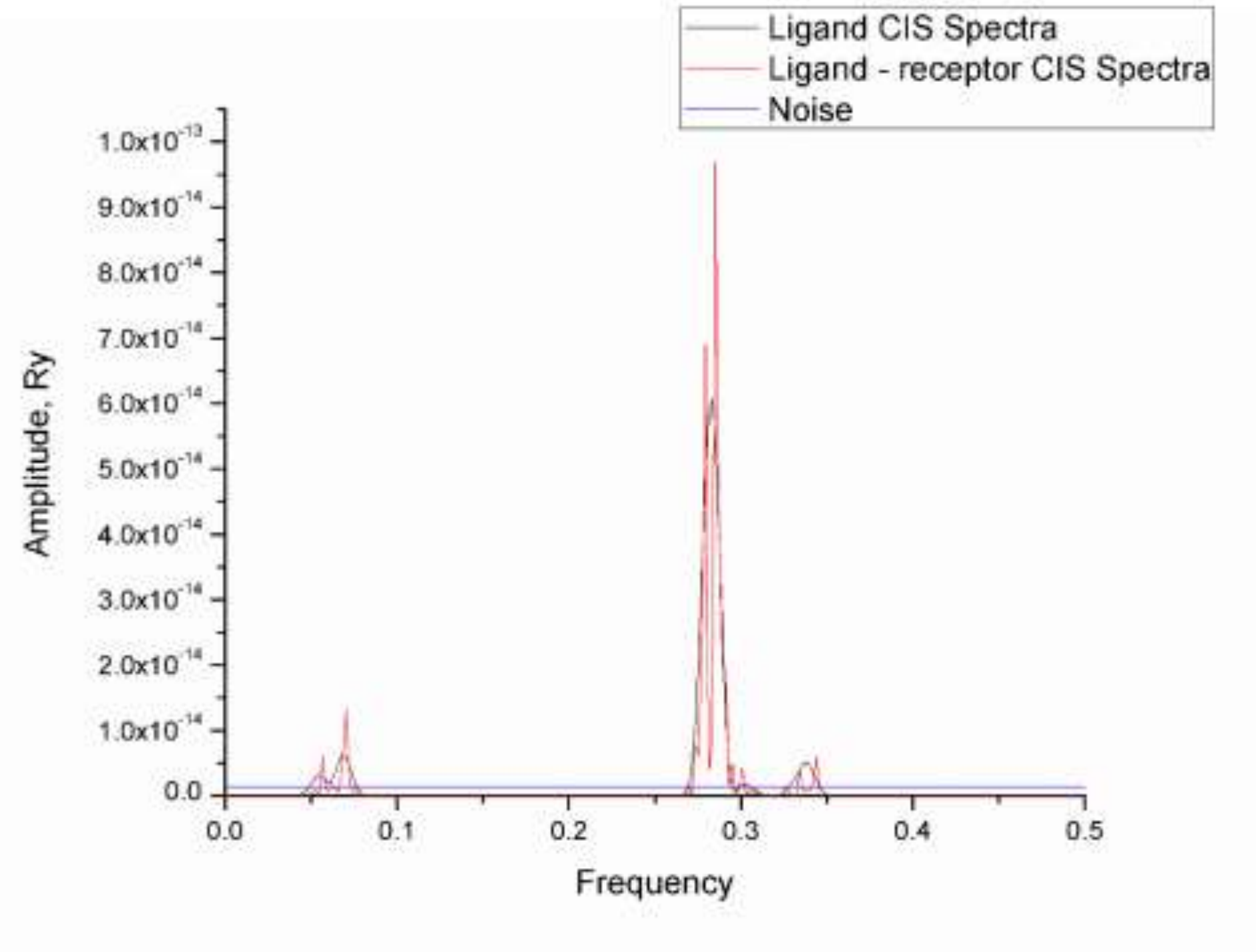

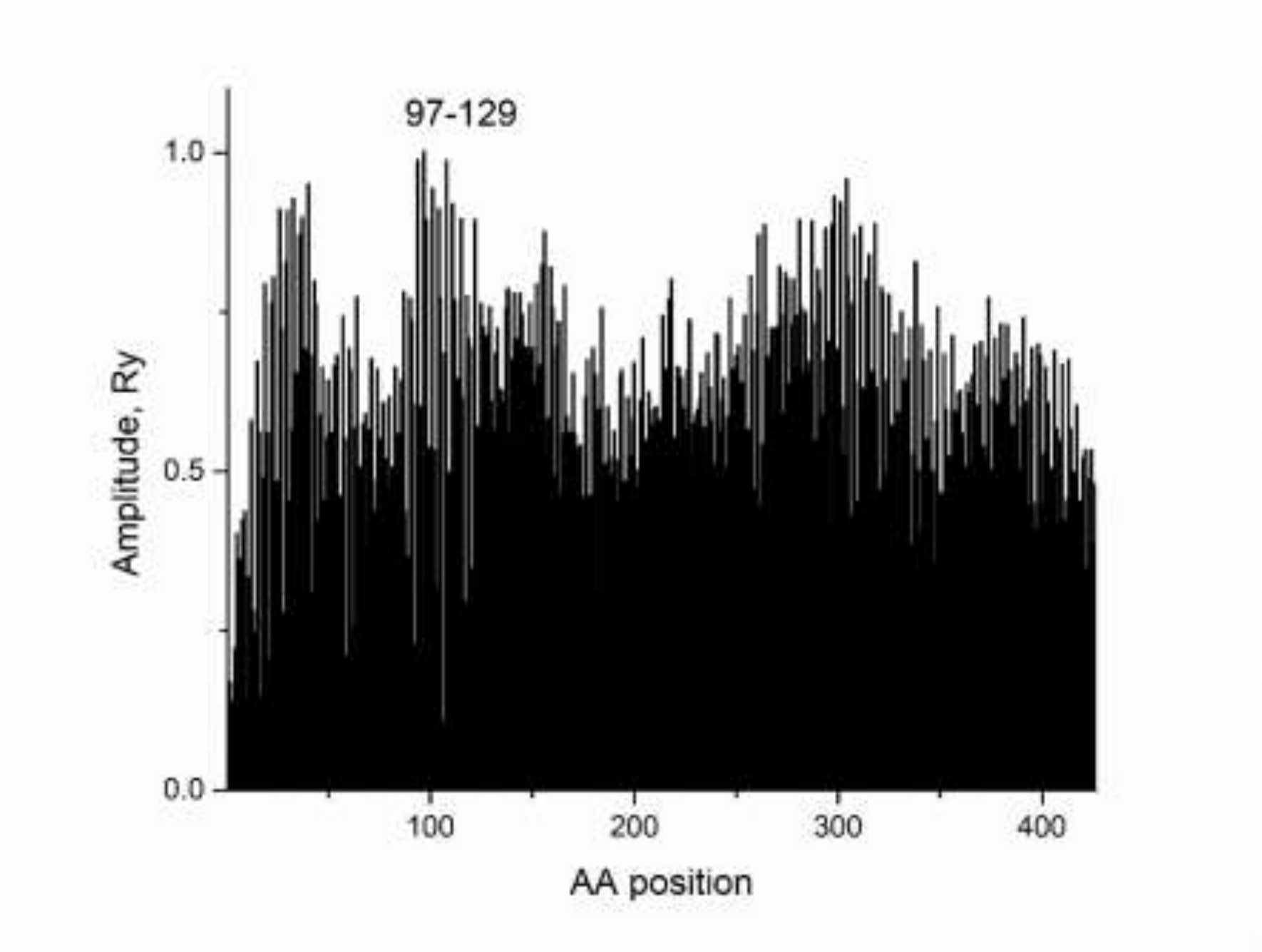

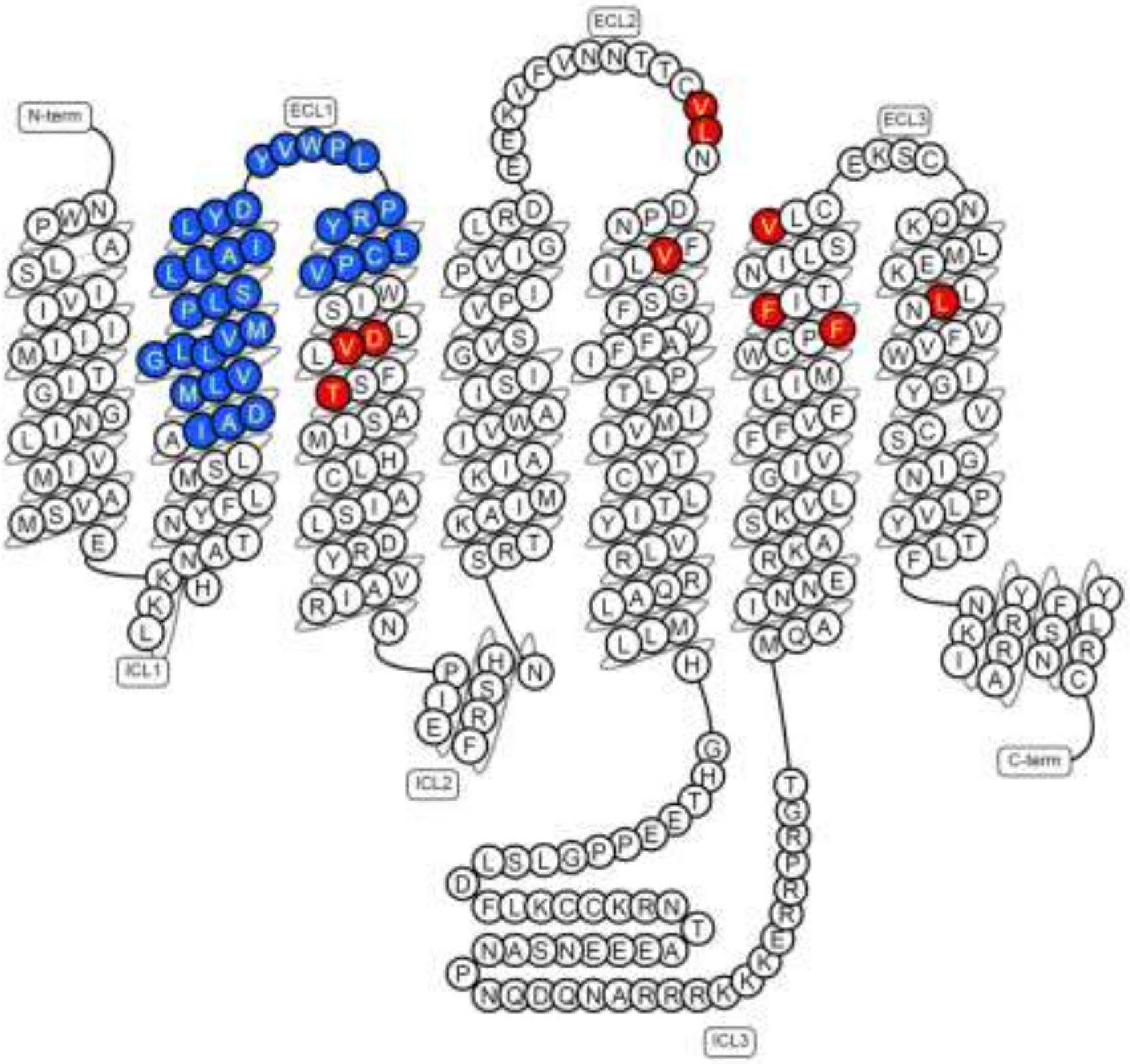

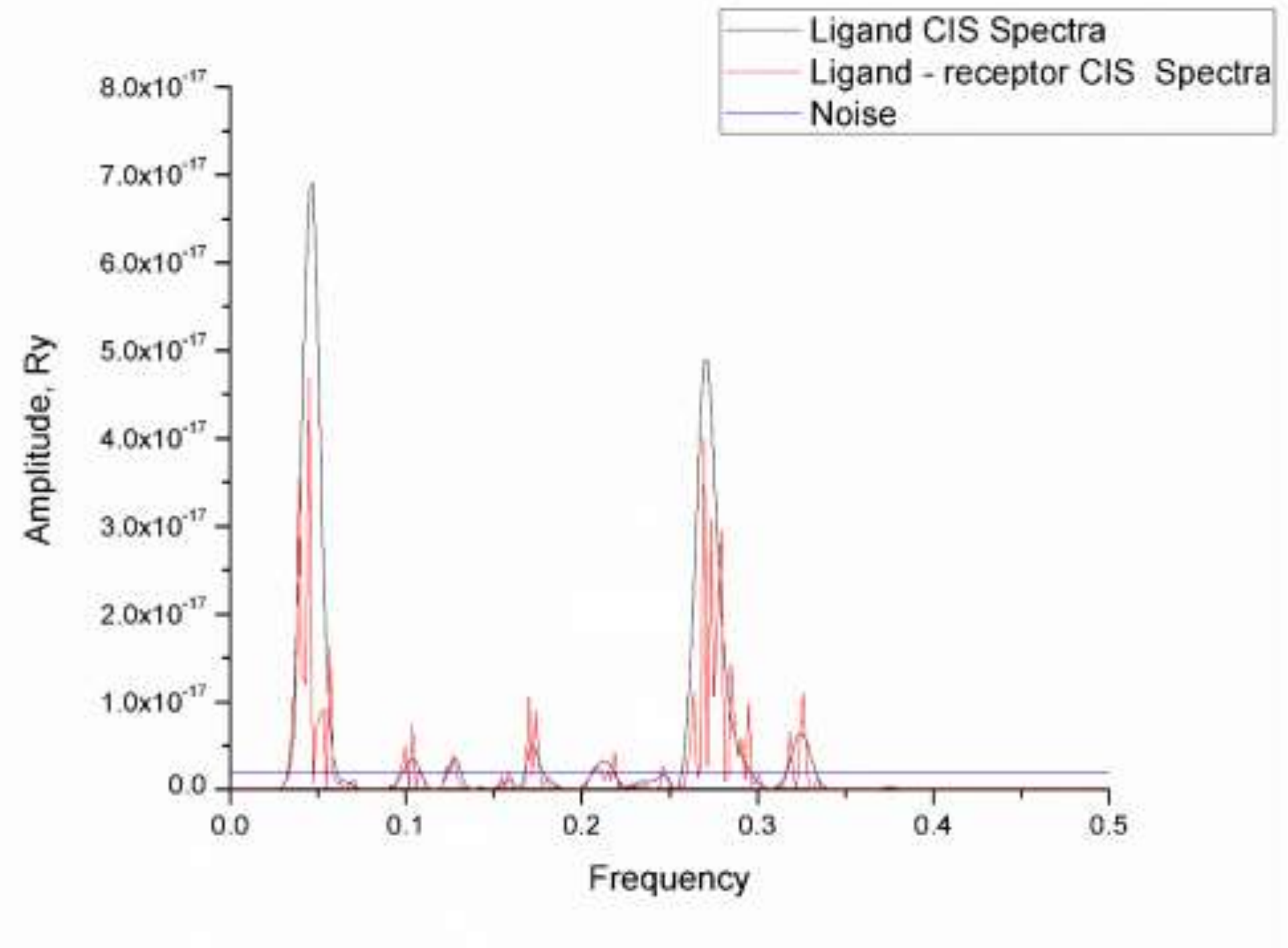

hsa:3577 high affinity interleukin-8 receptor

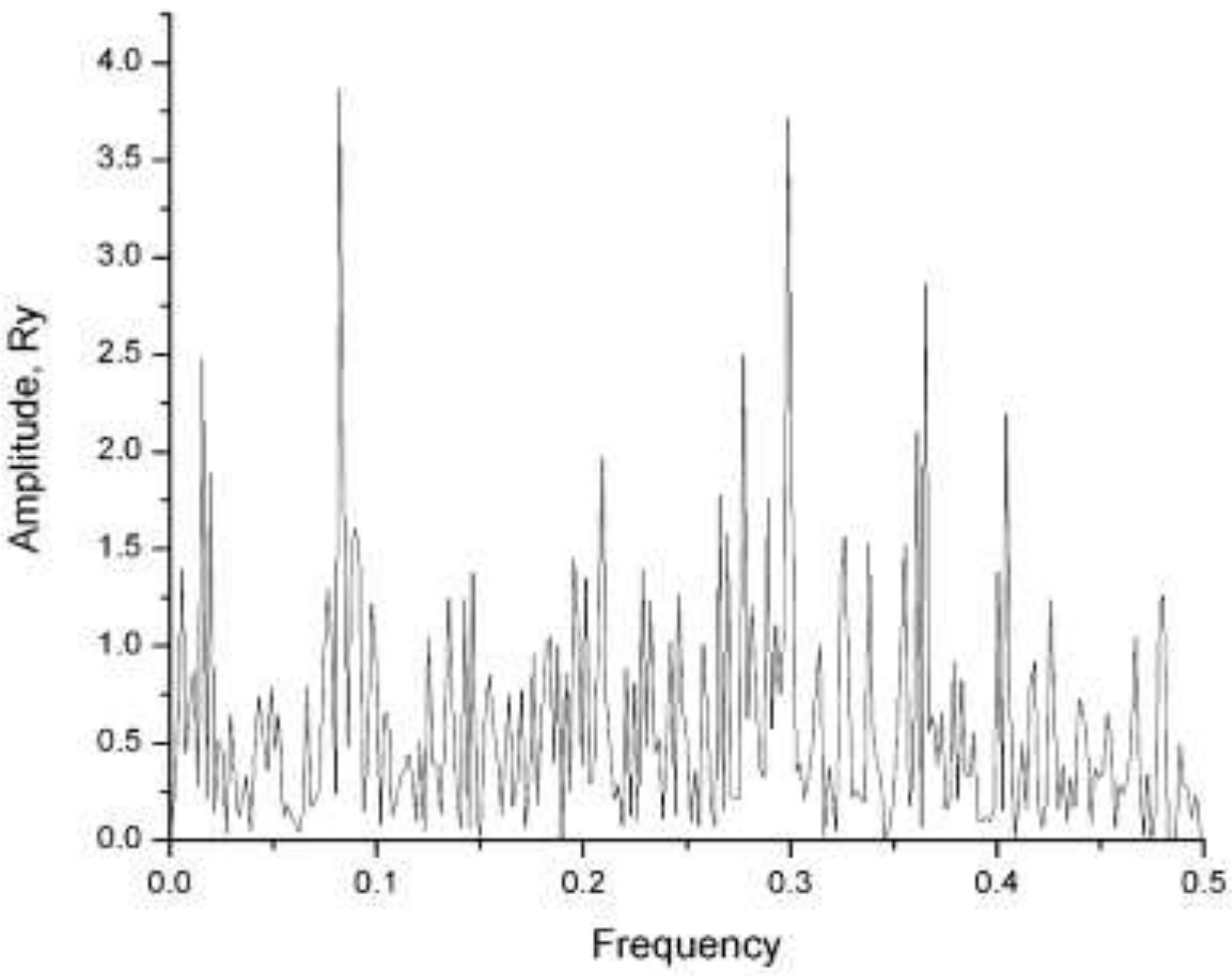

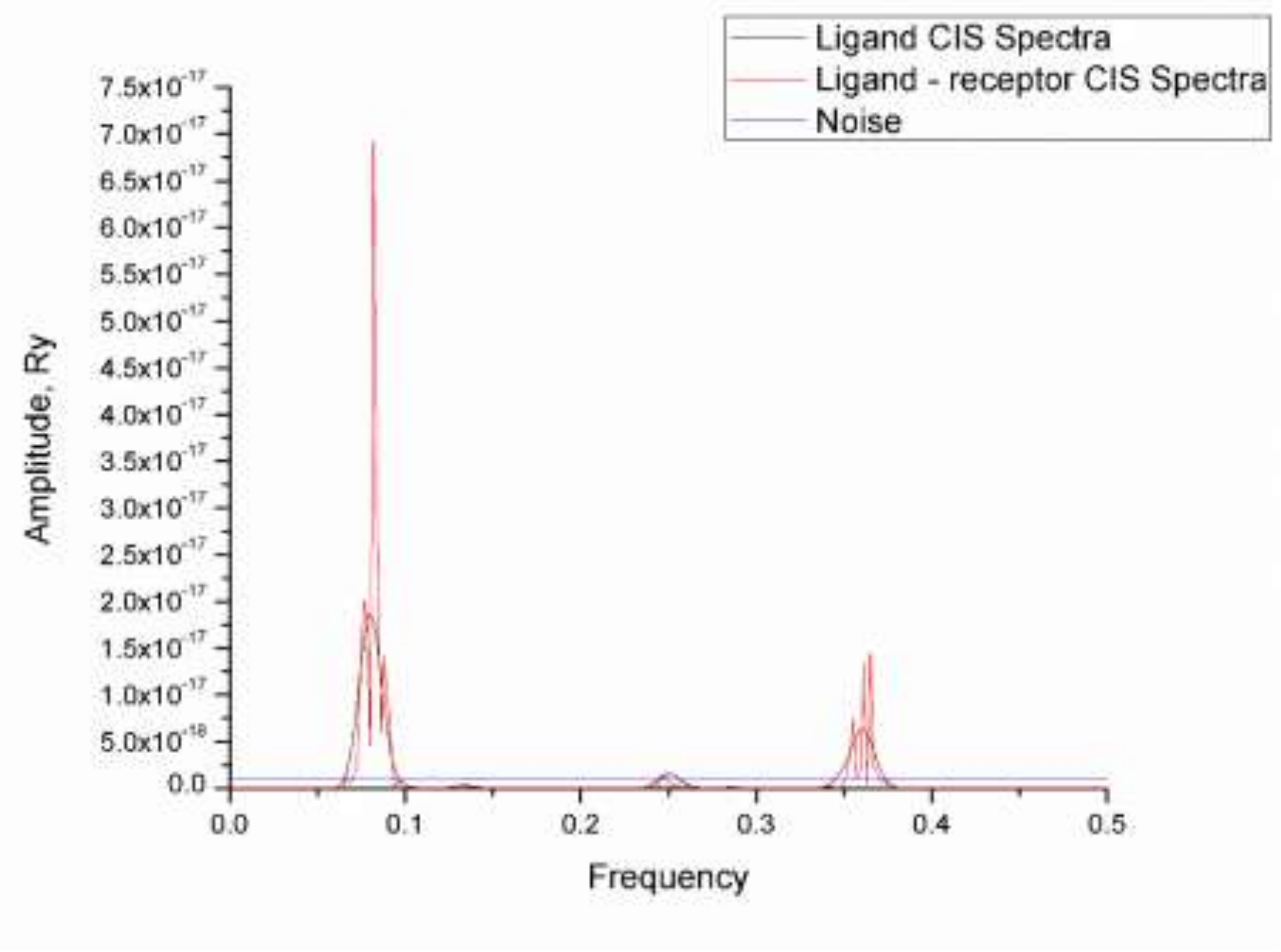

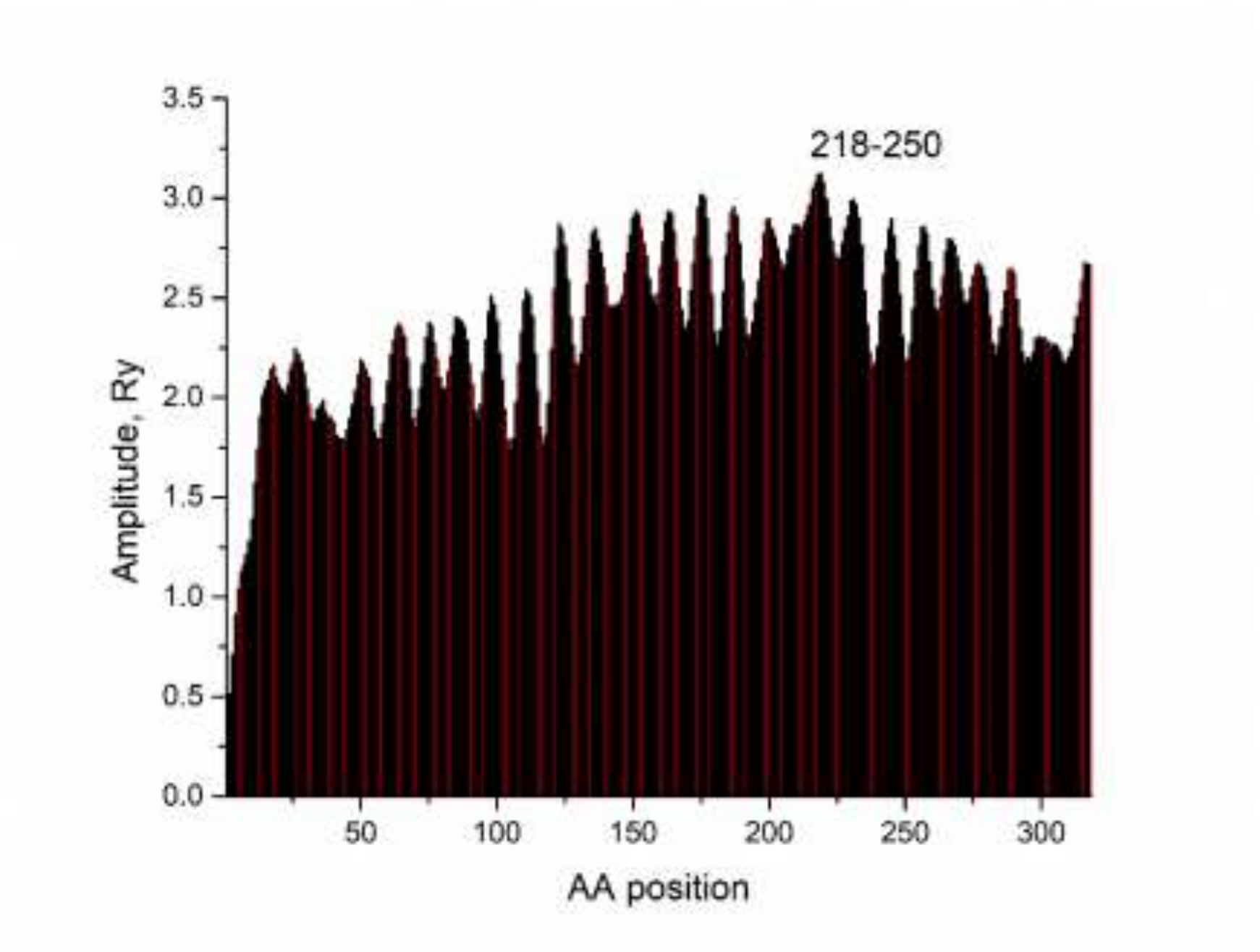

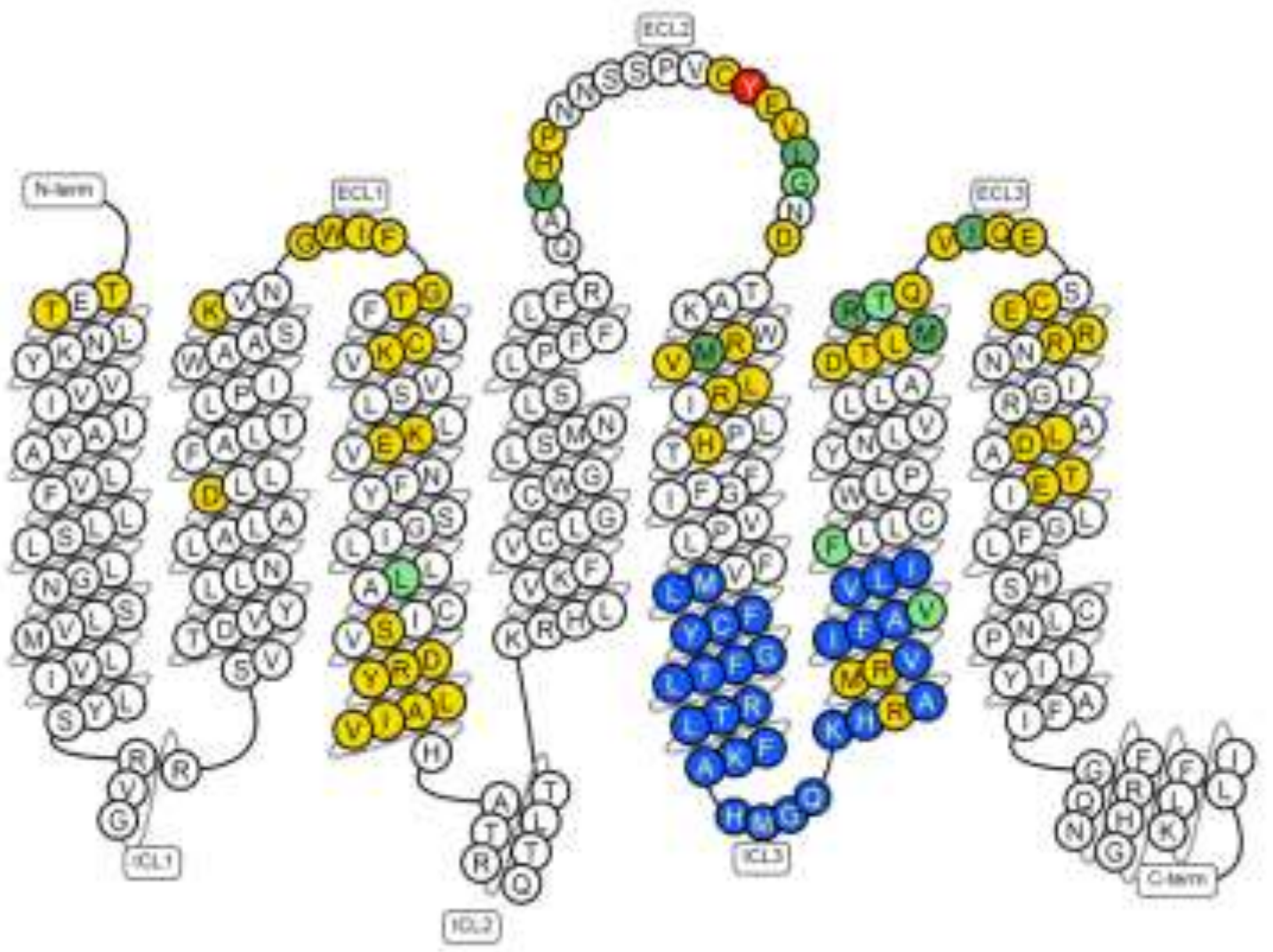

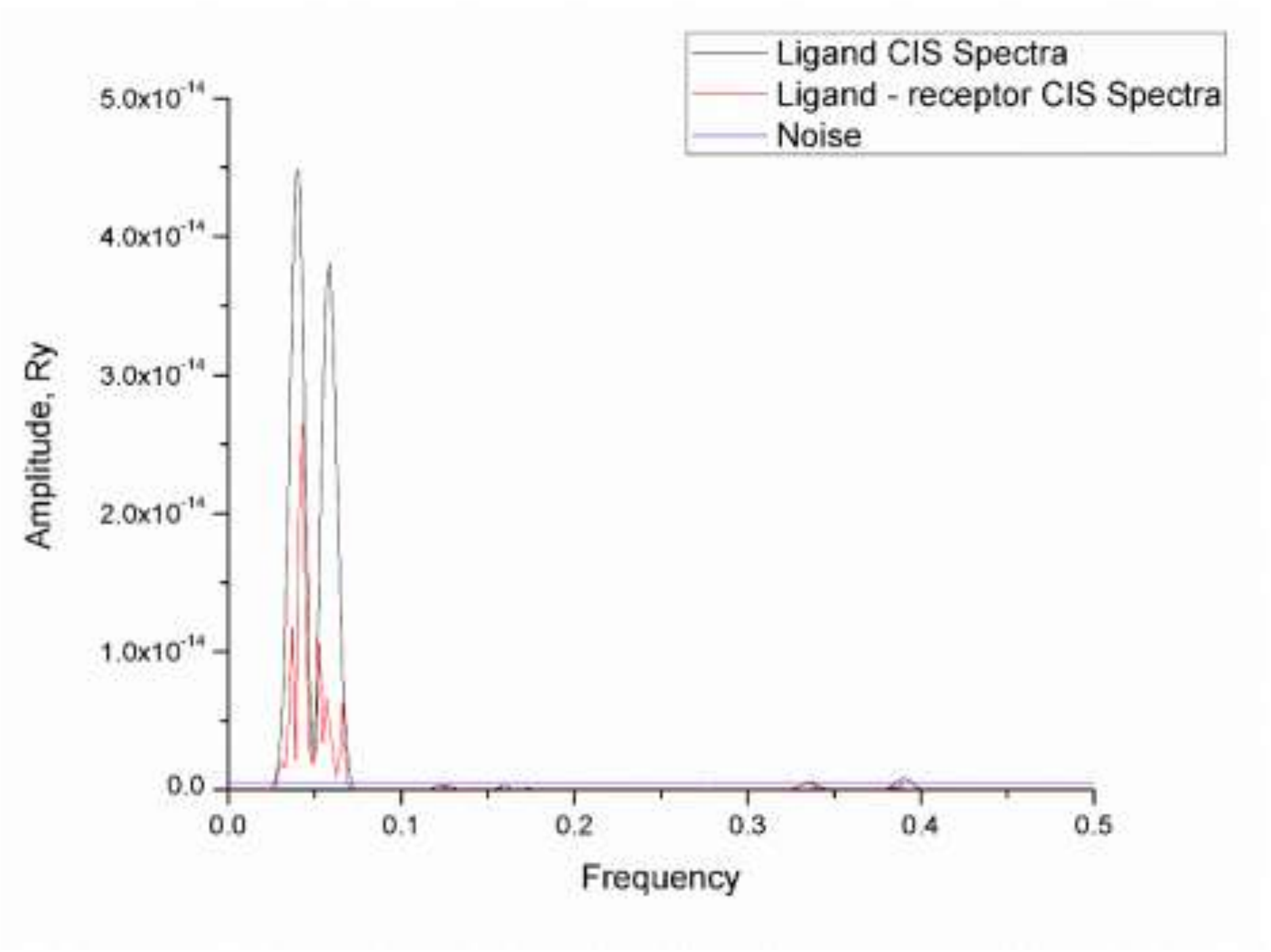

Training sets

| Positive set | KEGG Drug |
|---|---|
| 1 | D00255 |
| 2 | D00281 |
| 3 | D00283 |
| 4 | D00426 |
| 5 | D00437 |
| 6 | D00454 |
| 7 | D00509 |
| 8 | D00513 |
| 9 | D00604 |
| 10 | D00607 |

| Negative set | |
|---|---|
| 1 | D00049 |
| 2 | D00059 |
| 3 | D00079 |
| 4 | D00094 |
| 5 | D00095 |
| 6 | D00106 |
| 7 | D00110 |
| 8 | D00113 |
| 9 | D00136 |
| 10 | D00139 |

hsa 147

| Positive set | KEGG Drug |
|---|---|
| 1 | D00255 |
| 2 | D00281 |
| 3 | D00283 |
| 4 | D00426 |
| 5 | D00437 |
| 6 | D00454 |
| 7 | D00509 |
| 8 | D00513 |
| 9 | D00607 |
| 10 | D00609 |

| Negative set | |
|---|---|
| 1 | D00440 |
| 2 | D00443 |
| 3 | D00480 |
| 4 | D00493 |
| 5 | D00498 |
| 6 | D00514 |
| 7 | D00520 |
| 8 | D00521 |
| 9 | D00522 |
| 10 | D00523 |

hsa 148

| Positive set | KEGG Drug |
|---|---|
| 1 | D00095 |
| 2 | D00255 |
| 3 | D00281 |
| 4 | D00283 |
| 5 | D00426 |
| 6 | D00437 |
| 7 | D00454 |
| 8 | D00494 |
| 9 | D00503 |
| 10 | D00509 |

| Negative set | |
|---|---|
| 1 | D00440 |
| 2 | D00443 |
| 3 | D00480 |
| 4 | D00498 |
| 5 | D00499 |
| 6 | D00514 |
| 7 | D00520 |
| 8 | D00521 |
| 9 | D00522 |
| 10 | D00523 |

hsa 150

| Positive set | KEGG Drug |
|---|---|
| 1 | D00136 |
| 2 | D00255 |
| 3 | D00270 |
| 4 | D00281 |
| 5 | D00283 |
| 6 | D00332 |
| 7 | D00437 |
| 8 | D00454 |
| 9 | D00509 |
| 10 | D00513 |

Negative set

| | |
|---|---|
| 1 | D00394 |
| 2 | D00400 |
| 3 | D00410 |
| 4 | D00422 |
| 5 | D00440 |
| 6 | D00443 |
| 7 | D00480 |
| 8 | D00498 |
| 9 | D00499 |
| 10 | D00520 |

hsa 151

| Positive set | KEGG Drug |
|---|---|
| 1 | D00136 |
| 2 | D00255 |
| 3 | D00270 |
| 4 | D00281 |
| 5 | D00283 |
| 6 | D00437 |
| 7 | D00454 |
| 8 | D00509 |
| 9 | D00513 |
| 10 | D00563 |

Negative set

| | |
|---|---|
| 1 | D00760 |
| 2 | D00765 |
| 3 | D00769 |
| 4 | D00780 |
| 5 | D00845 |
| 6 | D00954 |
| 7 | D00965 |
| 8 | D00987 |
| 9 | D01071 |
| 10 | D01103 |

hsa 152

| Positive set | KEGG Drug |
|---|---|
| 1 | D00281 |
| 2 | D00509 |
| 3 | D00604 |
| 4 | D00606 |
| 5 | D00607 |
| 6 | D00609 |
| 7 | D00613 |
| 8 | D00996 |
| 9 | D01022 |
| 10 | D01603 |

| Negative set | |
|---|---|
| 1 | D01332 |
| 2 | D01346 |
| 3 | D01386 |
| 4 | D01441 |
| 5 | D01462 |
| 6 | D01652 |
| 7 | D01692 |
| 8 | D01699 |
| 9 | D01745 |
| 10 | D01782 |

hsa 153

| Positive set | KEGG Drug |
|---|---|
| 1 | D00095 |
| 2 | D00235 |
| 3 | D00255 |
| 4 | D00432 |
| 5 | D00437 |
| 6 | D00454 |
| 7 | D00483 |
| 8 | D00513 |
| 9 | D00598 |
| 10 | D00601 |

Negative set

| | |
|---|---|
| 1 | D00498 |
| 2 | D00499 |
| 3 | D00514 |
| 4 | D00520 |
| 5 | D00521 |
| 6 | D00522 |
| 7 | D00523 |
| 8 | D00525 |
| 9 | D00542 |
| 10 | D00559 |

hsa 154

| Positive set | KEGG Drug |
|---|---|
| 1 | D00095 |
| 2 | D00235 |
| 3 | D00255 |
| 4 | D00432 |
| 5 | D00437 |
| 6 | D00454 |
| 7 | D00483 |
| 8 | D00513 |
| 9 | D00598 |
| 10 | D00601 |

| Negative set | |
|---|---|
| 1 | D00301 |
| 2 | D00306 |
| 3 | D00336 |
| 4 | D00356 |
| 5 | D00364 |
| 6 | D00371 |
| 7 | D00380 |
| 8 | D00394 |
| 9 | D00400 |
| 10 | D00410 |

hsa 155

| Positive set | KEGG Drug |
|---|---|
| 1 | D00255 |
| 2 | D00432 |
| 3 | D00437 |
| 4 | D00454 |
| 5 | D00483 |
| 6 | D00513 |
| 7 | D00996 |
| 8 | D01390 |
| 9 | D01454 |
| 10 | D02066 |

| Negative set | |
|---|---|
| 1 | D00665 |
| 2 | D00666 |
| 3 | D00673 |
| 4 | D00674 |
| 5 | D00675 |
| 6 | D00676 |
| 7 | D00682 |
| 8 | D00683 |
| 9 | D00684 |
| 10 | D00687 |

hsa 1128

| Positive set | KEGG Drug |
|---|---|
| 1 | D00113 |
| 2 | D00232 |
| 3 | D00274 |
| 4 | D00283 |
| 5 | D00397 |
| 6 | D00454 |
| 7 | D00465 |
| 8 | D00494 |
| 9 | D00524 |
| 10 | D00525 |
| | D00540 |

| Negative set | |
|---|---|
| 1 | D00422 |
| 2 | D00674 |
| 3 | D00682 |
| 4 | D00683 |
| 5 | D00684 |
| 6 | D00687 |
| 7 | D00688 |
| 8 | D00760 |
| 9 | D00765 |
| 10 | D00769 |

hsa 1129

| Positive set | KEGG Drug |
|---|---|
| 1 | D00113 |
| 2 | D00397 |
| 3 | D00454 |
| 4 | D00494 |
| 5 | D01871 |
| 6 | D02070 |
| 7 | D02354 |
| 8 | D02361 |
| 9 | D00524 |
| 10 | D00540 |

| Negative set | |
|---|---|
| 1 | D01745 |
| 2 | D01782 |
| 3 | D01828 |
| 4 | D01891 |
| 5 | D01925 |
| 6 | D01964 |
| 7 | D01994 |
| 8 | D02007 |
| 9 | D02082 |
| 10 | D02147 |

hsa 1131

| Positive set | KEGG Drug |
|---|---|
| 1 | D00113 |
| 2 | D00232 |
| 3 | D00454 |
| 4 | D00494 |
| 5 | D01699 |
| 6 | D01871 |
| 7 | D02070 |
| 8 | D02354 |
| 9 | D02361 |
| 10 | D03621 |

| Negative set | |
|---|---|
| 1 | D00443 |
| 2 | D00480 |
| 3 | D00498 |
| 4 | D00499 |
| 5 | D00514 |
| 6 | D00520 |
| 7 | D00521 |
| 8 | D00522 |
| 9 | D00523 |
| 10 | D00525 |

hsa 1812

| Positive set | KEGG Drug |
|---|---|
| 1 | D00059 |
| 2 | D00110 |
| 3 | D00270 |
| 4 | D00283 |
| 5 | D00454 |
| 6 | D00493 |
| 7 | D00503 |
| 8 | D00560 |
| 9 | D00613 |
| 10 | D00790 |

| Negative set | |
|---|---|
| 1 | D00241 |
| 2 | D00295 |
| 3 | D00301 |
| 4 | D00306 |
| 5 | D00336 |
| 6 | D00356 |
| 7 | D00364 |
| 8 | D00371 |
| 9 | D00380 |
| 10 | D00394 |